\documentclass{aa}  

\bibpunct{(}{)}{;}{a}{}{,} % to follow the A&A style
\usepackage{graphicx}
\usepackage{txfonts}
\usepackage{placeins}

\usepackage{dsfont}

\usepackage{siunitx}

\DeclareSIUnit \h {\ensuremath{\mathit{h}}}
\DeclareSIUnit \parsec {pc}
\DeclareSIUnit \msol {\ensuremath{M_{\odot}}}
\DeclareSIUnit \au {AU}
\DeclareSIUnit \year {yr}

\newcommand{\rev}[1]{#1}
\newcommand{\revb}[1]{#1}

\newcommand{\myvec}[1]{\vec{#1}}

\newcommand{\mat}[1]{\tens{#1}}
\newcommand{\norm}[1]{\left\lVert {#1} \right\rVert}
\newcommand{\expect}[1]{\left\langle {#1} \right\rangle}
\newcommand{\expectgal}[1]{\left\langle {#1} \right\rangle_{\mathrm{g}} }
\newcommand{\trace}[1]{\mathrm{tr}\left( {#1} \right)}

\newcommand{\mul}[1]{\overset{(#1)}{\cdot}}
\newcommand{\myspan}[1]{\mathrm{span}\left( #1 \right)}

\newcommand{\citetgaus}[0]{\citet{stucker_2024_gaussian}}
\newcommand{\citepgaus}[0]{\citep{stucker_2024_gaussian}}

\newcommand{\quotes}[1]{``#1''}

\newcommand\superequiv{\mathrel{\rlap{\raisebox{\fontdimen22\textfont2}{$=$}}\raisebox{-0.5\fontdimen22\textfont2}{$ = $}}}

\begin{document} 
   \title{Probabilistic Lagrangian bias estimators and the cumulant bias expansion}

   \author{Jens Stücker
          \inst{1}
          \and
          Marcos Pellejero-Ibáñez \inst{2}
          \and
          Raul E. Angulo \inst{1,3}
          \and
          Francisco Maion
          \inst{1}
          \and
          Rodrigo Voivodic
          \inst{1}
          }

   \institute{Donostia International Physics Center (DIPC),
              Paseo Manuel de Lardizabal 4, 20018 Donostia-San Sebastian, Spain\\
              \email{jens.stuecker@univie.ac.at}
         \and
             Institute for Astronomy, University of Edinburgh, Royal Observatory, Blackford Hill, Edinburgh, EH9 3HJ , UK\\
             \email{mpelleje@ed.ac.uk}
         \and
             IKERBASQUE, Basque Foundation for Science, E-48013, Bilbao, Spain
             }

   \date{Received June 20, 2025; accepted May 1, 2025}

% \abstract{}{}{}{}{} 
% 5 {} token are mandatory
 
  \abstract
   {The spatial distribution of galaxies is a highly complex phenomenon currently impossible to predict deterministically. However, by using a statistical `bias' relation, it becomes possible to robustly model the average abundance of galaxies as a function of the underlying matter density field. Understanding the properties and parametric description of the bias relation is key to extract cosmological information from future galaxy surveys. Here, we contribute to this topic primarily in two ways: (1) We develop a set of \quotes{probabilistic} estimators for bias parameters using the moments of the Lagrangian galaxy environment distribution. These estimators include spatial corrections at different orders to measure bias parameters independently of the damping scale. We report robust measurements of a variety of bias parameters for haloes, including the tidal bias and its dependence with spin at a fixed mass. (2) We propose an alternative formulation of the bias expansion in terms of \quotes{cumulant bias parameters} that describe the response of the logarithmic galaxy density to large-scale perturbations. We find that cumulant biases of haloes are consistent with zero at orders $n > 2$. This suggests that: (i) previously reported bias relations at order $n > 2$ are an artefact of the entangled basis of the canonical bias expansion; (ii) the convergence of the bias expansion may be improved by phrasing it in terms of cumulants; (iii) the bias function is very well approximated by a Gaussian -- an avenue which we explore in a companion paper.
   }


   \maketitle
%
%________________________________________________________________

\section{Introduction}

\noindent Observations of the large-scale distribution of galaxies are among the most promising probes to accurately infer the cosmological parameters of our Universe. Past large-scale structure surveys have helped shaping our current model of the Universe by getting precise measurements of the angular diameter distance and growth rate at different epochs (see e.g. \citealt{eBOSS_2021}). Forthcoming surveys like the Dark Energy Instrument (DESI, \citealt{DESI}), or Euclid (\citealt{Euclid}) will measure sky positions and redshifts (both photometric and spectroscopic) of an unprecedented amount of galaxies. To reliably interpret these datasets it is of crucial importance to accurately model the spatial distributions of galaxies as a function of cosmology.

The evolution of the large-scale distribution of matter is predominantly driven by gravity and can be predicted very reliably through perturbation theory \citep[see][for a review]{Bernardeau_2002} or by gravity-only N-body simulations \citep[see e.g.][for a review]{AnguloHahn2022}. However, observed galaxies trace the matter distribution only in a `biased' way \citep{kaiser_1984, Mo_White_1996}. Optimally exploiting the information from galaxy surveys therefore requires not only accurately modelling gravity, but also to account for the formation and evolution of galaxies.

There are a variety of methodologies to model the formation of galaxies. The most detailed approach involves employing hydrodynamical simulations to explicitly follow gas dynamics and the formation and evolution of stars, black holes and galaxies (see e.g. \citealt{Vogelsberger_2014,Dubois_2014,Schaye_2015,Dave_2019}). In principle, such simulations would be the ideal method to account for the formation of galaxies if the underlying physics was modelled reliably and at an affordable cost. However, in practice these simulations are limited in volume and number due to their substantial computational requirements. Additionally, they necessitate the assumption of sub-grid physics e.g. to model the unresolved formation of stars, the growth of active galactic nuclei and their feedback processes. The results of such simulations depend strongly on the associated parameters and are therefore severely limited in their predictive ability \citep[e.g.][]{Genel_2019, camels_2021}.

To overcome the limitations of hydrodynamic simulations, a more agnostic approach can be pursued. Semi-analytic models, for example, follow a number of empirical and semi-empirical relations that provide more versatility to the modelling of galaxy formation physics (see e.g. \citealt{kauffmann_1999, Henriques_2015,Stevens_2016,Lacey_2016,Croton_2016,Lagos_2018}). An even more agnostic strategy involves techniques such as Sub-Halo Abundance Matching (see e.g. \citealt{Conroy_2006,Reddick_2013,Chaves-Montero_2016,Lehmann_2017,Dragomir_2018,Contreras_2021, ortega_2024}) or Halo Occupation Distribution (see e.g.\citealt{peacock_2000,Berlind_2002,Zheng_2005,Cacciato_2012,Salcedo_2022a}), which rely solely on gravity simulations and fundamental assumptions about populating collapsed structures (haloes and sub-haloes) based on their mass. These assumptions must be sufficiently flexible to accommodate various galaxy formation scenarios. Nonetheless, these techniques also have computational constraints, and the principles they rely on may oversimplify reality, necessitating extensions to account for dependencies beyond merely the mass of the collapsed structure (a phenomenon known as assembly bias, \citealt{Gao_2005,Weschler_2006,gao_2007,croton_2007,Dalal_2008,Faltenbacher_2010,Montero-Dorta_2017,Zehavi_2018,Ferreras_2019,Sato-Polito_2019,Tucci_2021,Salcedo_2022b, chaves_2023}, among others).

The most general approach currently known is a perturbative expansion of galaxy bias (see \citealt{Desjacques_2018} for a review). In this framework, the evolution of matter is assumed to be modelled accurately through purely gravitational effects whereas the galaxy field selectively populates the matter field. The galaxy field is then written in a perturbative manner as a function of the properties of the underlying matter distribution. In Eulerian bias schemes, the galaxy number density is expanded in terms of the final properties of the density field whereas in Lagrangian schemes this is done in terms of the initial properties of the linear density field. This approach offers great flexibility, allowing a single model to describe biased tracers with vastly different properties. Bias approaches based on effective field theory aim to describe the clustering behavior of galaxies down to $k\approx 0.2h/$Mpc \citep{Baumann_2012,Baldauf_2016,Vlah_2016}. They have been used extensively to extract robust cosmological constraints from surveys \citep{Ivanov_2020,d'Amico_2020,Colas_2020,Nishimichi_2020,Chen_2020,Philcox_2022}.

There exist different options for defining the biasing scheme. Traditionally, perturbation theory is used to describe the underlying gravitational evolution of the density field whereas in the recently developed \quotes{hybrid approaches} the gravitational evolution is treated exactly through N-body simulations \citep{Modi_2020,Kokron_2021,Zennaro_2023,Pellejero_2022,Pellejero_2023,DeRose_2023}.  However, in all bias methods the galaxy number density is expanded in a perturbative series with a set of free coefficients known as \quotes{bias parameters}. These parameters can be interpreted as the response of the galaxy number density to perturbations of the density field. While these parameters exhibit some physical insights \citep{lazeyras_2016,lazeyras_2019,Barreira_2021}, they are generally treated as nuisance parameters which are marginalized over to extract the cosmological information of interest. 

A quantitative understanding of the bias parameters is important for analysing future galaxy surveys. On the one hand, this is necessary to determine how well the bias expansion converges. Whether the truncation at a given order gives accurate results, depends on higher order terms being sufficiently small that they can be neglected. On the other hand, it is of significant interest to limit the prior volume that is used when fitting to cosmological surveys to maximize the extracted cosmological information. In particular, it may significantly benefit an analysis to fix bias parameters to so called \quotes{coevolution relations} that relate the parameters to each other. Therefore, significant effort has been put to measuring bias parameters and to constrain coevolution relations in simulations. 

One of the most accurate methods for measuring bias parameters is through separate universe simulations \citep{li_2014, wagner_2015}. In these, it is directly tested how the number of tracers responds to an increase in the large-scale density. This technique has been used to constraint density bias parameters of haloes to very high accuracy \citep{lazeyras_2016}. Further, similar ideas have been used to measure the response to non-Gaussian perturbations \citep{Barreira_2020} and to the changes in the Laplacian of the density field \citep{Lazeyras_2018}. Other bias parameters have been constrained through different methods. For example, the tidal bias has been measured through Fourier space correlations \citep{modi_2017, Lazeyras_2018} or all bias parameters can be constrained simultaneously through forward modeling of the power spectrum \citep[e.g.][]{zennaro_2022} or the field-level galaxy distribution \citep[e.g.][]{lazeyras_2021}. 

Another successful avenue of measuring and understanding the behavior of bias parameters is through peak theory \citep{bardeen_1986} where the number density of peaks of the initial density field is investigated as a function of a Lagrangian smoothing scale. If galaxies (and haloes) correspond to peaks of the initial density field, then bias parameters can be predicted through the response of peaks to large scale perturbations. While the mapping between structures and peaks bears significant uncertainty due to the necessary inclusion of a heuristic smoothing scale to define peaks and significant difficulty with treating the cloud-in-cloud problem \citep{bardeen_1986}, peak theory has proven as a useful tool in various aspects of biasing \citep{Desjacques_2018}. For example it predicts the importance of a Laplacian bias parameter \citep{bardeen_1986}, the scale dependence of the bias in the matter correlation function \citep{desjacques_2010, paranjape_2013a, paranjape_2013b} and non-zero velocity bias \citep{baldauf_2015}. All of these predictions have been established through measurements in simulations. Further, peak theory and excursion set approaches have motivated the possibility of measuring biases through correlations between density and haloes in Lagrangian space \citep{musso_2012, paranjape_2013a, paranjape_2013b, biagetti_2014}. In particular, \citet{paranjape_2013b} have shown that it is possible to recover precise measurements of (scale-independent) large scale bias parameters by mapping the scale-dependence of `naive' bias parameters under the assumption of Peak theory to their large scale limit.

In this paper, we propose a novel approach for accurately measuring Lagrangian bias parameters from simulations. In this approach we consider the distribution of the linear density field at the initial (Lagrangian) locations of galaxies -- which we call the \quotes{galaxy environment distribution}. We adopt a probabilistic approach to model this galaxy environment distribution and show that large-scale bias parameters have a simple relation to the moments of this distribution. We use this to derive estimators of the bias parameters that can \rev{take into account spatial corrections at any order to practically eliminate any scale dependence}. We derive such estimators for a large variety of bias parameters and provide corresponding measurements for the case of haloes. 
\revb{Operationally, our method is similar to aforementioned peak theory approaches. However, it is by construction independent of the assumption that galaxies form in smoothed Lagrangian density peaks and it can be used to correct bias measurements at arbitrary high spatial orders.}

Further, we propose a new set of \quotes{cumulant bias parameters} which are defined as the response of the logarithm of the galaxy number density to perturbations in the linear field. We show that these parameters have significantly improved properties when compared to their canonical counterparts. For the case of haloes we find that cumulant biases of order three and higher are consistent with zero. Therefore, the canonical coevolution relations for haloes at order order $3$ and beyond appear primarily as an effect of a sub-optimal parameterization. We suggest that rephrasing the bias expansion in terms of the cumulant bias parameters can significantly enhance its convergence.

The vanishing of cumulants for haloes at order three and beyond implies that the bias function can be well approximated through a Gaussian. We will explore such a Gaussian bias model in a companion paper \citetgaus.

This article is organized as follows: In Section \ref{sec:theory} we explain the probabilistic approach to measure Lagrangian bias parameters of scalar variables like the density and the Laplacian and we introduce the concept of the cumulant bias expansion. In Section \ref{sec:scalar_measurements} we present measurements of the corresponding parameters and demonstrate how cumulant bias parameters exhibit several practical advantages. In Section \ref{sec:tensor_theory} we show how to generalize the concept of the probabilistic estimators for tensorial bias variables, like for example the tidal bias $b_{K^2}$. In Section \ref{sec:tensorbiasmeas} we provide measurements of a few select tensorial quantities. In Section \ref{sec:relevance} we briefly discuss the quantitative importance of different bias terms. Finally, in Section \ref{sec:conclusions} we summarize the benefits of our novel estimators and we discuss under which circumstances it is advantageous to phrase the bias expansion in terms of the cumulant bias parameters.

\section{Theory} \label{sec:theory}

In this section, we introduce the necessary theory and (1) relate bias parameters to properties of the galaxy environment distribution, (2) express the moment and cumulant generating functions of the galaxy environment distribution in terms of the large-scale bias function, (3) introduce the concept of a cumulant bias expansion, (4) show how to derive estimators for both canonical and cumulant bias parameters. 

\subsection{Definitions}
The considerations here are based on the idea of the peak-background split (PBS) which states that `a long-wavelength density perturbation acts like a local modification of the background density for the purposes of the formation of halos and galaxies' \citep{kaiser_1984, bardeen_1986, Desjacques_2018}. Bias parameters describe the response of the galaxy number to such long-wavelength perturbations. \rev{In this article, we will exclusively focus on Lagrangian bias parameters as the response to perturbations in the linear density field.}

An exact implementation of the PBS is given by the separate universe approach. In this approach one considers some universe \rev{with background density $\rho_{\mathrm{bg},0}$ in which }\revb{galaxies form with an average number density $n_{g,0}$}. If one were to increase the \rev{initial} background density of the universe by a \rev{linear} amount $\delta_0$ (e.g. in a separate universe simulation, \citealt{frenk_1988, li_2014, wagner_2015}), then in the new universe \revb{galaxies form with a different average number density} $n_{g}(\delta_0)$.\footnote{\revb{Making this measurement in simulations may involve some uncertainty due to finite-size effects and cosmic variance. For simplicity, here we consider the limiting case of an infinite measurement volume so that the relation is completely deterministic.}} We call their ratio
\begin{align}
    F(\delta_0) &= \frac{n_g(\delta_0)}{n_{g,0}} = 1 + b_1 \delta_0 + \frac{1}{2} b_2 \delta_0^2 + \cdots + \frac{1}{n!}b_n\delta_0^n + \cdots
\end{align}
the \quotes{true bias function} or the \quotes{separate universe bias function}. \rev{Here, $\delta_0$ refers to a contrast in linear densities so that the new background density corresponds to}
\begin{align}
    \rho_{\mathrm{bg}} \approx \rho_{\mathrm{bg,0}} (1 + \delta_0 D(a))  \text{\quad for } a \rightarrow 0
\end{align}
\rev{where $D$ is the linear growth factor normalized to $D(a=1) = 1$.} $F$ can directly be measured with separate universe simulations \citep[e.g.][]{lazeyras_2016,baldauf_2016b} and we will refer to the coefficients of the indicated expansion as the \quotes{canonical bias parameters} or just \quotes{the bias parameters}:
\begin{align}
    b_n &=  \left. \frac{\partial^{n} F(\delta_0)}{\partial \delta_0^{n}} \right|_{\delta_0=0} 
    \label{eqn:pbs_bias}
\end{align}
Therefore, the bias parameters physically describe the response of the galaxy density to small perturbations at infinitely large scales. In this article we want to investigate galaxy bias from a probabilistic perspective.

We consider an infinitesimally small Lagrangian volume element which nothing is known about except the linear density contrast $\delta$ smoothed at some scale (and possibly other features of the linear field like the Laplacian $L$ or the tidal field). Neglecting primordial non-Gaussianity, the density contrast follows a Gaussian distribution 
\begin{align}
    p(\delta) &= \frac{1}{\sqrt{2 \pi} \sigma} \exp \left( - \frac{\delta^2}{2 \sigma^2} \right) \,\,. \label{eqn:p_of_delta}
\end{align}
For simplicity, we will assume throughout this article that the smoothed density contrast is defined with a sharp $k$-space filter. Most considerations here translate in a simple manner to differently filtered cases, but some additional care must be taken due to the more complicated correlation between large and small scales. This is discussed in more detail in Appendix \ref{app:filtering}.

\rev{If a sufficiently small volume element is considered, then it is only possible to have either 0 or 1 galaxy. We may therefore speak of a binary event `g' that a volume contains a galaxy. } We call the average probability that a galaxy forms in such a volume element ``$p(\mathrm{g})$'' and we call the conditional probability, given the knowledge of the linear density contrast, ``$p(\mathrm{g} | \delta)$''.\footnote{\rev{Our definitions here should not be confused with those in works that consider the 2d joint distribution function of the smoothed galaxy density $n_{\mathrm{g}}$ and the smoothed matter density $p(n_{\mathrm{g}}, \delta)$ and their conditional $p(n_{\mathrm{g}}| \delta)$ }  \citep{dekel_1999}. \rev{Describing these is significantly more complicated, as they are two-dimensional and they contain stochastic contributions with a complicated dependence on two smoothing operations (for defining the matter contrast and the galaxy density respectively).}} The excess probability
\begin{align}
    f(\delta) := \frac{p(\mathrm{g} | \delta)}{p(\mathrm{g})}
\end{align}
is parameterized through a function $f(\delta)$ which we will refer to as the \quotes{scale dependent bias function} or just the \quotes{bias function} throughout this article. The bias function depends in a predictable manner on the variance of $\delta$ at the considered scale, as we will show later.

Since densities at different scales add up linearly, a separate-universe style modification of the large-scale density contrast from 0 to $\delta_0$ will immediately translate to a modification of the linear density in our volume element $\delta \rightarrow \delta + \delta_0$. Therefore, $F$ and $f$ should be related through
\begin{align}
    F(\delta_0) &=  \langle f(\delta + \delta_0) \rangle \label{eqn:ngversusbias}
\end{align}
where the angled brackets indicate an expectation value taken over the Lagrangian volume \citep[see also][]{Desjacques_2018}. The relation indicates that in a separate universe experiment the number of galaxies should change according to the average change in probability of forming galaxies when changing the linear density contrast everywhere in space. Therefore, the canonical bias parameters are given in terms of the scale dependent bias function as
\begin{align}
    b_n &= \expect{ \frac{\partial^n f}{\partial \delta^n} } \,\,.\label{eqn:bn_renormalized}
\end{align}
We will see later that equation \eqref{eqn:ngversusbias} holds only approximately. This is so, since at any finite smoothing scale, the density contrast $\delta$ is correlated with other variables (e.g. the Laplacian) so that a change in the small scale density contrast at the location of all galaxies is not exactly equivalent to the \quotes{pure} density change in separate universe experiments. This introduces scale-dependencies that can be accounted for.

\begin{figure*}
    \centering
    \includegraphics[width=0.8 \textwidth]{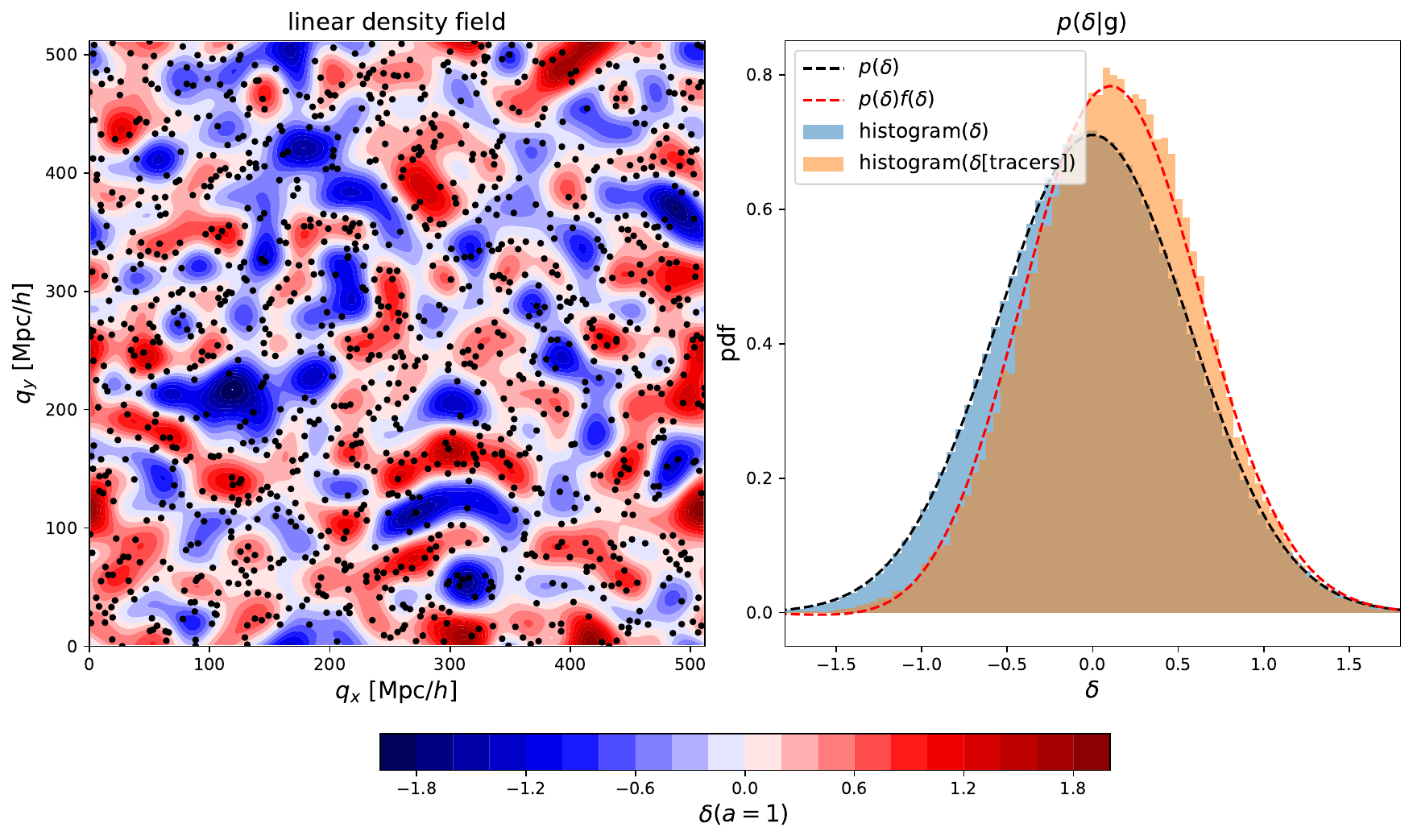}
    \caption{Illustration of the inference of the halo environment distribution $p(\delta|g)$. Left: we trace back galaxies to their origin in Lagrangian space (marked as black dots) and infer the (smoothed) linear density field $\delta$ at their Lagrangian locations. Right: The environment distribution (orange histogram) is given by the distribution of $\delta$ at the galaxy locations which is notably biased relative to the matter distribution $p(\delta)$ (blue histogram and a Gaussian represented as dashed line). The galaxy environment distribution is well approximated through $p(\delta) f(\delta)$ where here $f(\delta)$ is a quadratic polynomial bias function. }
    \label{fig:galaxy_env_distribution}
\end{figure*}

Finally, we introduce one further object which we call the \quotes{galaxy environment distribution} $p(\delta |g)$ or \quotes{halo environment distribution} when we talk specifically about haloes. This quantifies the probability of the linear density contrast in an infinitesimal volume element, given that there is a galaxy at the considered location. This function can easily be measured as a histogram of the linear density at the Lagrangian locations of galaxies as is illustrated in Figure \ref{fig:galaxy_env_distribution} -- where we have used a damping scale of $k_{\mathrm{d}} = 0.15 h \mathrm{Mpc}^{-1}$ here\footnote{Corresponding approximately to a sharp truncation scale in Fourier space, as explained in Section \ref{sec:measure_techinque}.}, leading to $\sigma = 0.56$.

Through Bayes' theorem, the relation between the galaxy environment distribution and the bias function is given by
\begin{align}
    p(\delta | g) &= \frac{p(\delta \cap g)}{p(g)} \\
                  &= \frac{p(g | \delta) p(\delta)}{p(g)} \\
                  &= f(\delta) p(\delta) \,\,.
\end{align}
This means for example, that the bias function $f(\delta)$ can be investigated in a non-parametric way by measuring $p(\delta | g)$ and dividing by the Gaussian background distribution $p(\delta)$ as we will investigate in the companion paper \citetgaus. 

In this section, we will use probability theory to investigate properties of the galaxy environment distribution. In particular, we will show that bias parameters are simply related to the moments of this distribution and that probability theory motivates the usage of better behaved \quotes{cumulant bias parameters}.

\subsection{Bias estimators}
Following up on equation \eqref{eqn:ngversusbias} we can write
\begin{align}
    F(\delta_0) &=  \langle f(\delta + \delta_0) \rangle \nonumber \\
                                  &= \int_{-\infty}^\infty  p(\delta) f(\delta + \delta_0) \mathrm{d} \delta  \nonumber \\
                                  &= \int_{-\infty}^\infty  p(\delta - \delta_0) f(\delta) \mathrm{d} \delta \,\,.  \label{eqn:ng_from_pf}
\end{align}
Where in the last line we have made the substitution $\delta \rightarrow \delta + \delta_0$, and $\delta$ represents the linearly extrapolated overdensity, hence it can assume values between negative and positive infinity. The latter equation can be interpreted as an alternative perspective onto the separate universe experiment: When increasing the background density, the probability of environments is changed by a factor $p(\delta - \delta_0) / p(\delta)$ whereas the likelihood of forming a galaxy when presupposing a given environment  stays constant.

Now, combining equations \eqref{eqn:pbs_bias} and \eqref{eqn:ng_from_pf}, we can evaluate the bias parameters as
\begin{align}
    b_n &= \left. \frac{\partial^n}{\partial \delta_0^n} \int_{-\infty}^\infty  p(\delta - \delta_0) f(\delta) \mathrm{d} \delta  \nonumber   \right|_{\delta_0 = 0} \\
        &= (-1)^n \int_{-\infty}^\infty  p^{(n)}(\delta) f(\delta) \mathrm{d} \delta   \nonumber \\
        &= (-1)^n \int_{-\infty}^\infty  \frac{p^{(n)}(\delta)}{p(\delta)} p(\delta |g) \mathrm{d} \delta   \nonumber \\
        &= (-1)^n \expectgal{ \frac{p^{(n)} (\delta)}{p(\delta)} } \label{eqn:bn_from_pderiv}
\end{align}
where the angled brackets with a `g' subscript indicate an expectation value evaluated over the locations of galaxies (rather than all of Lagrangian space). Further, we can use that $p$ is a Gaussian distribution for which derivatives are given by the (probabilisist's) Hermite polynomials:
\begin{align}
    \frac{\partial^n }{\partial \delta^n} \exp \left(- \frac{\delta^2}{2 \sigma^2}\right) &= (-1)^n \sigma^{-n} \exp \left(- \frac{\delta^2}{2 \sigma^2}\right) H_n \left( \delta / \sigma \right)
\end{align}
so that the bias estimators read
\begin{align}
    b_{n, \mathrm{so0}} &=  \expectgal{ \frac{H_n \left( \delta / \sigma \right)}{\sigma^n} } \,\,.\label{eqn:bn_o0}
\end{align}
\rev{Here the subscript `so0' indicates that these estimators, are of `spatial order 0', i.e. that they do not include corrections from higher spatial derivatives like e.g. the Laplacian -- as we explain in more detail later.} This expression has already been used by \citet{paranjape_2013a, paranjape_2013b} to measure bias parameters. It was motivated by its emergence in excursion set frameworks \citep{musso_2012} and peak statistics \citep{paranjape_2012}, but it is clearly also valid outside such frameworks, requiring only to assume the PBS. On a related note, it is worth mentioning that it has already been proposed by \citet{szalay_1988} to expand the field in terms of Hermite polynomials.

The estimators for the first four bias parameters read:
\begin{align}
    b_{1,\mathrm{so0}} &= \expectgal{ \frac{\delta}{\sigma^2} } \label{eqn:b1_o0} \\
    b_{2,\mathrm{so0}} &= \expectgal{ \frac{\delta^2 - \sigma^2}{\sigma^4} }\\
    b_{3,\mathrm{so0}} &= \expectgal{ \frac{\delta^3 - 3 \delta \sigma^2}{\sigma^6} } \\
    b_{4,\mathrm{so0}} &= \expectgal{ \frac{\delta^4 - 6 \delta^2 \sigma^2 + 3 \sigma^4}{\sigma^8} } \,\,. \label{eqn:b4_o0}
\end{align}

\subsection{The moment generating function}
Equation \eqref{eqn:bn_o0} shows that there exists a simple relation between moments of the galaxy environment distribution and the bias parameters. We can show this in a very general manner. Expanding the Gaussian background distribution we obtain
\begin{align}
    p(\delta - \delta_0) 
    &= \frac{1}{\sqrt{2 \pi} \sigma} \exp \left( - \frac{\delta^2 - 2\delta \delta_0 + \delta_0^2}{2 \sigma^2} \right) \nonumber \\
    &= p(\delta) \exp \left(\frac{\delta \delta_0}{\sigma^2} \right) \exp \left( - \frac{\delta_0^2}{2 \sigma^2} \right) \nonumber
\end{align}
and inserting this into equation \eqref{eqn:ng_from_pf} yields
\begin{align}
    F(\delta_0) &= \exp \left( - \frac{\delta_0^2}{2 \sigma^2} \right) \int_{-\infty}^\infty  f(\delta) p(\delta) \exp \left(\frac{\delta \delta_0}{\sigma^2} \right)  \nonumber \mathrm{d} \delta \\
    &= \exp \left( - \frac{1}{2} t^2 \sigma^2 \right) \int_{-\infty}^\infty  p(\delta |g) \exp \left(t \delta \right)  \nonumber \mathrm{d} \delta \\
    &= \exp \left( - \frac{1}{2} t^2 \sigma^2 \right) \expectgal{\exp \left(t \delta \right)} 
\end{align}
where we have labeled $t = \delta_0 / \sigma^2$. The last term can be identified with the moment generating function
\begin{align}
    M_g(t) = \expectgal{\exp \left(t \delta \right)}
\end{align}
of the galaxy environment distribution. Therefore, the moment generating function of the galaxy environment distribution and the separate universe bias function can be directly converted into each other
\begin{align}
    M_g(t) &= F(t \sigma^2) \exp \left(\frac{1}{2} t^2 \sigma^2 \right)  \label{eqn:moment_ng}
\end{align}
and the relation between the moments and the bias parameters as in equation \eqref{eqn:bn_o0} can equivalently be found by taking derivatives of the moment generating function
\begin{align}
    \mu_n &:= \expectgal{\delta^n} = \left. \frac{\partial}{\partial t^n} M_g(t) \right|_{t=0}  \,\,. \label{eqn:moments}
\end{align}
It is worth noting that this result may be related to the considerations in \citet{White_1979} where for the case of discrete tracers, the moment generating function of the galaxy count frequency distribution is expressed through the void probability function.
\subsection{Cumulant bias parameters}
We may further ask, how the bias parameters relate to the cumulants of the galaxy environment distribution. In probability theory the value of cumulants are generally thought to characterize a distribution more independently than its moments. For example, if a distribution has a large first moment $\expect{x}$, then we should also expect that it has large second and third moments $\expect{x^2}$ and $\expect{x^3}$. However, the first cumulant of a distribution $\expect{x}$ says very little about the second and third cumulants, $\expect{(x - \expect{x})^2}$ and $\expect{(x - \expect{x})^3}$ respectively. For example, for a Gaussian distribution the mean is independent of its variance and the third and higher order cumulants are actually zero.

The cumulant generating function is defined as
\begin{align}
    K_g(t) &= \log M_g(t) \nonumber \\
         &= \frac{1}{2} t^2 \sigma^2 + \log F(t \sigma^2) \,\,.  \label{eqn:cumulant_ng}
\end{align}
Cumulants of the galaxy environment distribution are simply given by the derivatives of the cumulant generating function
\begin{align}
    \kappa_n &= \left. \frac{\partial K_g}{ \partial t^n} \right|_{t=0} \,\,.\label{eqn:cumulants}
\end{align}
Therefore, they evaluate to
\begin{align}
    \kappa_n &= \begin{cases}
        \beta_n \sigma^{2n} & \mathrm{\quad if \quad} n \neq 2  \\
        \beta_2 \sigma^4 + \sigma^2 & \mathrm{\quad if \quad} n = 2
        \label{eqn:cum_to_beta}
    \end{cases}
\end{align}
where we have defined
\begin{align}
    \beta_n &= \left.  \frac{\partial^n}{\partial \delta_0^n}  \log F(\delta_0) \right|_{\delta_0=0} \label{eqn:cum_pbs_bias}. % \frac{n_g(\delta_0)}{n_{g,0}}
\end{align}
Notice that whether $\beta_2$ is above or below zero is a direct indication of whether the variance of the halo environment distribution is larger or smaller than that of the background. We will refer to these parameters as \quotes{cumulant bias parameters} and they are directly related to the canonical bias parameters $b_n$. Comparing their definition to the canonical bias parameters in equation \eqref{eqn:pbs_bias} shows that they relate to each other exactly in the same way that cumulants relate to moments (compare equations \eqref{eqn:moments} and \eqref{eqn:cumulants}). For example, at the first four orders we have
\begin{align}
    \beta_{1} &= b_{1} \label{eqn:beta1_of_b} \\
    \beta_{2} &= b_{2} - b_{1}^{2} \\
    \beta_{3} &=  b_{3} - 3 b_{1} b_{2}  + 2 b_{1}^{3} \label{eqn:beta3_of_b} \\
    \beta_{4} &= b_{4} - 4 b_{1} b_{3} - 3 b_{2}^{2} + 12 b_{1}^{2} b_{2}  - 6 b_{1}^{4} \,\,. \label{eqn:beta4_of_b}
\end{align}

\subsection{Interpretation}
Given a set of bias parameters, these relations allow us to directly find the cumulants of the galaxy environment distribution, or alternatively, they allow us to infer canonical bias parameters by measuring cumulants of the galaxy environment distribution. 

\rev{However, it is also possible to phrase the bias expansion directly in terms of the cumulant bias parameters.}
\begin{align}
    \log F &= \beta_1 \delta_0 + \frac{1}{2} \beta_2 \delta_0^2 + \frac{1}{6} \beta_3 \delta_0^3 + ... \label{eqn:logF_expansion}
\end{align}
There are several reasons to believe that they may form a better set of parameters than the canonical bias parameters:
\begin{itemize}
    \item The cumulant bias parameters are the derivatives of the logarithm of the galaxy density. Therefore, they presuppose the positivity of \rev{$F$} and may be better behaved, especially in low density regions. 
    \item The canonical bias parameters behave similarly to moments of the galaxy environment distribution. \rev{For example}, one may expect that if $b_1$ is large, automatically $b_4$ will also be large. On the other hand, we may expect $\beta_4$ to be rather independent of the value of $\beta_1$ -- just as cumulants are relatively independent of each other.
    \item If the galaxy environment distribution has the form of a Gaussian then $\beta_n = 0$ for all $n > 2$. On the other hand all $b_n$ would be non-zero in this case. Therefore, if all $\beta_n$ beyond degree 2 are small this motivates the usage of a Gaussian bias model \rev{-- where both $\log F$ and $\log f$ are quadratic polynomials}. We will show that this is indeed the case for haloes.
\end{itemize}

Further, we note that the cumulant bias parameters and the probabilistic considerations in this section motivate novel approaches to parameterize the bias function at finite smoothing scales. There are several possible well motivated approaches and we will just mention them here briefly, but we will leave a more thorough investigation to future studies. 

The first and most straight-forward proposition is to assume an expansion of the logarithm of the bias function, leading to an exponential of a polynomial
\begin{align}
    \log f(\delta) &= \gamma_0 + \gamma_1 \delta + \gamma_2 \delta^2 + \gamma_3 \delta^3 + ...\\
    f(\delta) &= \exp \left(\gamma_0 + \gamma_1 \delta + \gamma_2 \delta^2 + \gamma_3 \delta^3 + ... \right) \,\,.
\end{align}
Such a bias expansion would have to be truncated at an even order (if $n\geq 2$) and the highest order coefficient must be negative to guarantee a well normalized probability distribution. By definition this guarantees the positivity of the bias function which is arguably a desirable property -- especially if one intended to create mocks from a bias model. Further, the resulting distribution would be part of the exponential family which guarantees several desirable properties. The main limitation is that at orders $n > 2$ it might be difficult to properly renormalize the expression analytically\footnote{That means to write it in terms of scale independent bias parameters.}. 

This choice is the only general choice we know of that guarantees $f(\delta) > 0$ for all values of $\delta$ even if the expansion is truncated. However, there are a few other noteworthy options that do not fulfill this constraint. One option is motivated by the intriguing simplicity of the relation between cumulant generating function \eqref{eqn:cumulant_ng} and \rev{$F$}. This motivates to define the bias function directly through the cumulant generating function
\begin{align}
    K_g(t) &= K_{\mathrm{m}}(t) + \sum_n \frac{1}{n!} \beta_n t \sigma^{2n} \label{eqn:kumulant_expansion}
\end{align}
where $K_{\mathrm{m}}(t)$ is the cumulant generating function of the matter distribution. This general Ansatz might also be feasible in Eulerian space, but for the Lagrangian case it is simply $K_{\mathrm{m}}(t) = \frac{1}{2} t \sigma^2$.

If the expansion is truncated at an order $n = 2$ then this corresponds to a well defined Gaussian bias model, but if it is truncated at a higher order then it is not easy to find an analytical form of $f(\delta)$. However, a numerical expression can be obtained through a Fourier transformation. Unfortunately, such a truncation at higher orders does not lead to a well defined probability density, but instead may have negative function values. In fact the cumulant generating function does not yield a positive pdf for any finite order polynomial of degree $n>2$ \citep{Lukacs_1970}. However, this does not disfavour this approach over the canonical bias expansion which also yields negative pdf.

Further, we note that it is an option to phrase the bias expansion relative to a Gaussian distribution that has the correct cumulants up to order two:
\begin{align}
    p(\delta|g) &= \frac{1}{\sqrt{2 \pi} \sigma_g} \exp \left( - \frac{(\delta - \mu_g)^2}{2 \sigma_g^2} \right) (1 + \gamma_3 \delta^3 + \gamma_4 \delta^4 + ...)
\end{align}
where $\mu_g = \kappa_1 = \beta_1 \sigma^2$ and $\sigma_g^2 = \kappa_2 = \beta_2 \sigma^4 + \sigma^2$ and where the $\gamma_n$ start at order three, since the lowest two orders are already specified through the Gaussian. This type of expansion is similar to a truncated Edgeworth series (which is an expansion to approximate distributions with given cumulants). 

Finally, we note that it is possible to use the cumulant bias parameters also in a classical polynomial bias expansion. For example a third order expansion that only uses two parameters, assuming $\beta_3 = 0$  would read
\begin{align}
    f(\delta) %&= \beta_1 \delta + (\beta_2 + \beta_1^2) (\delta^2 - \sigma^2) + (\beta_3 + 3 \beta_1 (\beta_2 + \beta_1^2) - 2 \beta_1^3) (\delta^3 - 3 \delta \sigma^2) \\
              &= \beta_1 \delta + (\beta_2 + \beta_1^2) (\delta^2 - \sigma^2) + (3 \beta_1 \beta_2 - \beta_1^3) (\delta^3 - 3 \delta \sigma^2) + ...
\end{align}
which can be understood as lower order terms predicting the likely behavior of higher order terms. 

\subsection{Multivariate estimators}
We have so far only considered the bias function under the assumption that the only known aspect of the environment is the density $\delta$ at the considered location. However, other aspects of the environment may be known, such as the Laplacian
\begin{align}
    L &= \nabla^2 \delta
\end{align}
 or higher order derivatives of the density field
\begin{align}
    P &= \nabla^4 \delta \label{eqn:P_fourthderiv}
\end{align}
or the tidal field. Since the tidal field is inherently a tensorial quantity, its treatment is slightly more complicated and we will therefore discuss it in a dedicated manner in Section \ref{sec:tensor_theory}.

Assuming that we are dealing only with scalar quantities that describe the environment, we may summarize the environment through a single vector $\myvec{x}$ -- for example,  if we consider the density and Laplacian as variables we have $\myvec{x} = (\delta, L)^T$. The majority of the relations that we have derived in the previous section hold in a similar manner for the multivariate case as well. We will not derive all of these again, since their derivation follows in almost complete analogy, but we list all important relations.

Assuming that all considered environment variables follow from linear operators on the Gaussian random field, their distribution is given by a multivariate Gaussian
\begin{align}
    p(\myvec{x}) &= \frac{1}{2 \pi \sqrt{\det(\mat{C})}} \exp \left(- \frac{1}{2} \myvec{x}^T \mat{C}^{-1} \myvec{x}  \right) \,\,. \label{eqn:multivariate_gaussian}
\end{align}
For example, for the case of $\myvec{x} = (\delta, L)^T$ \rev{-- as is in particular well motivated by the peak model} \citep{bardeen_1986} \rev{--} we have the covariance matrix
\begin{align}
\mat{C}_{\delta,L} &= \left[\begin{matrix}\sigma_{0}^{2} & - \sigma_{1}^{2}\\- \sigma_{1}^{2} & \sigma_{2}^{2}\end{matrix}\right]\\
\mat{C}_{\delta,L}^{-1}  &= \left[\begin{matrix}\frac{\sigma_{2}^{2}}{\sigma_{0}^{2} \sigma_{2}^{2} - \sigma_{1}^{4}} & \frac{\sigma_{1}^{2}}{\sigma_{0}^{2} \sigma_{2}^{2} - \sigma_{1}^{4}}\\\frac{\sigma_{1}^{2}}{\sigma_{0}^{2} \sigma_{2}^{2} - \sigma_{1}^{4}} & \frac{\sigma_{0}^{2}}{\sigma_{0}^{2} \sigma_{2}^{2} - \sigma_{1}^{4}}\end{matrix}\right]
\end{align}
where $\sigma_0^2 = \expect{\delta^2}$, $\sigma_2^2 = \expect{L^2}$ and $\sigma_1^2 = \expect{-\delta L}$.

Now, we can conveniently write bias parameters in a matrix form
\begin{align}
    \myvec{b}_1 &= \left. \nabla_{\myvec{x}} F \, \right|_{\myvec{x} = 0} \\ %
    \mat{b}_2 &= \left. (\nabla_{\myvec{x}} \otimes \nabla_{\myvec{x}}) F \, \right|_{\myvec{x} = 0} \\
    \mat{b}_3 &= \left. (\nabla_{\myvec{x}} \otimes \nabla_{\myvec{x}} \otimes \nabla_{\myvec{x}} ) F \, \right|_{\myvec{x} = 0} \\
    \mat{b}_n &= \left. \nabla_{\myvec{x}}^{\otimes n} F \, \right|_{\myvec{x} = 0}
\end{align}
where $\otimes$ designates an outer product, the power notation in the last line designates a repeated outer product, $\myvec{b}_1$ is a vector,  $\mat{b}_2$ is a symmetric rank two matrix and $\mat{b}_3$ a symmetric rank three tensor and so on and $\nabla_{\myvec{x}}$ denotes a gradient with respect to the chosen variables, for example $\nabla_{\myvec{x}} = (\partial/\partial \delta, \partial/\partial L)^T$.

It is then straightforward to show
\begin{align}
    \mat{b}_n &= (-1)^n \expectgal{ \frac{\nabla_{\myvec{x}}^{\otimes n}  p}{p} } \label{eqn:bn_tensor}
\end{align}
which evaluates at the first two orders to
\begin{align}
    \myvec{b}_1 %&= \expectgal{ \frac{\nabla_x  p}{p} } \\
              &= \expectgal{\mat{C}^{-1} \myvec{x} } \\
    \mat{b}_2 %&= \expectgal{ \frac{\nabla_x  p}{p} } \\
              &= \expectgal{\mat{C}^{-1} (\myvec{x} \otimes \myvec{x} - \mat{C}) \mat{C}^{-1} } \label{eqn:measure_b2mat}
\end{align}
Importantly, these expressions do not lead to the same bias estimators for the density if a second variable is present that is correlated with the density. For example, if we use the density and Laplacian as variables, we find 
\begin{align}
b_{\delta, \mathrm{so2}} &= \expectgal{\frac{\delta \sigma_{2}^{2} + L \sigma_{1}^{2}}{\sigma_{0}^{2} \sigma_{2}^{2} - \sigma_{1}^{4}}} \label{eqn:b1_o2} \\
b_{L, \mathrm{so2}} &= \expectgal{\frac{L \sigma_{0}^{2} + \delta \sigma_{1}^{2}}{\sigma_{0}^{2} \sigma_{2}^{2} - \sigma_{1}^{4}}} \label{eqn:blap_estimator}
\end{align}
for the components of $\myvec{b}_1$ and
\begin{align}
b_{\delta\delta, \mathrm{so2}} &= \expectgal{\frac{\left(\delta \sigma_{2}^{2} + L \sigma_{1}^{2} \right)^{2} - \sigma_{2}^{2} \left(\sigma_{0}^{2} \sigma_{2}^{2} - \sigma_{1}^{4}\right)}{\left(\sigma_{0}^{2} \sigma_{2}^{2} - \sigma_{1}^{4}\right)^{2}}} \label{eqn:b2_o2} \\
b_{LL, \mathrm{so2}} &= \expectgal{\frac{\sigma_{0}^{2} \left(- \sigma_{0}^{2} \sigma_{2}^{2} + \sigma_{1}^{4}\right) + \left(L \sigma_{0}^{2} + \delta \sigma_{1}^{2}\right)^{2}}{\left(\sigma_{0}^{2} \sigma_{2}^{2} - \sigma_{1}^{4}\right)^{2}}} \\
b_{\delta L, \mathrm{so2}} &= \expectgal{\frac{\left(L \sigma_{0}^{2} + \delta \sigma_{1}^{2}\right) \left(L \sigma_{1}^{2} + \delta \sigma_{2}^{2}\right) - \sigma_{1}^{2} \left(\sigma_{0}^{2} \sigma_{2}^{2} - \sigma_{1}^{4}\right)}{\left(\sigma_{0}^{2} \sigma_{2}^{2} - \sigma_{1}^{4}\right)^{2}}} \label{eqn:bdl_o2}
\end{align} 
for the components of $\mat{b}_2$ where we have written e.g. $b_{\delta\delta}$ as a symbol for what was previously referred to as $b_2$. Further, we note that a general form for the $N$th density bias parameter is given by
\begin{align}
    b_{\delta^N, \mathrm{so2}} &= \expectgal{\frac{H_N((\delta + L \sigma_{1}^{2} / \sigma_{2}^{2})/\sigma_{*})}{\sigma_{*}^N}} \label{eqn:bn_so2}
\end{align}
with $\sigma_{*}^2 = \sigma_{0}^{2} - \sigma_{1}^{4} / \sigma_{2}^{2}$. We note \revb{that these} relations have already been derived through the response of density peaks to large scale perturbations \citep{bardeen_1986, Mo_White_1996, desjacques_2010} and have later been shown to apply also to more general tracers \citep{lazeyras_2016}.

The difference between \rev{equations} \eqref{eqn:b1_o2}-\eqref{eqn:bn_so2} and the estimators from equations \eqref{eqn:b1_o0}-\eqref{eqn:b4_o0} is that \rev{they} correspond to partial derivatives of the density at fixed values of the Laplacian. The \rev{previous} estimators are the derivatives of the projected distribution, which is not the same due to the correlation between density and Laplacian. As we will see in Section \ref{sec:meas_scaledep}, the new estimators are less scale dependent, since they are closer to pure partial derivatives with respect to the density (in the separate universe sense). 

We will refer to the \rev{estimators from equations} \eqref{eqn:b1_o2}-\eqref{eqn:bn_so2} as the \quotes{estimators with spatial corrections of order 2} whereas the previous estimators are without any spatial corrections (i.e. of spatial order 0). It is straightforward to obtain higher order spatial corrections e.g. by considering the covariance matrix of the distribution of $\myvec{x} = (\delta, L, P)^T$ with $P$ as in equation \eqref{eqn:P_fourthderiv}. \rev{For example, the estimators of density bias parameters at spatial order 4 can be phrased as}
\begin{align}
    b_{\delta^N} &= \expectgal{\frac{H_N(\delta_{*4}/\sigma_{*4})}{\sigma_{*4}^N}} \label{eqn:bn_so4} \\
    \delta_{*4} &= \delta + L \frac{\left(\sigma_{1}^{2} \sigma_{4}^{2} - \sigma_{2}^{2} \sigma_{3}^{2}\right)}{\sigma_{2}^{2} \sigma_{4}^{2} - \sigma_{3}^{4}} + P \frac{\left(\sigma_{1}^{2} \sigma_{3}^{2} - \sigma_{2}^{4}\right)}{\sigma_{2}^{2} \sigma_{4}^{2} - \sigma_{3}^{4}} \nonumber \\
    \sigma_{*4} &= \frac{\sigma_{0}^{2} \sigma_{2}^{2} \sigma_{4}^{2} - \sigma_{0}^{2} \sigma_{3}^{4} - \sigma_{1}^{4} \sigma_{4}^{2} + 2 \sigma_{1}^{2} \sigma_{2}^{2} \sigma_{3}^{2} - \sigma_{2}^{6}}{\sigma_{2}^{2} \sigma_{4}^{2} - \sigma_{3}^{4}} \nonumber
    %\label{eqn:bn_so4}
\end{align}
\rev{where $\sigma_3^2 = -\expect{\delta \cdot P}$ and $\sigma_4^2 = \expect{P^2}$. In practice it is easier to evaluate such high order estimators numerically, rather than deriving explicit expressions for them. The covariance matrix may easily be measured from a given linear density field and then used to evaluate the estimators as e.g. in equation} \eqref{eqn:measure_b2mat}, \rev{so that in principle it is not very difficult to obtain corrections at any spatial order. Throughout this article, we will mainly focus on the estimators of spatial order 2, since they are already sufficiently accurate to obtain good measurements of biases, but we will selectively show lower or higher order estimators throughout this article to demonstrate the convergence. }

We note that \rev{the} estimators with spatial corrections \rev{of order 2} are quite similar in spirit to the considerations by \citet{musso_2012, paranjape_2012, paranjape_2013a, paranjape_2013b} to map scale-dependent measurements of biases as in equation \eqref{eqn:bn_o0} under the assumption of excursion set (peaks) models to scale-independent large-scale parameters. However, here we can see that such a mapping to the large-scale limit can be done without the assumption of \rev{Peak or excursion set models and for spatial corrections of any order}. Consider for example that the relation between the spatial order 0 and spatial order 2 estimator of $b_1$ is given by 
\begin{align}
    \expectgal{\delta} = \mat{C} \myvec{b}_1 &= \sigma_0^2 b_{1, \mathrm{so}0} \nonumber \\
    &= \sigma_0^2 b_{1, \mathrm{so}2} - \sigma_1^2 b_{L, \mathrm{so}2} \,\,. \nonumber
\end{align}
Therefore, $b_{L, \mathrm{so}2}$ predicts the scale dependence of $b_{1,\mathrm{so}0}$ and $b_{1,\mathrm{so}2}$ posits a more scale-independent estimate:
\begin{align}
    b_{1,\mathrm{so}0} &= b_{1,\mathrm{so}2} - \frac{\sigma_1^2}{\sigma_0^2} b_{L,\mathrm{so}2} \,\, \label{eqn:bias_estimator_scaling}
\end{align}
\rev{Note that this argument can easily be pushed to higher spatial orders:}
\begin{align}
    b_{1,\mathrm{so}0} &= b_{1,\mathrm{so}4} - \frac{\sigma_1^2}{\sigma_0^2} b_{L,\mathrm{so}4} + \frac{\sigma_2^2}{\sigma_0^2} b_{P,\mathrm{so}4}  \,\,. \label{eqn:bias_estimator_scaling_o4}
\end{align}
The additional cost of our approach \citep[e.g. in comparison to][]{paranjape_2013b} is that, beneath the density, also the Laplacian has to be evaluated in Lagrangian space, but the benefits are model-independence and a much reduced mathematical complexity.

\subsection{Multivariate cumulants} \label{sec:multivariate_cumulants}
Analogously to the monovariate case, we define multivariate cumulant bias parameters as 
\begin{align}
    \myvec{\beta}_N = \left. \nabla_{\myvec{x}}^N \log(F) \right|_{\myvec{x} = 0}
\end{align}
which leads immediately to
\begin{align}
    \myvec{\beta}_1 &= \myvec{b}_1 \\
    \myvec{\beta}_2 &= \mat{b}_2 - \myvec{b}_1 \otimes \myvec{b}_1 \\
    \beta_{3, ijk} &= b_{3,ijk} - (b_{1,i} b_{2,jk} + b_{1,j} b_{2,ki} + b_{1,k} b_{2,ij}) + 2 b_{1,i} b_{1,j} b_{1,k}
 \end{align}
where we gave $\myvec{\beta}_3$ in index notation, since the central term is difficult to express in vectorial notation. Note that this leads to the same relations between the density cumulant biases and their canonical biases as in equations \eqref{eqn:beta1_of_b}-\eqref{eqn:beta4_of_b}, but it also includes additional relations like e.g.
\begin{align}
    \beta_{\delta \delta L} &= b_{\delta \delta L} - b_{\delta \delta} b_{L} - 2 b_{\delta L} b_{\delta} +2 b_{\delta}^2 b_L \,\,.
\end{align}

Finally, the relation between the separate universe bias function and the moment and cumulant generating functions is in analogy to equation to equations \eqref{eqn:moment_ng} and \eqref{eqn:cumulant_ng} given by
\begin{align}
    M(\myvec{t}) &= \exp \left(\frac{1}{2} \myvec{t}^T \mat{C} \myvec{t} \right) F(\myvec{x}_0 = \mat{C} \myvec{t}) \\
    K(\myvec{t}) &= \frac{1}{2} \myvec{t}^T \mat{C} \myvec{t} + \log \left(  F(\mat{C} \myvec{t}) \right) \label{eqn:multivariate_cumgen}
\end{align}
which can be differentiated to show that the cumulants are given in an easy form through the cumulant bias parameters,  e.g.
\begin{align}
    \myvec{\kappa}_1 &= \mat{C} \myvec{\beta}_1 \label{eqn:kappa1_multivar} \\ 
    \myvec{\kappa}_2 &= \mat{C} \myvec{\beta}_2 \mat{C}  + \mat{C} \\
    \kappa_{3,ijk} &= \sum_{abc} C_{ai} C_{bj} C_{ck} \beta_{abc} \,\,. \label{eqn:kappa3_multivar}
\end{align}

However, instead of measuring directly the cumulants of the galaxy environment distribution and inverting equations \eqref{eqn:kappa1_multivar} - \eqref{eqn:kappa3_multivar}, it is in practice simpler to instead define the variable 
\begin{align}
    \myvec{u} &= \mat{C}^{-1} \myvec{x}
\end{align}
and measure its cumulants $\kappa_u$ which relate to the cumulant biases as
\begin{align}
    \myvec{\beta}_{ijk...} &= \begin{cases}
        \kappa_{u,ijk...}  & \mathrm{\quad if \quad} i + j + k + ...  \neq 2  \\
        \kappa_{u,ijk...} - C^{-1}_{ab} & \mathrm{\quad if \quad} i + j + k + ...  = 2
    \end{cases}
\end{align}
where $a$ and $b$ indicate the indices of the non-zero variables in the second order case.

We show in Appendix \ref{app:filtering} that it is also possible to derive similar estimators if the density field was filtered with a function different than a sharp $k$-space filter. The only additional complication in that case that it is necessary to account for the correlation matrix between the smoothed large scales and unsmoothed small scales. However, for simplicity, we will focus in this article only on measurements with the sharp $k$-space filter.

\section{Density bias measurements} \label{sec:scalar_measurements}
In this Section we evaluate the Lagrangian density bias parameters for different sets of haloes. We will verify the consistency of our estimators by comparing them to the vast literature available on the subject; furthermore, we will also compare the canonical bias parameters $b_n$ with the cumulant bias parameters $\beta_n$.

\subsection{Simulation} \label{sec:simulation}
For the analysis through this paper, we use a single cosmological box simulation with high resolution. This simulation is part of the \quotes{BACCO simulation project} that was first introduced in \citet{angulo_2021}. It has a boxsize of $L = 1440 h^{-1}\mathrm{Mpc}$ with $4320^3$ particles leading to a mass resolution of $m_p = \SI{3.2e9}{} h^{-1} M_\odot$. The cosmological parameters are $\Omega_m = 0.307$, $\Omega_\Lambda = 0.693$, $\Omega_b = 0.048$., $n_s = 0.9611$, $\sigma_8 = 0.9$, $h=0.677$ which are similar to the \citet{planck_2020} cosmology except for the roughly $10\%$ larger value of $\sigma_8$. By default we use this simulation at a scale-factor of $a=1.08$ corresponding approximately to $a=1$. 

To identify haloes the simulation code uses a modified version of \textsc{subfind} \citep{Springel_2001} which first identifies haloes through a friends of friends (FoF) algorithm and subsequently calculates for each FoF group the mass $M_{200\mathrm{b}}$ in a region that encloses 200 times the mean density of the Universe.

\subsection{Bias measurements} \label{sec:measure_techinque}
We define sets of biased tracers by considering haloes selected by their mass. For this we consider 21 equally log-spaced halo masses between $10^{12} h^{-1} M_\odot$ and $10^{16} h^{-1} M_\odot$ and for each halo mass we consider all haloes with masses in the range
\begin{align}
    M_{200b} \in [M_i / 1.25, M_i \cdot 1.25 ] \,\,.
\end{align}

To evaluate our Lagrangian bias estimators we only need to know the linear field evaluated at the Lagrangian locations of the tracers. We approximate the Lagrangian location of each halo through the Lagrangian \rev{coordinate} of its most bound particle. Since, the simulation started from a Lagrangian grid, the Lagrangian origin of the most bound particle can easily be inferred from its id $i_{\mathrm{mb}}$ as 
\begin{align}
    \myvec{q}_{\mathrm{mb}} &= \frac{L}{N_{\mathrm{grid}}} \begin{pmatrix}
        i_x \\ i_y \\ i_z
    \end{pmatrix} \\
    i_{\mathrm{mb}} &= i_x N_{\mathrm{grid}}^2 + i_y N_{\mathrm{grid}} + i_z
\end{align}
where $N_{\mathrm{grid}} = 4320$ is the number of particles per dimension.

We know the linear density field of the simulation through the initial condition generator. To save computation time, we create a low resolution grid representation of the linear density field with $N_{\mathrm{lin}}^3$ grid points. For fields different than the linear density field we additionally multiply by the correct operator (e.g. $-k^2$ for the Laplacian) in Fourier space. We create a smoothed version of this field by multiplying with a sharp $k-$filter in Fourier space
\begin{align}
    \delta_k &= \delta_{\rm{lin},k} \cdot \Theta(k_{\mathrm{d}} - k),
\end{align}
\rev{where $\Theta$ is the heavy-side function} and we test in each measurement different damping scales $k_{\rm{d}} \in [0.1, 0.15, 0.2, 0.25, 0.3] h^{-1}\mathrm{Mpc}$. We then deconvolve this field with a linear interpolation kernel and interpolate it to the Lagrangian locations of our tracer set. We choose $N_{\mathrm{lin}}$ sufficiently large that the resulting interpolated values are virtually independent of this discretization, e.g. $N_{\mathrm{lin}} = 183$ at $k_{\rm{d}} = 0.1 h^{-1}\mathrm{Mpc}$ and $N_{\mathrm{lin}} = 549$ at $k_{\rm{d}} = 0.3 h^{-1}\mathrm{Mpc}$.

With the linear densities at the Lagrangian locations of tracers, it is easy to evaluate the spatial order 0 estimators of the biases by averaging as in Equation \eqref{eqn:bn_o0}. However, for higher spatial order estimators we also need to know the values of the Laplacian (equations \eqref{eqn:b1_o2} - \eqref{eqn:bdl_o2}) or of the fourth derivative (equation \eqref{eqn:bn_so4}). The Laplacian is inferred by multiplying the linear density field in Fourier space by $-k^2$ and then interpolated to the tracer locations. Other variables like the fourth derivatives can be evaluated in a similar manner. This allows us to evaluate any of the bias estimators from Section \ref{sec:theory} directly as simple expectation values over tracers. Note that this is a major difference compared to the measurement process in \citet{paranjape_2013b} where instead the spatial order 0 estimators as in equation \eqref{eqn:bn_o0} were evaluated, but then mapped to scale-independent parameters through peak theory arguments.

To estimate the covariance of a set of measured bias parameters we use a Jackknife technique. For this we divide the box in Lagrangian space into $N_{\mathrm{jk}}^3$ subboxes with $N_{\mathrm{jk}} = 4$. We perform $64$ measurements of the vector of bias parameters $\myvec{b}_i$ by subsequently leaving out all tracers in one of the subboxes. Then we estimate the covariance through
\begin{align}
    \mat{C}_{\myvec{b}} &= \frac{1}{N_{\mathrm{jk}}^3 -1} \sum_i (\myvec{b}_i - \myvec{b}_0) \otimes (\myvec{b}_i - \myvec{b}_0) \\
    \myvec{b}_0 &= \frac{1}{N_{\mathrm{jk}}^3} \sum_i \myvec{b}_i \,\,.
\end{align}

It is worth noting that the Jackknife estimator results in a more reliable estimate of the uncertainty of the measurement than e.g. a simple bootstrap would yield. When comparing these estimators we found that the Jackknife gives larger (more conservative) error estimates, since it also accounts for the uncertainty induced by cosmic variance, which is sampled by leaving out spatially correlated parts of the data sets.

\subsection{$b_1$ and scale dependence of estimators} \label{sec:meas_scaledep}

\begin{figure}
    \centering
    \includegraphics[width=\columnwidth]{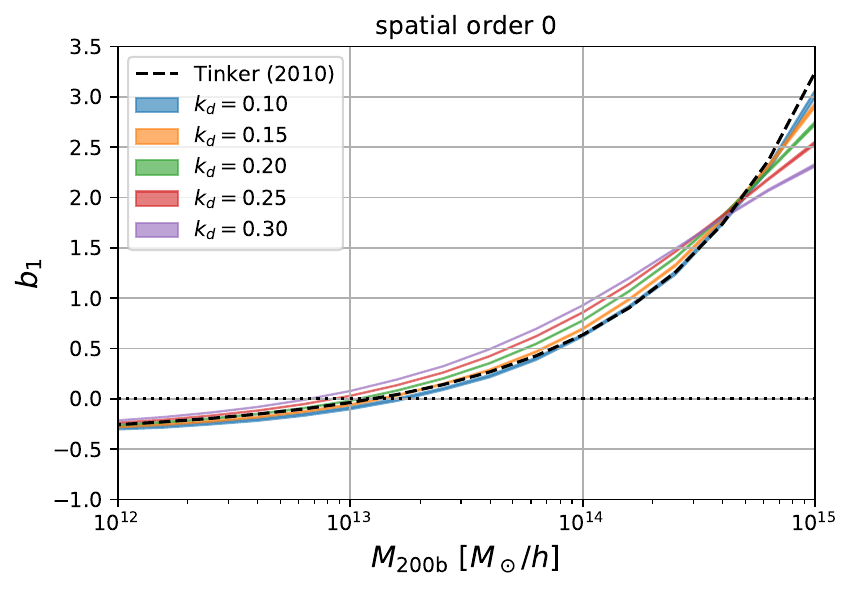}
    \includegraphics[width=\columnwidth]{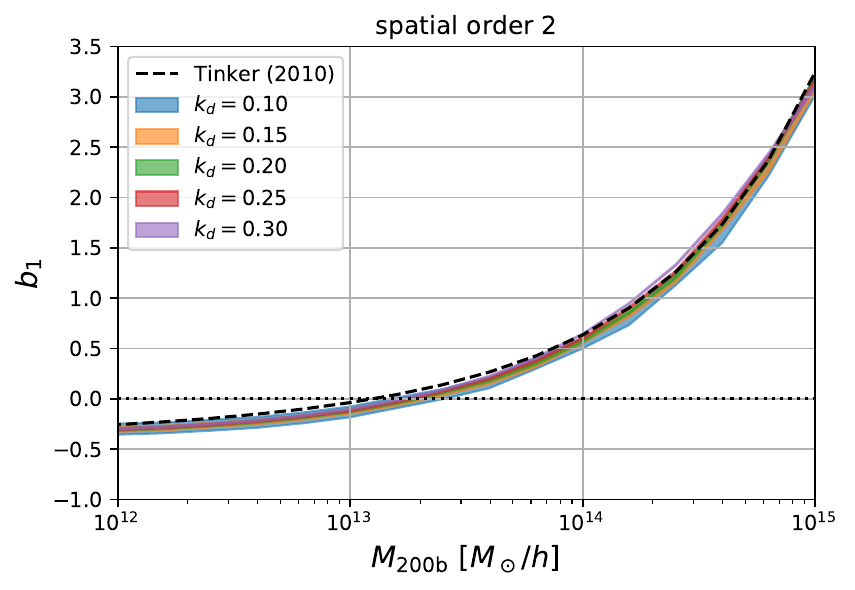}
    \caption{$b_1$ as a function of halo mass using the estimators from equation \eqref{eqn:b1_o0} (top) and equation $\eqref{eqn:b1_o2}$ (bottom) measured at different damping scales (different coloured regions). The shaded regions indicate the $1 \sigma$ certainty region of the estimators. Using the $b_1$ estimator that includes the Laplacian correction increases the uncertainty of the $b_1$ estimates, but reduces the dependence on the damping scale, leading to a good agreement across different scales.}
    \label{fig:b1_measurements}
\end{figure}

In Figure \ref{fig:b1_measurements} we show measurements of $b_1$ for haloes as a function of $M_{200b}$ using the estimator from equation \eqref{eqn:b1_o0} without spatial corrections (top panel) versus the estimator from equation \eqref{eqn:b1_o2} with spatial corrections of order 2 (bottom panel). The dashed lines indicates the fitting function from \citet{Tinker_2010} evaluated for the cosmology of our simulation.

As expected, haloes of large masses $M \sim 10^{15} h^{-1} M_\odot$ are highly biased $b_1 \sim 3$ whereas low-mass haloes $M \sim 10^{12} h^{-1} M_\odot$ have even a slightly negative (Lagrangian) bias.  The spatial order 0 estimators have a very small degree of statistical uncertainty, but exhibit a significant scale dependence and are inconsistent across different damping scales. On the other hand the estimators with spatial order 2 have a larger statistical uncertainty, but seem consistent across different damping scales, except at very large $k_{\mathrm{d}}$ and large halo masses where also some scale dependence becomes visible, probably indicating the effect of higher spatial order terms. 

Our spatial order two measurements agree at all scales well with the fitting function from \citet{Tinker_2010}.  While they lie systematically about $5 \%$ below this fit, this is within the quoted uncertainty of the \citet{Tinker_2010} fit and a similar difference can be seen in \citet{lazeyras_2016}. In comparison to other studies at $z=0$ \citep[e.g.][]{lazeyras_2016}, we have lower bias values at the same masses (e.g. $b_1 \sim 0.5$ at $10^{14} h^{-1} M_\odot$ instead of $b_1 \sim 1$), because of the later time ($z=-0.08$) and relatively high variance ($\sigma_8 = 0.9$) of our simulation. 

\begin{figure}
    \centering
    \includegraphics[width=\columnwidth]{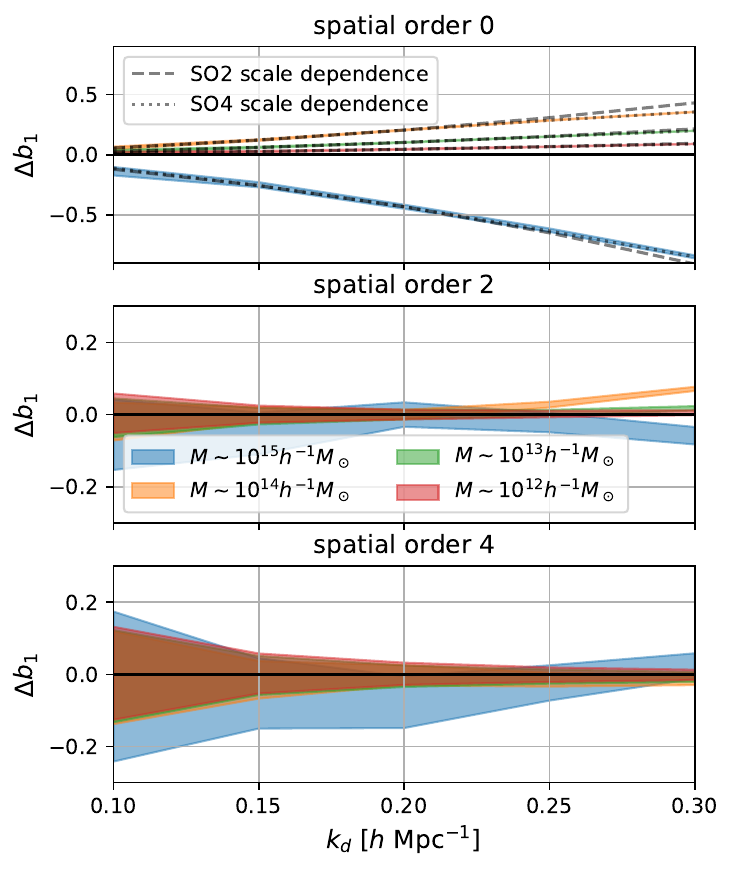}
    \caption{Scale dependence of the error in $b_1$ estimates for different halo mass selections. The error is phrased relatively to the measurement with spatial order $2$ at $k_{\rm{d}} = 0.2 h \mathrm{Mpc}^{-1}$. On large scales (small $k_{\rm{d}}$) the spatial order 0 estimators from equation \eqref{eqn:b1_o0} converge well to this estimate, whereas the spatial order 2 estimators from equation \eqref{eqn:b1_o2} agree well up to $k_{d} \gtrsim 0.25 h \mathrm{Mpc}^{-1}$ where slight disagreements arise. The spatial order 4 estimates remain scale-independent even beyond this scale.
    }
    \label{fig:b1_scale_dependence}
\end{figure}

\begin{figure*}
    \centering
    \includegraphics[width=0.72\textwidth]{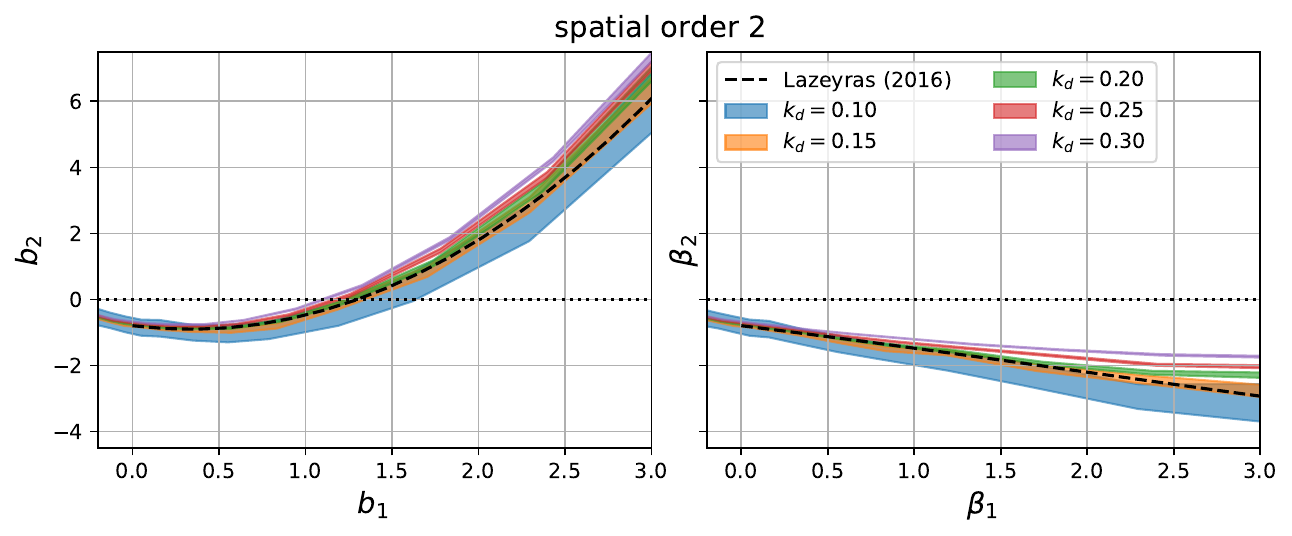}
    \caption{$b_2$ and $\beta_2$ as a function of $b_1 = \beta_1$ using the spatial order 2 estimators. The black dashed line is the  $b_2(b_1)$ relation inferred by \citet{lazeyras_2016} from separate universe simulations. The spatial order 2 estimators seem consistent with the literature coevolution relation down to damping scales of $k_{\mathrm{d}} \sim 0.15 - 0.2 h \mathrm{Mpc}^{-1}$. Noteworthy, it seems that $\beta_2 < 0$ holds always, which means that the width of the galaxy environment distribution $p(\delta | \mathrm{g})$ is always smaller than that of the background $p(\delta)$. Therefore, the cumulant bias parameter $\beta_2 - \beta_1$ relation appears slightly simpler than the $b_2 - b_1$ relation.}
    \label{fig:b1_b2_coev}
\end{figure*}

To further highlight the difference in the scale dependence of the estimators, we show in Figure \ref{fig:b1_scale_dependence} the error $\Delta b_1$ in $b_1$ measurements as a function of damping scale for a few different halo masses. Here we phrase the error $\Delta b_1 = b_1 - b_{1,\mathrm{ref}}$ relative to the spatial order 2 estimator at $k_{\mathrm{d}} = 0.2 h \mathrm{Mpc}^{-1}$ as  $b_{1,\mathrm{ref}}$. As expected, the order zero estimator seems to approach the selected reference value on large scales.  However, the spatial order 0 has a significant scale dependence  whereas the order 2 estimators are almost scale independent except for the largest  $k_{d} \gtrsim 0.25 h \mathrm{Mpc}^{-1}$. We also show measurements with the spatial order 4 estimator from equation \eqref{eqn:bn_so4} which are scale-independent to even smaller scales, showing that the measurements converge well with adding higher order spatial corrections. However, for getting a good estimate of the bias parameters it seems sufficient to use the spatial order 2 estimator at scales $k_{d} \lesssim 0.2 h \mathrm{Mpc}^{-1}$ which bears a similar accuracy in measurement, e.g. to evaluating the spatial order 4 estimator at $k_{d} \sim 0.3 h \mathrm{Mpc}^{-1}$.

We can estimate the scale dependence of the spatial order 0 estimator as in equations \eqref{eqn:bias_estimator_scaling} and \eqref{eqn:bias_estimator_scaling_o4}. We mark the estimated scale dependence of the spatial order 0 estimators using the spatial order 2 bias parameters measured at at $k_{\mathrm{d}} = 0.2 h \mathrm{Mpc}^{-1}$ as dashed lines in Figure \ref{fig:b1_scale_dependence} and using the spatial order 4 parameters from $k_{\mathrm{d}} = 0.3 h \mathrm{Mpc}^{-1}$ as dotted lines. The scale dependence of the spatial order 0 estimator is predicted quite well, showing that the consideration of the Laplacian introduces a correction of order $\sigma_1^2 / \sigma_0^2$ which scales approximately as $k_{\mathrm{d}}^2$ and including of the fourth derivatives provides additional corrections that become relevant on even smaller scales. 

Our measurements therefore confirm the results from previous studies that including the Laplacian is vital to recover scale-independent bias on large scales \citep{paranjape_2013b, lazeyras_2016}. Further, if the modelling is pushed to  small scales $k_{d} \gtrsim 0.25 h$ the inclusion of additional higher spatial derivative terms may prove beneficial.

\subsection{$b_2$ versus $\beta_2$} \label{sec:measureb2}

The first cumulant bias parameter that is different from the canonical bias parameter of the same order is $\beta_2 = b_2 - b_1^2$ where $b_1 = \beta_1$. Since there is a one to one relation between $(b_1, b_2)$ and $(\beta_1, \beta_2)$ it might seem that there should not be any advantage by using $\beta_2$ over $b_2$ as a parameter when fitting datasets. However, here we will show that $\beta_2$ seems to be more independent of $\beta_1$ than $b_2$ of $b_1$, especially when their covariance matrix is considered.

In Figure \ref{fig:b1_b2_coev}, we show the measured coevolution relation between $b_2$ and $b_1$ (left) versus the coevolution between $\beta_2$ and $\beta_1$ (right) for the spatial order 2 estimators. In comparison we also show the coevolution relations measured by \citet{lazeyras_2016} as a dashed line, which seems to match our measured relation well up to $k_{\mathrm{d}} \lesssim 0.15 - 0.2 h \mathrm{Mpc}^{-1}$, showing that our method for measuring the bias parameters is indeed able to recover the correct large-scale limit of the bias parameters.

We notice that for halos we find $\beta_2<0$ across all masses, and with high statistical significance. This means that the width of the halo environment distribution $p(\delta | g)$ is smaller than the width of the background distribution $p(\delta)$ -- showing that halo formation is more selective than a random distribution. Possibly, the assumption $\beta_2 < 0$ could be used to limit the considered prior range when fitting certain galaxy surveys. However, it is difficult to anticipate whether this should hold for every possible set of galaxies. 

Further, we note that the coevolution relation between $\beta_2$ and $\beta_1$ is monotonic and roughly linear. For $b_1 \gtrsim 2$ it satisfies $|\beta_2| < |b_2|$. Therefore, $\beta_2$ versus $\beta_1$ seems slightly simpler than the $b_2$ to $b_1$ relation. 

\begin{figure}
    \centering
    \includegraphics[width=\columnwidth]{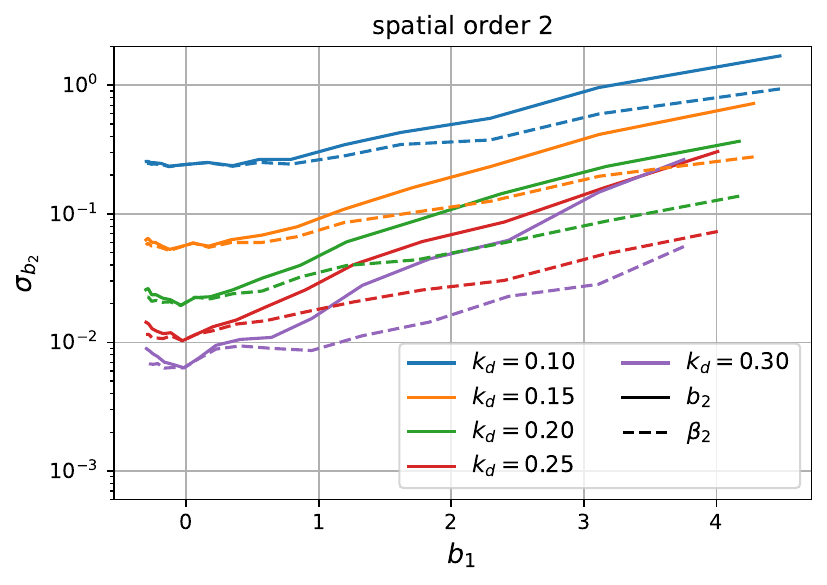}
    \caption{Uncertainty of $b_2$ measurement (solid lines) and $\beta_2$ measurements (dashed lines) as a function of $b_1$ for different damping scales. The uncertainty of $\beta_2$ is significantly smaller than of $b_2$ for high values of $b_1$. }
    \label{fig:uncertainty_b2}
\end{figure}

\begin{figure}
    \centering
    \includegraphics[width=\columnwidth]{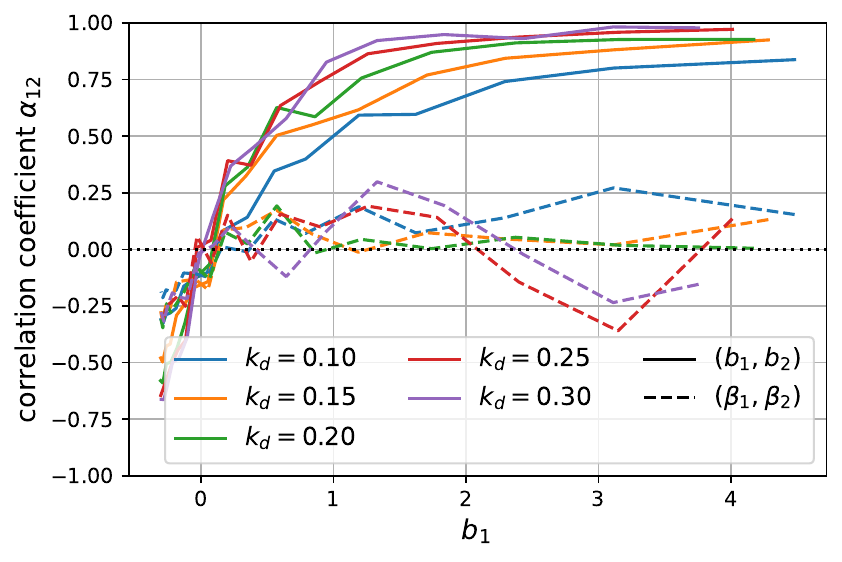}
    \caption{Correlation coefficient $\alpha_{12}$ between the $b_1$ and $b_2$ measurement (solid) and the $\beta_1$ and $\beta_2$ measurement (dashed). Note that $\beta_2$ as a parameter is much more independent of the value of $b_1$ than $b_2$. }
    \label{fig:correlation_coeff}
\end{figure}

The benefit of using $\beta_2$ as a parameter is even more evident when considering the covariance matrix of the measurements. In Figure \ref{fig:uncertainty_b2} we show the uncertainty of the $b_2$ measurements (solid lines), versus the uncertainty of the $\beta_2$ measurements. Comparing different damping scales shows that the inclusion of smaller scales in the measurements decreases the statistical uncertainty (but increases systematic error). In all cases and especially at high values of $b_1$ the uncertainty of $\beta_2$ is significantly smaller than that of $b_2$. This is so, since $b_2$ and $b_1$ are more correlated than $\beta_2$ and $\beta_1$. %

To highlight this we show in Figure \ref{fig:correlation_coeff} the correlation coefficient 
\begin{align}
    \alpha_{12} &:= \frac{C_{12}}{\sqrt{C_{11} C_{22}}}
\end{align}
where $\mat{C}$ is the covariance matrix between $b_1$ and $b_2$ or $\beta_1$ and $\beta_2$. For $b_1 \gg 0$ the correlation coefficient of $(b_1,b_2)$ is quite large, even close to $1$ in some cases. On the other hand, the correlation coefficient of $(\beta_1,\beta_2)$ is quite small and seems on average consistent with $0$. Therefore, measurements of $b_1$ are quite entangled with $b_2$, but not so much with $\beta_2$. While we do not show it here, we have found that this is also the case for higher order correlations, e.g. of $(b_1, b_3)$ versus $(\beta_1, \beta_3)$.

This can be understood when considering the difference between moments and cumulants of a probability distribution. E.g. if we knew about some distribution $p(x)$ that it has a large first moment $\expect{x}$, then we might also expect that the second moment $\expect{x^2}$ is large. On the other hand, knowledge about the mean of a distribution, is very uninformative about its width -- that is its second cumulant $\sigma_x^2 = \expect{(x - \expect{x})^2} = \expect{x^2} - \expect{x}^2$. In the same sense, we expect cumulant bias parameters to be more independent of each other than canonical bias parameters.

\subsection{Higher order biases}

\begin{figure*}
    \centering
    \includegraphics[width=0.72\textwidth]{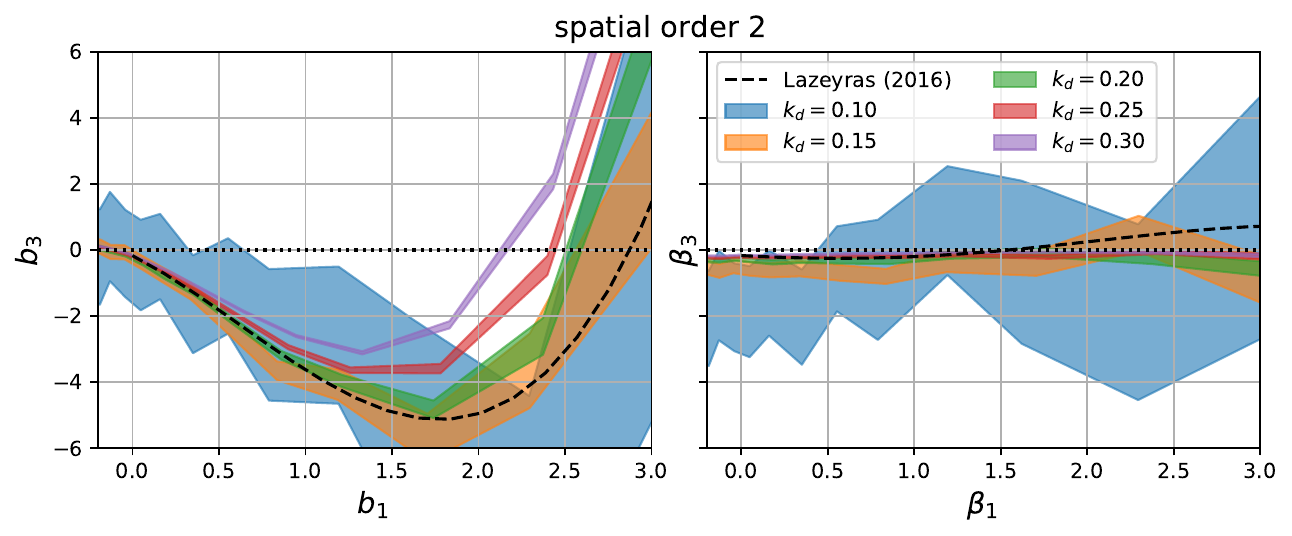}
    \includegraphics[width=0.72\textwidth]{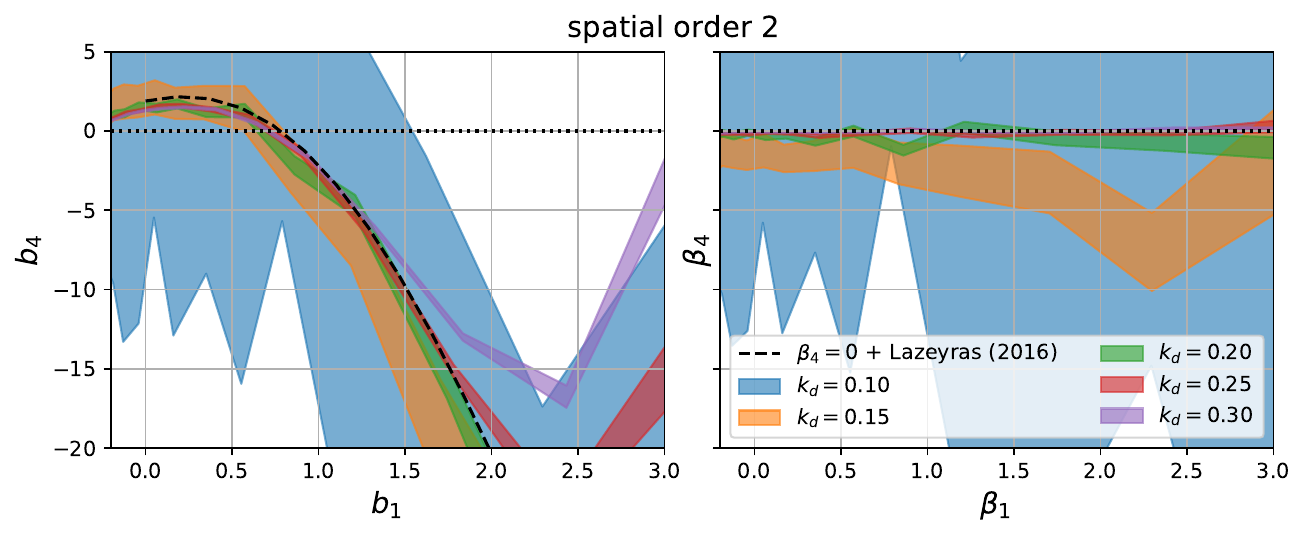}
    \caption{Coevolution relations of higher order bias parameters $b_3$ and $\beta_3$ (top) and $b_4$ and $\beta_4$ (bottom) for the spatial order 2 bias estimators. For $b_4$ we indicate as a dashed line a prediction that follows from combining the \citet{lazeyras_2016} measurements of $b_1$, $b_2$ and $b_3$ with equation \eqref{eqn:beta4_of_b} when using $\beta_4 = 0$. Strikingly, $\beta_3$ and $\beta_4$ are extremely close to $0$ -- independently of the value of $b_1$.}
    \label{fig:b3b4_vs_b1}
\end{figure*}

While the second order cumulant bias $\beta_2$ has already slight advantages over $b_2$ as highlighted in the last section, the benefits are even more significant at higher orders. Here, we will compare the behavior of $b_3$ and $b_4$ with the behavior of $\beta_3$ and $\beta_4$. 

In Figure \ref{fig:b3b4_vs_b1}, we show $b_3$, $b_4$, $\beta_3$ and $\beta_4$ as a function of $b_1$ for the spatial order two estimator. The uncertainty of the measurements of $b_3$ and $b_4$ are much larger than for $b_2$ so that we hardly find any signal at $k_{\mathrm{d}} \sim 0.1 h \mathrm{Mpc}^{-1}$. However, at $k_{\mathrm{d}} \sim 0.15 h \mathrm{Mpc}^{-1}$ we can find a meaningful signal and it is consistent with the $b_3$ measurements of \citet{lazeyras_2016}.

Strikingly, the value of $\beta_3$ is extremely small in comparison to $b_3$ and seems approximately consistent with 0. Noteworthy this is not only the case for scales where $b_3$ is reasonably scale independent ($k_{\mathrm{d}} \lesssim 0.15 h \mathrm{Mpc}^{-1}$), but it is also so for much smaller scales. There seems to be no noteworthy relation between $\beta_3$ and $\beta_1$. Further, the coevolution relation measured by \citet{lazeyras_2016} seems consistent with $\beta_3 = 0$. Therefore, we could summarize the third order coevolution relation through the relation we get by using $\beta_3 = 0$ in equation \eqref{eqn:beta3_of_b}:
\begin{align}
    b_3 &= 3 b_1 b_2 - 2 b_1^3 \,\,.
\end{align}

Similarly, we find for $\beta_4$ an apparent independence of $\beta_1$ and consistency with zero across different scales. Assuming $\beta_4 = 0$ in equation \eqref{eqn:beta4_of_b} leads to a coevolution relation between $b_4, b_3, b_2$ and $b_1$ given by
\begin{align}
     b_4 &= 4 b_1 b_3 + 3 b_2^2 - 12 b_1^2 b_2 - 6 b_1^4
\end{align}
We mark this relation with the measurements of $b_1$, $b_2$ and $b_3$ from \citet{lazeyras_2016} as a dashed line in the bottom left panel of Figure \ref{fig:b3b4_vs_b1}. This seems in good agreement with the measured $b_4$.

We conclude that at order $n \geq 3$ the cumulant bias parameters are very close to zero. Coevolution relations at these orders can be summarized simply through $\beta_n = 0$. We might therefore suggest that high order coevolution relations do not represent any physical insights, but rather highlight that the canonical bias parameters form a suboptimal, deeply entangled basis.

Further, the fact that all the cumulant biases beyond order two are close to zero  indicates that the galaxy environment distribution is very well approximated by a Gaussian distribution. We will investigate the possibility of a Gaussian bias model in \citetgaus{}  where we also explain how this fact may arise naturally from the Gaussianity of the background distribution. 

\subsection{Laplacian bias}

\begin{figure}
    \centering
    \includegraphics[width=\columnwidth]{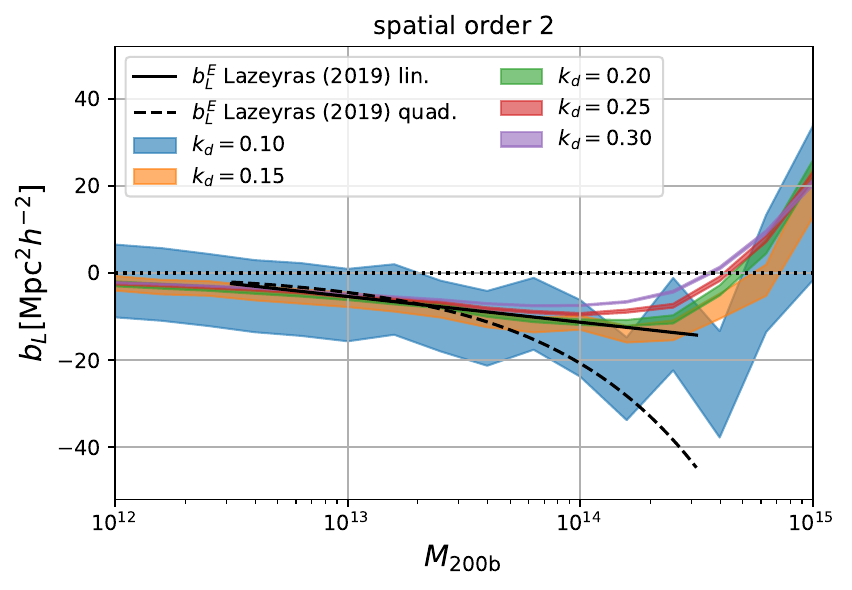}
    \caption{The Lagrangian Laplacian bias $b_L$ as a function of halo mass. At masses $M \lesssim 3 \times 10^{13} h^{-1} M_\odot$ the scale dependence of the $b_L$ measurements disappears and they agree well with the fits of the Eulerian Laplacian bias from \citet{lazeyras_2019}.
    }
    \label{fig:bL_measurements}
\end{figure}

In Figure \ref{fig:bL_measurements}, we show the Lagrangian Laplacian bias parameter as a function of halo mass, as inferred from the estimator in equation \eqref{eqn:blap_estimator}. The Laplacian measurements have a much stronger scale dependence than our previous measurements of density bias parameters. However, it seems that at masses $M \lesssim 3 \times 10^{13} h^{-1} M_\odot$ the scale dependence disappears and a reliable measurement is obtained. 

We compare our Lagrangian measurements with the Eulerian measurements of \citet{lazeyras_2019}. In general, the relation between the Lagrangian and Eulerian Laplacian bias depends on velocity bias \citep[e.g.][]{Desjacques_2018}. The velocity bias $b_s$ quantifies the difference between the displacement field of matter $\myvec{s}$ and galaxies $\myvec{s}_{\mathrm{g}}$ and gets as first order contributions of the form
\begin{align}
   \myvec{s}_{\mathrm{g}} - \myvec{s} \approx b_{s} \nabla \delta.
\end{align}
Therefore, the galaxy density can get an extra contribution proportional to the divergence of this field, that scales as $L = \nabla^2 \delta$. However, for our set of tracers $\myvec{s}_{\mathrm{g}} = \myvec{s}$ by definition, so that $b_{s} = 0$ and we simply assume that their Lagrangian and Eulerian Laplacian biases are identical or at least very similar.

In the reliable range $M \lesssim 3 \times 10^{13} h^{-1} M_\odot$ it seems that our measurements agree well with the linear and the quadratic (in halo radius) fit that was inferred by \citet{lazeyras_2019} for the Eulerian Laplacian bias. Therefore, we confirm that the Laplacian bias is negative for haloes and that it can plausibly have amplitudes of order $b_L \sim -20 \mathrm{Mpc}^2 h^{-2}$. However, we leave more detailed considerations to future studies. In principle, more accurate measurements could be obtained by consideration of higher order corrections and through measurements at higher redshifts.

\subsection{The scale-split break-down scale} \label{sec:theory_pbs_breakdown} \label{sec:pbs_breakdown}

The core assumption of the bias formalism is the separation of scales: The formation of galaxies and haloes is assumed to depend only on the properties of the Lagrangian environment on some small length scale. Larger scale perturbations are only relevant as far as they determine the distribution of small scale environments. We will refer to this as the `scale-split' assumption.

This assumption makes it possible to define scale-independent bias parameters. For example, $b_1$ describes how the likelihood of forming a galaxy responds when changing only the linear density contrast $\delta$ of the relevant smaller scale environment, while keeping other aspects (e.g. $L$) constant. As all larger scale density perturbations affect this aspect in the same way $b_1$ is independent of scale.

Physically, the scale-split assumption has to become invalid for sufficiently small smoothing scales. For example, halo formation may respond differently to perturbations that are smaller than their Lagrangian radius, than to those which are larger. 

Here, we want to show that the scale-split assumption also becomes mathematically inconsistent beyond some scale for a given set of variables: Recall that we have shown in equation \eqref{eqn:cum_to_beta}, how $\kappa_2$, the variance of the galaxy environment distribution, changes with scale. The predicted variance is only well defined, if 
\begin{align}
    \sigma^2 \leq \sigma_{\mathrm{max}, \delta}^2 &=  \frac{1}{-\beta_2} \label{eqn:smax_delta}
\end{align}
and becomes zero if $\sigma = \sigma_{\mathrm{max}, \delta}$. Mathematically, at this scale the environment distribution and the bias function have to become Dirac-delta functions. We may understand $\sigma_{\mathrm{max}}$ as the scale where (formally) all information that is relevant for galaxy formation has been accounted for and the biasing becomes deterministic in density. 

Beyond this scale, the PBS predicts a galaxy environment distribution with negative variance -- making any density-only bias model mathematically inconsistent. It may seem as if it was possible to set up expansion bias models for arbitrary high $\sigma$, but this is only so, since the negativity of the bias function makes it formally possible to have a negative variance. Actual galaxies will strictly obey $\kappa_2 >0$ and therefore, the response of density-only models has to become scale-dependent latest at $\sigma_{\mathrm{max}, \delta}$ -- but likely already earlier. We call this the \quotes{scale-split break-down scale}.

The break-down scale is different if additional variables are considered. For the multivariate distribution of density and Laplacian $(\delta, L)$, we have to require that the covariance matrix of the galaxy environment distribution is positive semi-definite. For this it is at least necessary that
\begin{align}
    \det(\myvec{\kappa}_2) &\geq 0 \nonumber \\
    %1 & 2
    \Leftrightarrow \det(\mat{1} + \mat{C} \myvec{\beta}_2) &\geq 0 \label{eqn:Cmax_deltaL}
\end{align}
which appears slightly more complicated, but it will also be violated at some finite damping scale.

\begin{figure}
    \centering
    \includegraphics[width=\columnwidth]{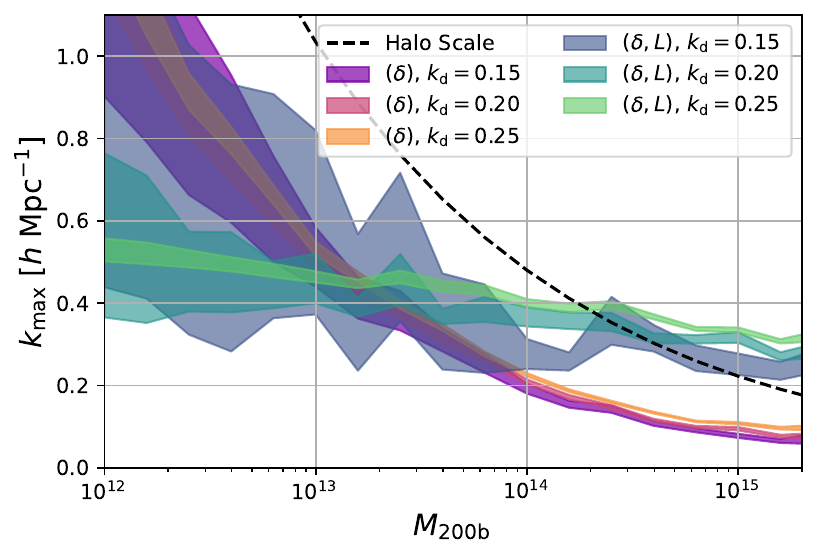}
    \caption{The maximum damping scale where the PBS can be valid and bias scale-independent. The three reddish contours show the break-down scale of density-only bias models, estimated with bias parameters measured at different scales, and the three green-blueish contours show it for ($\delta$, $L$) bias models. The black dashed line shows the wavenumber associated with the Lagrangian radius of haloes. The break-down scale is consistent across different measurements and only seems to scale strongly with halo radius for the $(\delta)$ case.}
    \label{fig:pbs_breakdown}
\end{figure}

To measure the break-down scale, we infer for each halo mass the bias parameters at three different scales $k_{\mathrm{d}} \in (0.15, 0.2, 0.25) h \mathrm{Mpc}^{-1} $. Then, we evaluate the co-variance matrix of the background distribution at 500 different equally log-spaced damping scales between $10^{-3} h \mathrm{Mpc}^{-1}$ and $10^{1} h \mathrm{Mpc}^{-1}$ and determine the earliest damping scale $k_{\mathrm{max}}$ where the covariance is large enough to violate equation \eqref{eqn:smax_delta} or equation \eqref{eqn:Cmax_deltaL}. We show the corresponding results as shaded contours in Figure \ref{fig:pbs_breakdown}, where we additionally mark the characteristic wave-number of haloes
\begin{align}
    k_{\mathrm{halo}} &= \frac{\pi}{R_{\mathrm{halo}}}
\end{align}
where $R_{\mathrm{halo}}$ is the Lagrangian radius that encloses the halo mass $M_{200\mathrm{b}}$.

Comparing the measurements at different damping scales, we find that the inferred scale is reasonably converged with the scale that we measured the bias parameters at. The density-only break-down scale is typically a factor two smaller than $k_{\mathrm{halo}}$ and seems roughly proportional to it. We therefore conclude that any Lagrangian local in matter density (LLIMD) bias model \rev{with scale-independent bias parameters} has to break down at a length scale roughly a factor two larger than Lagrangian radii of the considered haloes\footnote{\rev{Models that explicitly allow for a scale-dependent density response might still work.}}.

On the other hand, the $(\delta, L)$ case shows a notably different break-down scale. It scales only weakly with halo mass and it ranges only between $k_{\mathrm{max}} \sim 0.3 - 0.5 h \mathrm{Mpc}^{-1}$. Note that this corresponds roughly to the Lagrangian scale of haloes with $M \sim 2 \times 10^{14} h^{-1} M_{\odot}$, which have $b_1 \sim 1$. For haloes above $M \gtrsim 3 \times 10^{13} h^{-1} M_\odot$ including the Laplacian increases $k_{\mathrm{max}}$ relative to the density-only case, but for lower masses it decreases it. 

Comparing with the right panel of Figure \ref{fig:b1_b2_coev}, we notice that the bias parameters become already scale-dependent at notably smaller damping scales than $k_{\mathrm{max}}$. For example, for $M_{200\mathrm{b}} \sim 10^{15} h^{-1} M_{\odot}$ with $b_1\sim 3$, the measurements of $\beta_2$ are already scale-dependent beyond $k_{\mathrm{d}} \gtrsim 0.15 h \mathrm{Mpc}^{-1}$, whereas the mathematical break-down scale is $k_{\mathrm{max}} \sim 0.3 h \mathrm{Mpc}^{-1}$. We therefore suggest that considerable care should be taken when setting up bias models close to the break-down scale. This is particularly relevant for hybrid methods, which might in principle allow to describe galaxy clustering at notably smaller scales than are usually considered in perturbative schemes \citep[e.g.][]{Modi_2020, zennaro_2022}.

\section{Estimators for tensorial bias parameters}  \label{sec:tensor_theory}
In Section \ref{sec:theory} we have shown how to infer general estimators of bias parameters associated with scalar variables (like the density and Laplacian) with spatial corrections of any order, and in Section \ref{sec:scalar_measurements} we have shown that these can be used to obtain reliable bias measurements from a single simulation. 

However, the theory in Section \ref{sec:theory} does not explain yet, how to obtain estimators for parameters like the tidal bias $b_{K^2}$ which is defined as the response to
\begin{align}
    K^2 &= \trace{\mat{K} \cdot \mat{K}} \\
    \mat{K} &=  (\myvec{\nabla} \otimes \myvec{\nabla}) \phi - \frac{\delta}{3} \mat{I}_2
\end{align}
where $\phi$ is the displacement potential, $\mat{I}_2$ is the unit matrix and $\mat{K}$ is the traceless tidal tensor. In this section, we will present a general scheme to measure the related bias parameter $b_{K^2}$ and any other bias parameters that follow from contractions of derivatives of the potential field.

To achieve this, it is not optimal to consider directly the distribution of such scalar contracted quantities, since these distributions may get quite complicated. For example, $p(K^2)$ is not a Gaussian distribution, but rather a $\chi^2$ distribution with five degrees of freedom. Further, it is not immediately obvious how to define partial derivatives with respect to such variables, since partial derivatives may depend on what other terms are kept fixed. For example, it is not clear how a term like $K^3 = \trace{\mat{K} \cdot \mat{K} \cdot \mat{K}}$ would be derived with respect to $K^2$. It seems therefore difficult to generalize equation \eqref{eqn:bn_from_pderiv} in this manner.

However, a clear and general framework for measuring such \quotes{tensorial} bias terms can be developed by instead considering the full (quite high dimensional) distribution of the tidal tensor and its derivatives. Since the potential field of the early universe is a Gaussian random field, these must follow a multivariate Gaussian distribution so that it is easy to compute derivatives of the distribution function in a general manner. The resulting \quotes{bias tensors} can be decomposed into isotropic tensors that each have a one to one correspondence with traditionally used bias parameters. In this section, we will introduce the needed mathematical notions step by step and will provide estimators for a few selected bias terms.

\subsection{Tensorial bias expansion}
We write the bias expansion in tensorial form as 
\begin{align}
    F &= 1 + \mat{B}_{\mat{T}} \mat{T} + \frac{1}{2} \mat{T} \mat{B}_{\mat{T} \mat{T}} \mat{T} + ... + \mat{B}_{\mat{S}} \mat{S} + ... + \mat{B}_{\mat{R}} \mat{R} + ... \label{eqn:tensorbiasexp}
\end{align}
where again $F = n_{\mathrm{g}}/n_{\mathrm{g},0}$ (in a \quotes{separate universe} sense) and
\begin{align}
    \mat{T} &= (\nabla \otimes \nabla) \phi \\
    \mat{S} &= (\nabla \otimes \nabla \otimes \nabla) \phi \\
    \mat{R} &= (\nabla \otimes \nabla \otimes \nabla \otimes \nabla) \phi
\end{align}
where $\mat{B}_{\mat{T}}$ is a canonical bias tensor of rank 2, $\mat{B}_{\mat{T} \mat{T}}$ of rank 4, $\mat{B}_{\mat{S}}$ of rank 3 etc. and where an omitted product sign indicates a product over the indices of the last fully symmetric part of the first tensor and the first fully symmetric part of the second tensor. 
For example:
\begin{align}
    \mat{B}_{\mat{T}} \mat{T} &= \mat{B}_{\mat{T}} \mul{2} \mat{T}  = \sum_{ij} B_{\mat{T}, ij} T_{ij} \nonumber \\
    \mat{T} \mat{B}_{\mat{T} \mat{T}} \mat{T} &= \mat{T} \mul{2} \mat{B}_{\mat{T} \mat{T}} \mul{2} \mat{T} = \sum_{ijkl} T_{ij} B_{\mat{T} \mat{T}, ijkl} T_{kl}  \nonumber \\
    \mat{B}_{\mat{S}} \mat{S} &= \mat{B}_{\mat{S}} \mul{3} \mat{S} = \sum_{ijk} B_{\mat{S}, ijk} S_{ijk} \nonumber
\end{align}
where write $\mul{n}$ to explicitly denote the number of dimensions that are contracted. 

Note that, just as before, the bias tensors correspond to derivatives of the large-scale bias function, e.g.
\begin{align}
    \mat{B}_{\mat{T}} &= \left. \frac{\partial }{\partial \mat{T}} F \, \right|_{\mat{T}=0} \\
    \mat{B}_{\mat{T} \mat{T}} &= \left. \left( \frac{\partial }{\partial \mat{T}} \otimes \frac{\partial }{\partial \mat{T}} \right) F \, \right|_{\mat{T}=0} 
\end{align}
\subsection{Isotropic tensors}

Due to isotropy of the universe, each of the bias tensors has to be isotropic. That means that a bias tensor should be identical when measured from a rotated frame of reference. For example, it has to be
\begin{align}
    \mat{U}^T  \mat{B}_{\mat{T}} \mat{U} &= \mat{B}_{\mat{T}}
\end{align}
for any rotation matrix $\mat{U}$. From this it follows immediately that $\mat{B}_{\mat{T}}$ has to be proportional to the unit matrix $\mat{I}_2$ (where the proportionality constant is equal to $b_1$). In general a rank $n$ tensor $\mat{A}$ is isotropic if it holds for any rotation matrix $\mat{U}$:
\begin{align}
    A_{abc...} U_{ai} U_{bj} U_{ck} ... &= A_{ijk...}
\end{align}
where we have used Einstein's sum convention and where the rotation matrix is applied to each index of $\mat{A}$ individually. To express whether some tensor is isotropic, we define $\mathds{U}_{n}$ as the space of all tensors that are isotropic and of rank $n$.

In general, all isotropic tensors can be decomposed in index notation through different combinations of the Kronecker-delta symbol $\delta_{ij}$ and the Levi-Civita symbol $\epsilon_{ijk}$. For example
\begin{align}
    \mat{A} \in \mathds{U}_{2} &\Rightarrow A_{ij} = a \delta_{ij} \\
    \mat{A} \in \mathds{U}_{3} &\Rightarrow A_{ijk} = a \epsilon_{ijk} \\
    \mat{A} \in \mathds{U}_{4} &\Rightarrow A_{ijkl} = a \delta_{ij} \delta_{kl} + b \delta_{ik} \delta_{jl} + c \delta_{il} \delta_{jk}
\end{align}
where $a,b,c \in \mathds{R}$. A compact way of writing the same type of statement is e.g.
\begin{align}
    \mathds{U}_{4} = \myspan{\{ \delta_{ij}\delta_{kl}, \delta_{il}\delta_{jk}, \delta_{ik}\delta_{jl} \}}
\end{align}
which signifies that $\mathds{U}_{4}$ is the space of tensors that can be reached through linear combinations of the indicated tensors and we say that these are basis tensors of $\mathds{U}_{4}$.

It is easy to see that we should be able to decompose each of the bias tensors into a small number of independent scalars that multiply the basis tensors and that correspond to traditional bias parameters. However, before performing such a decomposition we further need to consider the symmetries of the tensors.

\subsection{Symmetric isotropic tensors}
Since the bias tensors correspond to derivatives of a scalar function with respect to symmetric tensors, they have themselves to obey the same symmetries. For example, the tensor $\mat{B}_{\mat{T} \mat{T}}$ has to be symmetric in the first two and last two indices\footnote{And additionally symmetric to exchanges between the first two and last two indices. However, this symmetry follows automatically here, so that we do not consider it explicitly.}:
\begin{align}
    B_{\mat{T} \mat{T}, ijkl} &= B_{\mat{T} \mat{T}, jikl} = B_{\mat{T} \mat{T}, ijlk} = B_{\mat{T} \mat{T}, jilk} \,\,.
\end{align}
Therefore, $\mat{B}_{\mat{T} \mat{T}}$ cannot be every tensor from the tensorspace $\mathds{U}_{4}$, but only from the subspace $\mathds{V}_{22}  	\subseteq \mathds{U}_{4}$ of isotropic rank four tensors that have the $2,2$ symmetry. 

To formalize this, we define $\mathds{V}_{n} \subset \mathds{U}_{n}$ as the space of all isotropic tensor of rank $n$ that are additionally fully symmetric under any permutation of indices. 
\begin{align}
    \mat{B}_{\mat{T}} &\in \mathds{V}_2\\
    \mat{B}_{\mat{S}} &\in \mathds{V}_3 \\
    \mat{B}_{\mat{R}} &\in \mathds{V}_4 \,\,.
\end{align}
Note that $\mathds{V}_3 = \{ \mat{0}\}$, i.e. the only symmetric rank three isotropic tensor is $\mat{0}$. Further, we define $\mathds{V}_{nm}$ as the space of all isotropic tensors of rank $n+m$ that are symmetric in the first $n$ indices and the last $m$ indices. For example,
\begin{align}
    \mat{B}_{\mat{T} \mat{T}} &\in \mathds{V}_{22}.
\end{align}
To find a basis for some symmetric isotropic tensor space of rank $n$, we can consider all basis tensors of $\mathds{U}_n$, symmetrize these and then discard any tensors that are duplicate or zero. We define the symmetrization operator $S_n$ which symmetrizes a tensor in $n$ indices. E.g.
\begin{align}
    S_{2}(M_{ij}) &= \frac{1}{2}(M_{ij} + M_{ji}) \\
    S_{3}(M_{ijk}) &= \frac{1}{6}(M_{ijk} + M_{ikj} + M_{jik} + M_{jki} + M_{kij} + M_{jki})
\end{align}

\noindent Further, we define a double symmetrization operator $S_{nm}$ that  symmetrizes the first $n$ indices and the last $m$ indices. For example, the effect of $S_{22}$ onto a rank four tensor $\mat{M}$ is given in index notation by
\begin{align}
    S_{22}(M_{ijkl}) &= \frac{1}{4} \left( M_{ijkl} + M_{jikl} + M_{ijlk} + M_{jilk} \right)
\end{align}
and it  acts as follows on the basis tensors of $\mathds{U}_{4}$:
\begin{align}
    S_{22}(\delta_{ij} \delta_{kl}) &= \delta_{ij} \delta_{kl} =: I_{22, ij}\\
    S_{22}(\delta_{ik}\delta_{jl}) &= \frac{1}{2} ( \delta_{ik}\delta_{jl} + \delta_{il}\delta_{jk}) =: I_{2=2,ijkl}\\
    S_{22}(\delta_{il}\delta_{jk}) &= \frac{1}{2} ( \delta_{il}\delta_{jk} + \delta_{ik}\delta_{jl}) = I_{2=2, ijkl}
\end{align}
where we have used that $\delta_{ij}$ is itself symmetric. The symmetrized second and third basis tensor are identical. Therefore, $\mathds{V}_{22}$ has (unlike $\mathds{U}_{4}$) only two basis tensors
\begin{align}
    \mathds{V}_{22} &= \myspan{\{\mat{I}_{22}, \mat{I}_{2=2}\}} \,\,.
\end{align}
\begin{figure}
    \centering
    \includegraphics[width=0.6 \columnwidth]{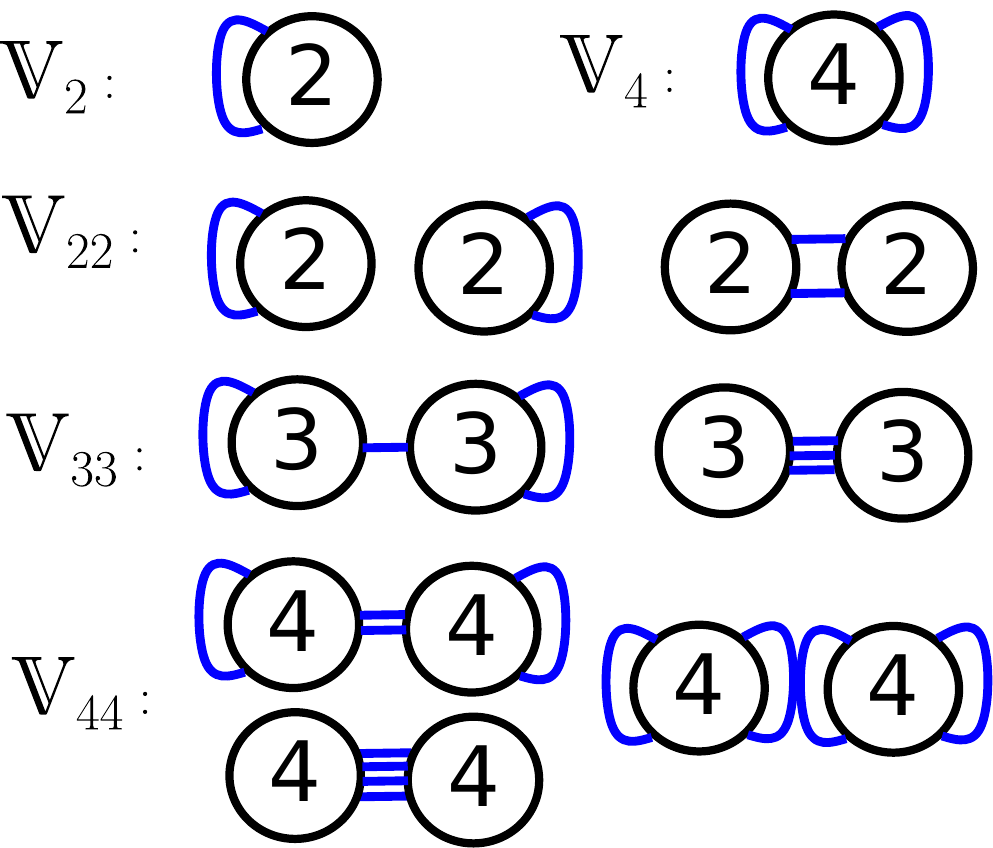}
    \caption{Graphic representation of the isotropic tensors that form a basis for a few selected isotropic tensor spaces with symmetries. All basis tensors of a space with given symmetry can be constructed by considering the number of different ways that the symmetry groups can be connected. In this figure each circle with number $n$ represents a group of $n$ fully symmetric indices and each connection represents one delta symbol (that can either connect two indices from the same group or from two different groups).}
    \label{fig:tensorspaces}
\end{figure}
\noindent The main difference between $\delta_{ij}\delta_{kl}$ and the two tensors $\delta_{ik}\delta_{jl}$ and $\delta_{il}\delta_{jk}$ is that the delta symbols used to define $\delta_{ij}\delta_{kl}$ each connect internally inside the groups of symmetric indices and have no connections between the groups. (Recall that $i \leftrightarrow j$ and $k  \leftrightarrow l$ are to be symmetrized.) However, $\delta_{ik}\delta_{jl}$ and $\delta_{il}\delta_{jk}$ each have zero group internal connections, but two connections between the symmetry groups. In fact, the symmetrization operation identifies all terms that have the same number of intra-/inter- group connections and therefore we only need to consider how many independent possibilities exist to connect the symmetry groups to identify the basis tensors of any space $\mathds{V}_{mn}$. We can therefore represent the basis tensors of each tensor space through a simple diagram as illustrated in Figure \ref{fig:tensorspaces}. We label these tensors through a symbol that shows the number of inter group connections in the index:
\begin{align}
    \mathds{V}_{33} &= \myspan{\{\mat{I}_{3-3}, \mat{I}_{3\equiv3}\}} \\
    \mathds{V}_{44} &= \myspan{\{\mat{I}_{44}, \mat{I}_{4=4}, \mat{I}_{4 \superequiv 4}\}} \\
    \mathds{V}_{24} &= \myspan{\{\mat{I}_{24}, \mat{I}_{2=4}\}}
\end{align}
We explain in Appendix \ref{app:tens3sym}, how to construct tensors with three or more symmetry groups. For these cases the procedure is mostly analogous, but it additionally needs to be considered whether there is symmetry with respect to permutations of the different symmetry groups.

\subsection{Orthogonal basis} \label{sec:orthbasis}
\begin{table}
    \caption{Orthogonal basis tensors that we consider here.}
    \label{tab:isotropictensors}
    \centering
    \begin{tabular}{c|l|c|c}
         Tensor & Index Representation & Norm$^2$ & $\phi$ contr. \\
         \hline
$       J_{2}$ & $                                                                                  S_{2}(\delta_{ij})$ & $                   3$ & $\delta$\\
$      J_{22}$ & $                                                                      S_{22}(\delta_{ij}\delta_{kl})$ & $                   9$ & $\delta^2$\\
$     J_{2=2}$ & $                         S_{22}(\delta_{ik}\delta_{jl}) - \frac{1}{3} S_{22}(\delta_{ij}\delta_{kl})$ & $                   5$ & $K^2$\\
$       J_{4}$ & $                                                                       S_{4}(\delta_{ij}\delta_{kl})$ & $                   5$ & $L$\\
$     J_{3-3}$ & $                                                           S_{33}(\delta_{ij}\delta_{kl}\delta_{mn})$ & $        \frac{25}{3}$ & $(\nabla \delta)^2$\\
$J_{3\equiv3}$ & $  S_{33}(\delta_{ik}\delta_{mj}\delta_{ln}) - \frac{3}{5} S_{33}(\delta_{ij}\delta_{kl}\delta_{mn})$ & $                   7$  & $S_{3\equiv3}$\\
$      J_{24}$ & $                                                           S_{24}(\delta_{ij}\delta_{kl}\delta_{mn})$ & $                  15$  & $\delta L$\\
$     J_{2=4}$ & $  S_{24}(\delta_{ik}\delta_{jl}\delta_{mn}) - \frac{1}{3} S_{24}(\delta_{ij}\delta_{kl}\delta_{mn})$ & $        \frac{35}{6}$ & $\phi_{2=4}$
    \end{tabular}
    \tablefoot{ The first column indicates the label we use for the tensor, the second column one possible way of defining it in index notation, the third column the square of its norm (as the full contraction with itself) and the last column the term that is obtained when fully contracting it with the corresponding derivatives of the potential as in the bias expansion. }
\end{table}
The basis tensors highlighted in the last section are sufficient to uniquely decompose the bias tensors. However, it is advantageous to define the basis tensors in such a way that they are orthogonal to each other.

For example, we can decompose an isotropic tensor $\mat{M} \in \mathds{V}_{22}$ as
\begin{align}
     \mat{M} &= A \mat{I}_{22} + B \mat{I}_{2=2} \,\,.
\end{align}
Its full contraction with $\mat{I}_{22}$ is
\begin{align}
     \mat{M} \mul{4} \mat{I}_{22} &= A (\mat{I}_{22} \mul{4} \mat{I}_{22}) + B (\mat{I}_{2=2} \mul{4} \mat{I}_{22}) \\
           &= 9 A + 3 B
\end{align}
which includes contributions of both $A$ and $B$. We could infer $A$ and $B$ from $\mat{M}$ by additionally contracting with $\mat{I}_{2=2}$ and then solving the resulting system of equations.

However, it would be desirable to have a simple way of finding the coefficients. Therefore, we define an orthogonal basis such that
\begin{align}
    \mat{J}_a \mul{r} \mat{J}_b &= \begin{cases} \norm{\mat{J}_a}^2 \quad \textrm{if } a = b \\ 0 \quad \textrm{else} \end{cases}
\end{align}
where the contraction goes over the full rank $r$ of $\mat{J}_a$ and $\mat{J}_b$ (which must have equal rank). Such a basis can be found through Gram-Schmidt orthogonalization and is for example given for $\mathds{V}_{22}$ by:
\begin{align}
    \mat{J}_{22} &= \mat{I}_{22} \\
    \mat{J}_{2=2} &= \mat{I}_{2=2} - \frac{1}{3} \mat{I}_{22} \,\,.
\end{align}
Then we can easily find the coefficients as
\begin{align}
     \mat{M} &= \alpha \mat{J}_{22} + \beta \mat{J}_{2=2} \\
    \alpha &= \frac{\mat{M} \mul{4} \mat{J}_{22}}{\norm{\mat{J}_{22} }^2} 
           = \frac{1}{9} \mat{M} \mul{4} \mat{J}_{22}\\
    \beta  &= \frac{\mat{M} \mul{4} \mat{J}_{2=2}}{\norm{\mat{J}_{2=2}}^2} 
           = \frac{1}{5} \mat{M} \mul{4} \mat{J}_{2=2} \,\,.
\end{align}
Note that the thus inferred coefficients are different than the $A$ and $B$ and the chosen basis will also affect the inferred bias parameters.

It is worth noting that a decomposition of a bias tensor into different bases
\begin{align}
    \mat{B}_{\mat{T}\mat{T}} &= b_{\mat{I}_{22}} \mat{I}_{22} + b_{\mat{I}_{2=2}} \mat{I}_{2=2} \nonumber \\
                             &= b_{\mat{J}_{22}} \mat{J}_{22} + b_{\mat{J}_{2=2}} \mat{J}_{2=2}
\end{align}
corresponds to different independent scalar variables appearing in the bias expansion. For example, if we track the corresponding term from equation \eqref{eqn:tensorbiasexp}, we find
\begin{align}
    \mat{T} \mat{B}_{\mat{T}\mat{T}} \mat{T} &= b_{\mat{I}_{22}} \trace{\mat{T}}^2 + b_{\mat{I}_{2=2}} \trace{\mat{T}^2} \nonumber \\ &= b_{\mat{I}_{22}} \delta^2 + b_{\mat{I}_{2=2}} T^2\\
    \mat{T} \mat{B}_{\mat{T}\mat{T}} \mat{T} &= b_{\mat{J}_{22}} \trace{\mat{T}}^2 + b_{\mat{J}_{2=2}} \left(\trace{\mat{T}^2} - \frac{\delta^2}{3} \right) \nonumber \\ &= b_{\mat{J}_{22}}\delta^2 + b_{\mat{J}_{2=2}} K^2
\end{align}
so that the non-orthogonal basis leads to $\delta$ and $T^2$ as independent terms, whereas the orthogonal basis leads to using $\delta$ and $K^2$ as independent terms. The corresponding bias parameters can be directly identified $b_{\mat{J}_{22}} = b_2$ and $\frac{1}{2} b_{\mat{J}_{2=2}} = b_{K^2}$.

We list an overview of the orthogonal basis tensors that we will use here in Table \ref{tab:isotropictensors}. While it is possible to derive the full algebra of products between isotropic tensors from the index representations, this is rather cumbersome. Therefore, we have written a short code that creates explicit numerical representations of these tensors and with which it is easy to evaluate (and decompose) different types of products of these tensors. We use this together with the symbolic computer algebra system \textsc{sympy} \citep{sympy} to systematically compute symbolic representations of expressions in the following sections. This code is openly available\footnote{\url{https://github.com/jstuecker/probabilistic-bias}}.

\subsection{Bias estimators}
Given the orthogonal basis tensors, we define the tensorial bias parameters through 
\begin{align}
    b_{\mat{J}_{X}} &= \left. \frac{\partial^N F}{\partial \mat{T}_0^N} \mul{2N} \frac{\mat{J}_{X}}{\norm{ \mat{J}_X }^2} \right|_{\mat{T}_0=0}
\end{align}
where the rank of $\mat{J}_{X}$ is $2N$. While for scalar terms, as in equation \eqref{eqn:pbs_bias}, the bias parameters are simply given by derivatives of the galaxy number with respect to the corresponding scalar, more generally we define bias parameters for any tensorial terms as derivatives of the galaxy number with respect to the tidal tensor (or higher spatial order tensors) that are contracted with the corresponding isotropic tensor.

In complete analogy to equation \eqref{eqn:bn_tensor} it follows that these parameters can be estimated as
\begin{align}
    b_{\mat{J}_{X}} &= (-1)^N \expect{\frac{1}{p} \frac{\partial p}{\partial \mat{T}^N} \mul{2N} \frac{\mat{J}_{X}}{\norm{ \mat{J}_X }^2}}_g \label{eqn:tensorbiasestimators}
\end{align}
where $p$ is the full distribution of the tidal tensor $\mat{T}$.

Bias terms corresponding to higher spatial derivatives like $\mat{R}$ and $\mat{S}$ can be defined analogously.

\subsection{Tidal bias}
We show in Appendices \ref{app:covariances} and \ref{app:tidal_distribution} that the distribution of the tidal tensor is given by
\begin{align}
    p(\mat{T}) &= N \exp \left(- \frac{1}{2} \mat{T}^T \mat{C}_{\mat{T}}^+ \mat{T} \right)
\end{align}
where $\mat{C}_{\mat{T}}^+$ (the generalized inverse of the covariance matrix, as explained in the Appendix) is given by
\begin{align}
    \mat{C}_{\mat{T}}^+ &= \frac{1}{\sigma_0^2} \mat{J}_{22} + \frac{15}{2 \sigma_0^2} \mat{J}_{2=2} \,\,.
\end{align}
Taking derivatives yields
\begin{align}
    \frac{1}{p} \frac{\partial p}{\partial \mat{T}} &=  - \mat{C}_{\mat{T}}^+ \mat{T} \\
    \frac{1}{p} \frac{\partial^2 p}{\partial \mat{T}^2} &= (\mat{T} \mat{C}_{\mat{T}}^+) \otimes (\mat{C}_{\mat{T}}^+ \mat{T}) - \mat{C}_{\mat{T}}^+ \,\,.
\end{align}
and we find
\begin{align}
   b_{\mat{J}_2} = b_1 &=  \expectgal{\frac{\mat{J}_{2} \mat{C}^+ \mat{T}}{\norm{\mat{J}_{2}}^2}} =  \expectgal{\frac{\mat{J}_{2} \mat{T}}{\sigma_{0}^{2}}} = \expectgal{\frac{\delta}{\sigma_{0}^{2}}} \label{eqn:bj2_estimator}.
\end{align}
Here, it appeared a product between isotropic tensors $\mat{J}_{2} \mul{2} \mat{J}_{22} = 3 \mat{J}_{2}$. We evaluate terms of this type systematically with numerical representations as described in Section \ref{sec:orthbasis}.
Naturally, this estimator of $b_{\mat{J}_2} = b_1$ is consistent with the one that we have inferred previously in equation \eqref{eqn:b1_o0}.

At second order we find the estimators
\begin{align}
    b_{\mat{J}_{22}} = b_2 &= \expectgal{\frac{\mat{T}^T \mat{J}_{22} \mat{T} - \sigma_{0}^{2}}{\sigma_{0}^{4}}} \nonumber \\
                     &= \expectgal{\frac{\delta^2 - \sigma_0^2}{\sigma_0^4}} \\
    b_{\mat{J}_{2=2}} = 2 b_{K^2} &=  \frac{15}{4 \sigma_{0}^{4}} \expectgal{3 \mat{T}^T \mat{J}_{2=2} \mat{T} - 2 \sigma_{0}^{2} }\\
    &= \frac{15}{4 \sigma_{0}^{4}} \expectgal{3 K^2 - 2 \sigma_{0}^{2}} \label{eqn:bk2_o0_estimator}
\end{align}
We note that conventionally the bias parameter $b_{K^2}$ is defined so that it appears with a pre-factor $K^2$ in the bias expansion whereas our parameter $b_{\mat{J}_{2=2}}$ appears with a factor $\frac{1}{2} K^2$ in the expansion (after contracting the corresponding tensors) so that it is $b_{K^2} = \frac{1}{2} b_{\mat{J}_{2=2}}$. Although we think that our fore-factor convention makes in principle more sense, since $b_{\mat{J}_{2=2}}$ corresponds to a second derivative term, we will still present results in terms of the conventional notation (e.g. $b_{K^2}$) throughout this paper.

We note that just as the large-scale value of $\delta / \sigma^2$ at an object's location is -- motivated by equation \eqref{eqn:b1_o0} -- sometimes called the \quotes{bias-per-object} \citep[e.g.][]{paranjape_2018, contreras_2023}, one could refer in a similar manner to the value of
\begin{align}
    b_{K^2, 1h} &= \frac{15}{8 \sigma_{0}^{4}} (3 K^2 - 2 \sigma_{0}^{2})
\end{align}
as the value of the \quotes{tidal bias-per-object}.

Just as estimators with spatial corrections could be obtained for the density biases by considering the joint distribution of $p(\delta, L)$, we can obtain higher spatial order estimators for the tidal bias by considering the joint distribution of second and fourth potential derivatives $p(\mat{T}, \mat{R})$ and evaluating equation \eqref{eqn:tensorbiasestimators} for this distribution. We show how to derive the bias estimator for this case in appendices \ref{app:distr_fourthderiv} and \ref{app:tidal_estim_with_corr}. The resulting estimator is
\begin{align}
    b_{\mat{J}_{2=2}}   &= \frac{15}{4 \sigma_{*}^{8}} \expectgal{3 K^2 \sigma_{2}^{4} + 6 \phi_{2=4} \sigma_{1}^{2} \sigma_{2}^{2} + 3 \phi_{4=4} \sigma_{1}^{4} - 2 \sigma_{2}^{2} \sigma_{*}^{4}} \label{eqn:bk2_o2_estimator}
\end{align}
where
\begin{align}
    \phi_{2=4} &:= \mat{T} J_{2=4} \mat{R} = \sum_{ij} (\partial_i \partial_j \phi) (\partial_i \partial_j \delta) - \frac{1}{3} \delta \nabla^2 \delta \\
    \phi_{4=4} &:= \mat{T} J_{4=4} \mat{R} = \sum_{ij} (\partial_i \partial_j \delta)^2 - \frac{1}{3} (\nabla^2 \delta)^2
\end{align}
and $\sigma_{*}^2 = \sigma_{0}^{2} - \sigma_{1}^{4} / \sigma_{2}^{2}$.

\subsection{Estimators for third derivative terms}

The distribution of third derivatives of the potential $\mat{S}$ is derived in Appendix \ref{app:distr_thirdderiv} and is given by a multivariate Gaussian with the generalized inverse covariance matrix
\begin{align}
   \mat{C}_{\mat{S}}^+ &= \frac{3}{\sigma_1^{2}} \mat{J}_{3-3} + \frac{35}{2 \sigma_1^{2}} \mat{J}_{3\equiv3}
\end{align}
Note that it is fine to consider the third spatial derivatives independently of the second derivatives, since the joint distribution factorizes
\begin{align}
    p(\mat{T}, \mat{S}) &= p(\mat{T}) p(\mat{S})
\end{align}
We find bias estimators by evaluating equation \eqref{eqn:tensorbiasestimators} as:
\begin{align}
    b_{\mat{J}_{3-3}} &= \frac{3}{\sigma_1^4} \expectgal{\mat{S} \mat{J}_{3-3} \mat{S} -  \sigma_{1}^{2}} \nonumber \\
                      &= \frac{3}{\sigma_1^4}  \expectgal{(\nabla \delta)^2 - \sigma_{1}^{2}} \label{eqn:j3-3_estimator} \\
    b_{\mat{J}_{3\equiv3}} &= \frac{35}{\sigma_1^4} \expectgal{5 \mat{S} \mat{J}_{3\equiv3} \mat{S}  - 2 \sigma_{1}^{2}} \nonumber \\
                      &= \frac{35}{\sigma_1^4} \expectgal{5 \phi_{3\equiv3} - 2 \sigma_{1}^{2}} \label{eqn:j3---3_estimator}
\end{align}
where 
\begin{align}
    \phi_{3\equiv3} &= \sum_{ijk} (\partial_{ijk} \phi)^2 - \frac{3}{5} \sum_i (\partial_{i} \delta)^2
\end{align}
and where we note again that conventions may differ by a factor $2$ so that e.g. $b_{\mat{J}_{3-3}} = 2 b_{(\nabla \delta)^2}$ if $b_{(\nabla \delta)^2}$ is the bias parameter that would multiply $(\nabla \delta^2)$ without any forefactor in a bias expansion.

\subsection{Tensorial cumulant biases} \label{sec:tenscumulants}
As for the scalar case, it is possible to define tensorial cumulant bias parameters instead of tensorial canonical bias parameters. Unfortunately, interesting differences arise only at orders that are higher than the ones that we have considered here. However, for the benefit of future studies, we still want to briefly line out, how to define tensorial cumulant biases.

Analogously to equation \eqref{eqn:tensorbiasexp}, we consider a tensorial expansion of $\log F$
\begin{align}
    \log F &= 1 + \mat{\beta}_{\mat{T}} \mat{T} + \frac{1}{2} \mat{T} \mat{\beta}_{\mat{T} \mat{T}} \mat{T} + ... + \mat{\beta}_{\mat{S}} \mat{S} + ... + \mat{\beta}_{\mat{R}} \mat{R} + ... 
\end{align}
Taking derivatives at zero and identifying coefficients yields the relations
\begin{align}
    \mat{\beta}_{\mat{T}} &= \mat{B}_{\mat{T}} \\
    \mat{\beta}_{\mat{T} \mat{T}} &= \mat{B}_{\mat{T} \mat{T}} - \mat{B}_{\mat{T}} \otimes \mat{B}_{\mat{T}}
\end{align}
To decompose this into isotropic tensors we may contract these equations with the isotropic tensors to find the relations
\begin{align}
    \beta_1 = \beta_{\mat{J}_2} &= b_{\mat{J}_2} = b_1 \\
    \beta_{2} = \beta_{\mat{J}_{22}} &= b_{\mat{J}_{22}} - \beta_{\mat{J}_2}^2 = b_2 - b_1^2 \\
    2 \beta_{K^2} = \beta_{\mat{J}_{2=2}} &= b_{\mat{J}_{2=2}} = 2 b_{K^2}
\end{align}
Which is consistent with what we have already derived earlier. Tensorial cumulant biases become more interesting at third order where third derivative of the bias Function yields
\begin{align}
    \beta_{\mat{T}\mat{T}\mat{T}} &= \mat{B}_{\mat{T}\mat{T}\mat{T}} - 3 S_{222}(\mat{B}_{\mat{T}} \otimes \mat{B}_{\mat{T}\mat{T}}) + 2 S_{222} (\mat{B}_{\mat{T}} \otimes \mat{B}_{\mat{T}} \otimes \mat{B}_{\mat{T}})
\end{align}
If we contract this with the isotropic tensor $S_{222}(J_{2=2} \otimes J_2)$, we have
\begin{align}
    \beta_{\mat{J}_{2=22}} &= b_{\mat{J}_{2=22}} - 3 b_{\mat{J}_{2}} b_{\mat{J}_{2=2}}
\end{align}
Therefore, if the full multivariate bias function is close to Gaussian, also tensorial bias parameters of this form should be expected to be close to zero. Dealing with such objects goes beyond this study and we will leave this as an interesting proposition for future work.

\section{Measurements of tensorial bias parameters} \label{sec:tensorbiasmeas}
In this section we will use the estimators that we have derived in Section \ref{sec:tensor_theory} to perform measurements of tensorial bias parameters. We will focus here on the tidal bias $b_{K^2}$ and on the biases associated with third potential derivatives $b_{(\nabla \delta)^2} = \frac{1}{2} b_{\mat{J}_{3-3}}$ and $b_{\mat{J}_{3\equiv3}}$. For the tidal bias we have additionally derived the estimator with spatial corrections of order two, so that we can use it to validate that the presented theory indeed behaves as expected. Further, we will briefly check the effect of assembly bias on $b_{K^2}$ which has previously been reported to be quite significant. For the measurements in this section we use the same simulations and jackknife technique as described in sections \ref{sec:simulation}  and \ref{sec:measure_techinque}.

\subsection{Tidal bias} \label{sec:measure_tidalbias}

\begin{figure}
    \centering
    \includegraphics[width=\columnwidth]{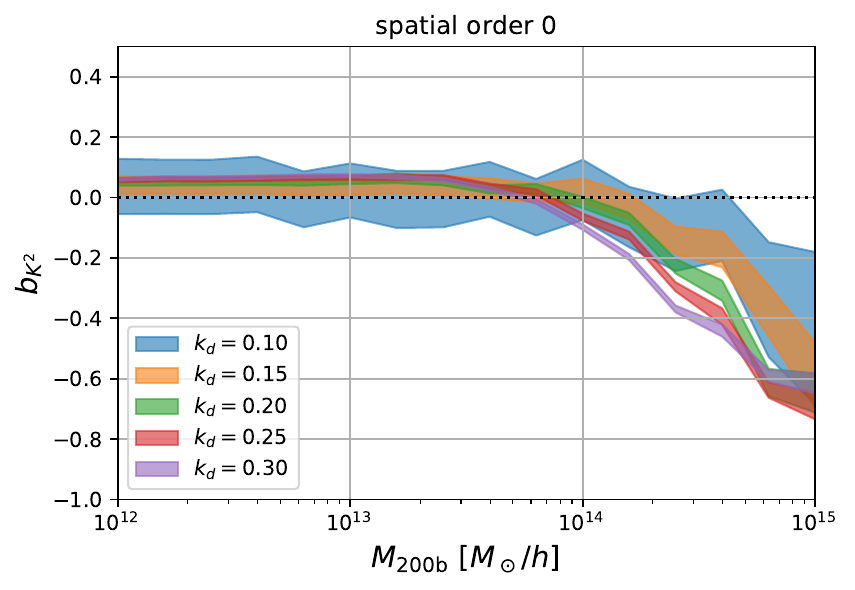}
    \includegraphics[width=\columnwidth]{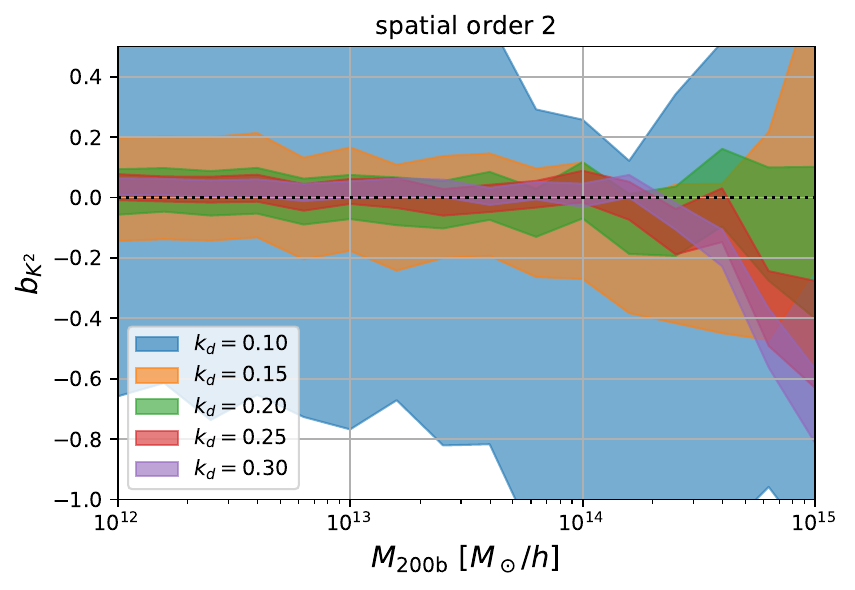}
    \caption{Tidal bias as a function of mass and for different damping scales measured with the spatial order 0 estimator (top) versus the spatial order 2 estimator (bottom). The spatial order 2 estimator exhibits significantly reduced scale dependence at the price of increased errorbars.}
    \label{fig:bk2_vs_m}
\end{figure}

\begin{figure}
    \centering
    \includegraphics[width=\columnwidth]{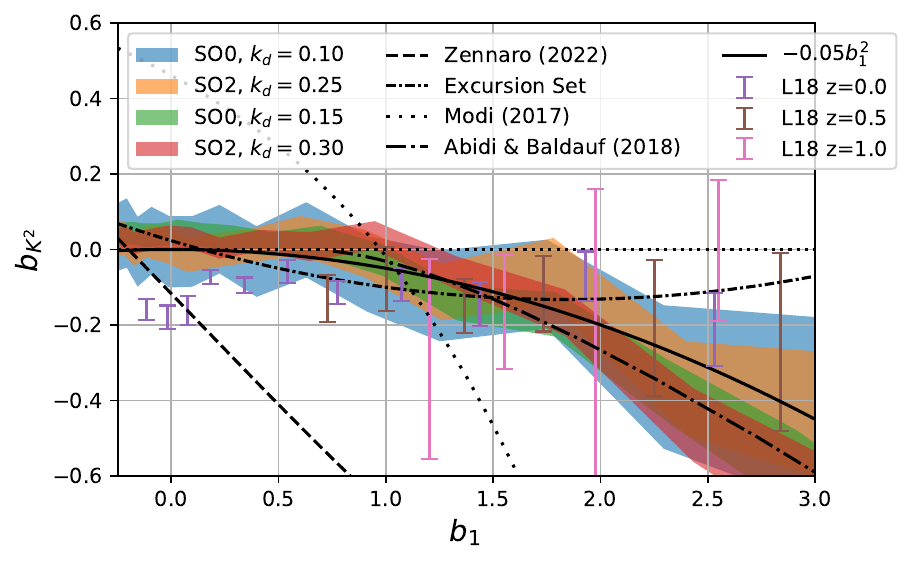}
    \caption{Coevolution relation $b_{K^2}$ versus $b_1$. Coloured shaded regions are our measurements and errorbars show the measurements from \citet{Lazeyras_2018}. Black lines include measurements from \citet{zennaro_2022}, \citet{modi_2017} and \citet{abidi_baldauf_2018}, an excursion set prediction from \citet{sheth_2013} and  an approximation that we suggest $b_{K^2} = -0.05 b_1^2$.}
    \label{fig:bk2_vs_b1}
\end{figure}

We measure the tidal bias $b_{K^2} = \frac{1}{2} b_{\mat{J}_{2=2}}$ with the estimators from equation \eqref{eqn:bk2_o0_estimator} at spatial order zero and equation \eqref{eqn:bk2_o2_estimator} at spatial order two and show the results as a function of mass in Figure \ref{fig:bk2_vs_m}. The order zero estimator exhibits a strong scale dependence at scales $k_{\mathrm{d}} \gtrsim 0.15 h \mathrm{Mpc}^{-1}$. However, overall it seems that the tidal bias is very small in amplitude and that a slightly negative Lagrangian tidal bias is preferred at high halo masses. On the other hand, the spatial order two estimator has a much lower scale dependence and seems mostly consistent across different damping scales, with maybe a slight inconsistency at very high halo masses and large $k_{\mathrm{d}}$. This comes at the price of a significantly increased uncertainty in the measured bias values. However, this demonstrates that the estimators that we have derived in Section \ref{sec:tensor_theory} behave indeed in the way that we would expect. More precise measurements of the tidal bias could easily be obtained by employing a larger simulation volume or by averaging over several realizations.

In Figure \ref{fig:bk2_vs_b1} we show the coevolution relation between $b_{K^2}$ and $b_1$. Here, we show for each estimator only the smallest two scales that appear reasonably unbiased in Figure \ref{fig:bk2_vs_m}. These are $k_{\mathrm{d}} = 0.1 h \mathrm{Mpc}^{-1}$ and $k_{\mathrm{d}} = 0.15 h \mathrm{Mpc}^{-1}$ for the spatial order zero estimator and $k_{\mathrm{d}} = 0.25 h \mathrm{Mpc}^{-1}$ and $k_{\mathrm{d}} = 0.3 h \mathrm{Mpc}^{-1}$ for the spatial order two estimator. Noteworthy, the estimators of different orders agree well with each other. For the larger $k_{\mathrm{d}}$ values the estimators have smaller errorbars, but appear slightly biased at large values of $b_1 \gtrsim 2$. A rough approximation to the observed functional shape is given by
\begin{align}
    b_{K^2} = -0.05 b_1^2 \label{eqn:bk2_coev}
\end{align}
though the uncertainty is of course quite high. However, in comparison to some previous literature estimates, we find very small values of $b_{K^2}$. The first measurement that we compare with is from \citet{modi_2017}. Their relation is stated for Eulerian bias parameters $b_{ K^2}^\mathrm{E}(b_{1}^\mathrm{E})$ and we convert it to a Lagrangian relation through $b_1 = b_{1}^\mathrm{E} - 1$ and $b_{K^2} = b_{K^2}^\mathrm{E} + 2/7 b_1$ \citep{sheth_2013, Desjacques_2018}. Similar to other studies \citep{Lazeyras_2018, zennaro_2022}, we find a strong disagreement between our measurements and the \citet{modi_2017} measurements. Further, we show as a dashed line in Figure \ref{fig:bk2_vs_b1} the coevolution relation of the tidal bias that was presented in \citet{zennaro_2022}. This relation has been measured at very small scales $k \sim 0.7 h \mathrm{Mpc}^{-1}$ and is not really expected to accurately represent the large-scale tidal bias. However, it is interesting to see that the large-scale tidal response that we measure here is so much weaker than the response that can be measured at small scales. Next, we compare with the excursion set prediction from \citet{sheth_2013} parameterized as $b_{1}^{\mathrm{E}} = 0.524 - 0.547 b_{1}^{\mathrm{E}} + 0.046 (b_{1}^{\mathrm{E}})^2 $. This prediction seems to be in better agreement with our measurements, since it generally predicts smaller amplitudes for the tidal bias and a slight negative tendency. However, the predicted upturn in the function  at $b_1 \gtrsim 2$ does not seem to be present in any of our measurements.

Further, we compare our measurements with the Lagrangian tidal bias inferred by \citet{abidi_baldauf_2018}. For this, we use their fitted $b_{K^2}(M)$ relation and map masses to $b_1$ values through the \citet{Tinker_2010} bias relation evaluated for their simulated cosmology. Our measurements are in good qualitative and quantitative agreement with the results from \citet{abidi_baldauf_2018} when compared at the same value of $b_1$. 

Finally, we compare our data with the measurements from \citet{Lazeyras_2018}, again transformed from Eulerian to Lagrangian parameters, indicated through error-bars in Figure \ref{fig:bk2_vs_b1}. Our measurements agree roughly with that study, in that we favor a very small tidal bias with a preferentially slightly negative amplitude. We suggest therefore, that high Lagrangian tidal biases $|b_{K^2}| > 0.5$ for mass-selected haloes are strongly disfavored for lowly biased objects $b_1 \ll 2$. 

However, we also note that there are some statistically significant differences at $b_1 \sim 0$, where we find tidal bias consistent with zero, whereas the data of \citet{Lazeyras_2018} clearly favor a negative value of $b_{K^2} \sim -0.2$. Given \revb{our agreement with} \citet{abidi_baldauf_2018}\revb{,} the fact that past erroneous measurements have rather predicted too large absolute values of the tidal bias rather than too small ones, and that the three lowest $b_1$ data points of \citet{Lazeyras_2018} do not seem really consistent with the overall behaviour of the curve, we suggest that our measurements are more reliable here and that the Lagrangian tidal bias becomes indeed very small $|b_{K^2}| \ll 0.1$ for $b_1 \lesssim 0.5$.

We conclude that overall the amplitude of the Lagrangian $b_{K^2}$ is quite low for mass selected haloes. Given the challenges of the measurement and the history of errors, it would be desirable to have a clear measurement that does not suffer from the usual scale dependencies and large uncertainties. This could be achieved through anisotropic separate universe simulations as they have been presented in \citet{stuecker_2021,masaki_2020, akitsu_2021}. We checked whether the simulations that were presented in \citet{stuecker_2021} are sufficient to measure the tidal bias at the level that we observe here. The simulation parameter choices were not optimal for measuring it, since those simulations do not have sufficient statistics of high mass objects -- which are the objects that would have a significant tidal bias according to equation \eqref{eqn:bk2_coev}. However, a reliable measurement could be obtained by running with larger tidal fields that are tuned to measure the tidal bias at higher redshift, where most objects have a larger $b_1$ and likely a larger $b_{K^2}$. However, that said, we find that the simulations have significant enough statistics to constraint an upper bound  $|b_{K^2}| < 0.1$  (roughly 1$\sigma$ level) at low masses $M_{\mathrm{200b}} \sim 10^{13} h^{-1}M_{\odot}$, consistent with our observations here.

\subsection{Assembly bias in $b_{K^2}$}

\begin{figure}
    \centering
    \includegraphics[width=\columnwidth]{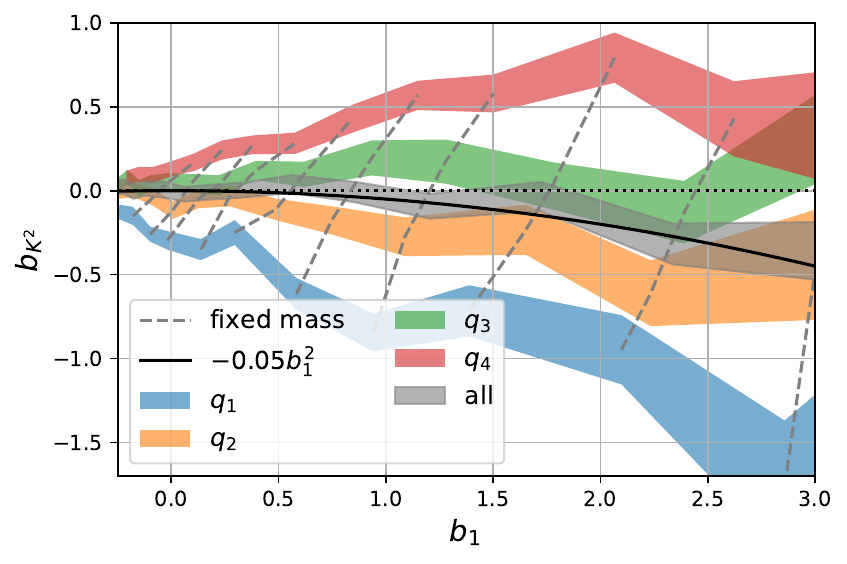}
    \caption{Assembly bias in the Lagrangian $b_{K^2}(b_1)$ relation for haloes split into four quartiles by spin (at fixed mass). All lines use the spatial order 2 estimator and $k_{\mathrm{d}} = 0.25 h \mathrm{Mpc}^{-1}$ as damping scale. Dashed lines show the variation of $b_1$ and $b_{K^2}$ at fixed mass. The $b_{K^2} - b_1$ relation shows a strong degree of assembly bias -- with larger spin selections yielding larger $b_{K^2}$ and larger $b_1$.}
    \label{fig:bk2_assembly}
\end{figure}

While the tidal bias seems to be quite low for mass selected haloes, it is not at all obvious that this should also be the case for galaxies. The formation and selection of galaxies may depend on other properties that are more sensitive to the tidal field, like the spin, the formation time and the halo concentration. For example, in tidal torque theory the spin of an object should depend on the Lagrangian tidal tensor \citep{white_1984}.
Therefore, it is important to investigate whether there exists assembly bias with respect to such properties. \citet{lazeyras_2021} have shown that the tidal bias $b_{K^2}^{\mathrm{E}}$ and the associated $b_{K^2}(b_1)$ coevolution relation depend strongly on secondary halo properties. Here, we will briefly check whether we can reproduce these results.

We measure the spin of each halo in our simulation following \citet{bullock_2001} as
\begin{align}
    \lambda &= \frac{|\myvec{L}|}{\sqrt{2} v_{200c} r_{200c}}
\end{align}
where 
\begin{align}
    \myvec{L} &= \expect{\myvec{r} \times \myvec{v}}_{\mathrm{particles}}
\end{align}
is the specific angular momentum averaged over all particles that are within the bound component of the main subhalo according to \textsc{subfind} and $r_{200c}$ is the radius within which the density is 200 times the critical density of the Universe and $v_{200c}$ the circular velocity at that radius.

We then group haloes in each mass bin by their spin into four quartiles $q_1, q_2, q_3$ and $q_4$, $q_1$ corresponding to the $25\%$ haloes with lowest spin and $q_4$ to the $25\%$ with highest spin and measure the bias parameters independently in each quartile $q_1$. We show the resulting $b_{K^2} (b_1)$ relation in Figure \ref{fig:bk2_assembly}. Here, we present only results for the spatial order 2 estimator at a damping scale of $k_{\mathrm{d}} = 0.25 h \mathrm{Mpc}^{-1}$, but we have checked that other choices lead to the same results. 

The $b_{K^2} - b_1$ relation exhibits a strong degree of assembly bias with halo spin.  For example, at $b_1 \sim 1$ the tidal bias of mass selected haloes seems very small $|b_{K^2}| \ll 0.1$, but after the selection on halo spin, values may range between $-1$ and $0.5$. In individual mass bins, the spin selection increases both the value of $b_1$ and the value of $b_{K^2}$.

The sign, the $b_1$ dependence and the overall amplitude of the spin-assembly bias are all in good agreement with the measurements of \citet{lazeyras_2021}. For example, at $b_{1} = 2$ we find assembly bias variations of $\Delta b_{K^2} \sim 2$ (between low and high spins) consistent with their variations at $b_{1}^{E} = 3$ of around $\Delta b_{K^2}^{E} \sim 2$ and the assembly bias decreases towards lower $b_1$ and increases towards higher $b_1$. Therefore, our measurements provide a validation of the results of that study, but it also shows that our method allows to reliably measure the tidal bias. 

While we do not present it here, we have also tested whether we can reproduce the concentration assembly bias that was measured by \citet{lazeyras_2021}. We have found that if we also define concentration through $v_{\mathrm{max}}/v_{200c}$, then we find an assembly bias of a similar amplitude. However, if concentrations are defined through fits of the halo density profile, the assembly bias is smaller (possibly consistent with 0) and prefers the opposite sign. The concentration assembly bias seems therefore to be a bit more uncertain than that of the spin.

We conclude that even if $b_{K^2}$ seems quite small for mass selected haloes, the tidal bias may not so easily be neglected for tracers that follow more general selection criteria. Values of the Lagrangian $b_{K^2}$ of order unity (and slightly larger) are a plausible possibility for realistic galaxy catalogues.

\subsection{Biases of third derivative terms}

\begin{figure}
    \centering
    \includegraphics[width=\columnwidth]{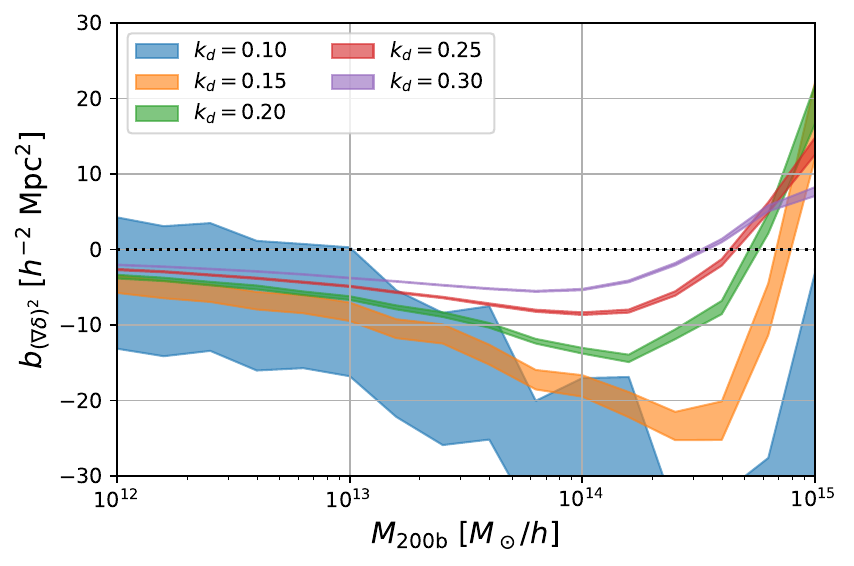}
    \includegraphics[width=\columnwidth]{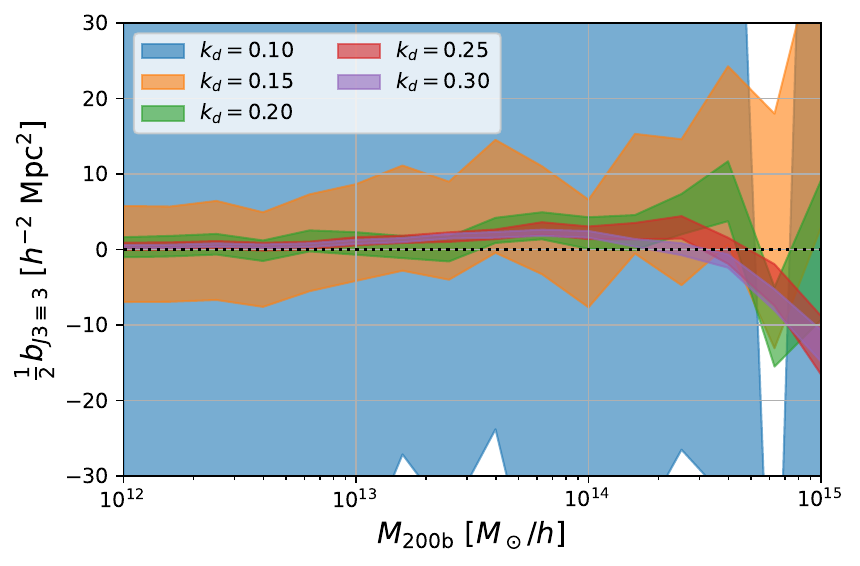}
    \caption{Measurements of biases terms of third derivatives of the potential. The bias of the density gradient $b_{(\nabla \delta)^2}$ (top) has a very strong scale dependence so that we cannot reliably measure it here. Bottom: the fully contracted third derivatives of the potential seem to have a very small associated bias parameter. }
    \label{fig:bj3---3}
\end{figure}

In Figure \ref{fig:bj3---3} we show the measurements that we obtain by evaluating the bias estimators from equations \eqref{eqn:j3-3_estimator} and \eqref{eqn:j3---3_estimator} and where we again divide by a factor $2$ to get $b_{(\nabla \delta)^2} = b_{\mat{J}_{3-3}} / 2$ in accordance with the literature convention.

The value of $b_{(\nabla \delta)^2}$ appears to depend strongly on the damping scale and we can therefore not claim a reliable measurement here. However, it seems mostly $b_{(\nabla \delta)^2} < 0$ which seems consistent with the idea that haloes should preferentially form at peaks where density gradients should be small. However, at very large masses $b_{(\nabla \delta)^2}$ seems to become positive. We are not quite sure how to interpret this behavior, but it might just be an artefact of the uncorrected scale dependencies. It is worth noting that the negative amplitude of $b_{(\nabla \delta)^2}$ is quite significant $b_{(\nabla \delta)^2} \sim 10 h^{-2} \mathrm{Mpc}^2$ and even grows towards larger scales. We will therefore discuss later whether this (usually neglected) bias term is worth including under some circumstances. We note that $b_{(\nabla \delta)^2}$ has also been measured by \citet{biagetti_2014}. However, since we have not properly corrected the scale dependence here, we abstain from a comparison.

On the other hand $b_{\mat{J}_{3\equiv3}}$ appears to be quite small $|b_{\mat{J}_{3\equiv3}}/2| \lesssim 5 h^{-2} \mathrm{Mpc}^2$ for everything except the very high mass end. Therefore, the contributions of $b_{\mat{J}_{3\equiv3}}$ to the bias expansion should be significantly lower than $b_{(\nabla \delta)^2}$.

\section{Relevance of bias parameters} \label{sec:relevance}

\begin{table}
    \caption{Sets of bias parameters that are shown in Figure \ref{fig:bias_relevance}. }
    \label{tab:bias_relevance_choices}
    \centering
    \begin{tabular}{c|c|c|c}
     parameter &   $10^{14} h^{-1} M_\odot$ &  $10^{15} h^{-1} M_\odot$ & unit\\
    \hline
            $b_1$               &       0.54 &       3.08 &\\
            $b_2$               &      -0.89 &       7.20 &\\
            $b_3$               &      -1.93 &       7.81 &\\
            $b_4$               &       7.81 &     -32.30 &\\
            $b_L$               &     -13.92 &      17.76 & $h^{-2} \mathrm{Mpc}^2$\\
        $b_{K^2}$               &       0.04 &      -0.17 &\\
        $b_{(\nabla \delta)^2}$ &     -13.81 &      22.87 & $h^{-2} \mathrm{Mpc}^2$\\
   $\frac{1}{2} b_{3 \equiv 3}$ &       2.10 &      -1.08 & $h^{-2} \mathrm{Mpc}^2$\\
   \hline
            $\beta_2$           &      -1.13 &    -2.25   &\\
            $\beta_3$           &      -0.33 &    -0.66   &\\
            $\beta_4$           &      0.19  &    -1.11   &\\
    \end{tabular}
    \tablefoot{
       The first set corresponds to $M \sim 10^{14} h^{-1} M_\odot$ and has moderate bias values whereas the second set with $10^{15} h^{-1} M_\odot$ is strongly biased. The last two lines show cumulant bias parameters that can be used instead of $b_3$ and $b_4$.
    }
\end{table}

\begin{figure}
    \centering
    \includegraphics[width=\columnwidth]{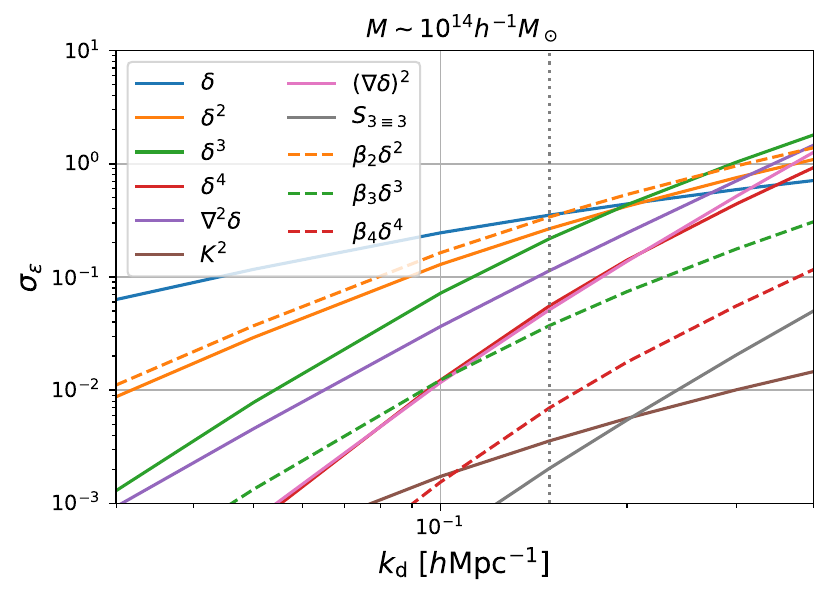}
    \includegraphics[width=\columnwidth]{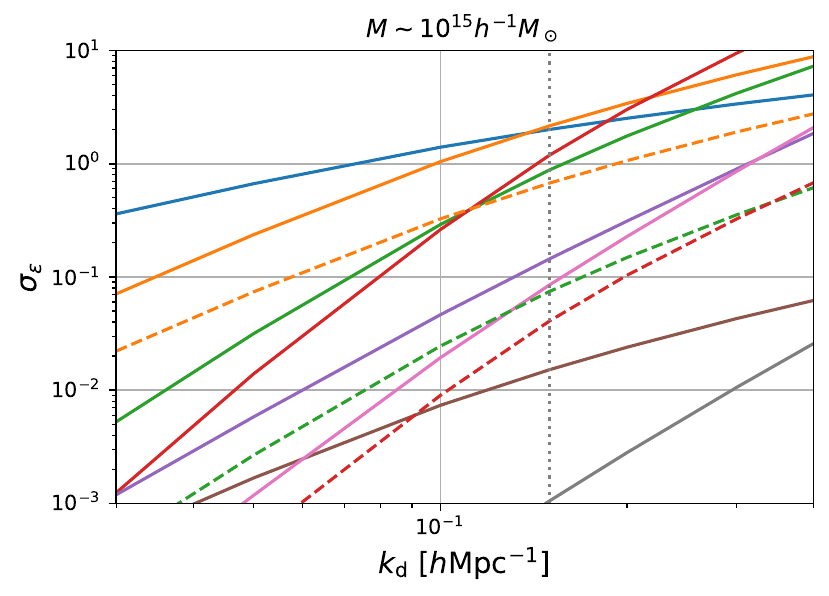}
    \caption{Quantitative contribution of different bias terms to the bias expansion as a function of damping scale and for two different sets of tracers. To avoid cluttering, we omit the bias fore-factors of some terms in the legend. The vertical dotted line indicates an example scale of interest $k_{\mathrm{d}} = 0.15 h \mathrm{Mpc}^{-1}$. The dashed lines show the contributions of cumulant biases at order three and four which can directly be compared to the contributions of $b_3$ and $b_4$ (labeled as $\delta^3$ and $\delta^4$).}
    \label{fig:bias_relevance}
\end{figure}

We aim to estimate the relevance of different contributions to the bias expansion. If we have a term of the form $\frac{1}{n!} b_X \cdot X$ of the renormalized bias expansion, then we can estimate its relative relevance by comparing the typical difference in the predicted galaxy density field between a bias function $f_X$ that includes the term versus an expansion $f_0$ that neglects it 
\begin{align}
    \sigma_{\epsilon,X}^2 &= \expect{(f_X - f_0)^2 } \nonumber \\
                      &= \left( \frac{b_X}{n!} \right)^2 \expect{X^2}
\end{align}
where the value of $\expect{X^2}$ depends on the adopted damping scale and where $\sigma_{\epsilon,X}$ is a dimensionless quantity that can be compared across different bias terms. For example, for the contribution of $b_3$ we have
\begin{align}
     \sigma_{\epsilon,\delta^3}^2 &= \left( \frac{b_3}{6} \right)^2 \expect{(\delta^3 - 3\delta\sigma^2)^2}
\end{align}
The value of $\sigma_{\epsilon,X}$ quantifies the typical error that is made at the field level (in Lagrangian space) when neglecting the corresponding term. 

We choose two halo sets $M \sim 10^{14} h^{-1} M_\odot$ and $M \sim 10^{15} h^{-1} M_\odot$ as examples and measure all their bias parameters with a damping scale $k_{\mathrm{d}} = 0.2 h \mathrm{Mpc}^{-1} $ and use the spatial order two estimators for all parameters (except $b_{(\nabla \delta)^2}$ and $b_{3 \equiv 3}$). We list the corresponding parameters in Table \ref{tab:bias_relevance_choices}. The first set represents a moderately biased population whereas the second set represents an example of strongly biased tracers. We then keep these parameters fixed and evaluate $\expect{X^2}$ at many different damping scales to obtain $\sigma_{\epsilon,X}$ which we show in Figure \ref{fig:bias_relevance}.

For the moderately biased tracers the relevance of parameters decreases reasonably between different orders. However, this statement depends strongly on the considered damping scale. E.g. at $k_{\mathrm{d}} = 0.4 h \mathrm{Mpc}^{-1}$  many different terms reach similar amplitudes, whereas at $k_{\mathrm{d}} = 0.05 h \mathrm{Mpc}^{-1}$ there is a strong relevance order between different terms. If we aimed to fit a bias expansion at the field level at $k_{\mathrm{d}} = 0.15 h \mathrm{Mpc}^{-1}$ (marked as a vertical dotted line), then we would find that for the moderately biased set $b_1 > b_2 > b_3 > b_4$ all lie within a factor two of each other and for the strongly biased set we even have a changed order $b_2 > b_1 > b_4 > b_3$ and all of these lie within a factor of three with each other. This is of course very problematic, since it does not at all seem clear that higher order terms are smaller than lower order terms here, which casts doubts on the convergence of the expansion. Of course this problem can be controlled by simply going to larger smoothing scales, but at the cost of a reduced constraining power. However, another possibility is to make use of the advantages of the cumulant bias parameters. With these parameters significant parts of higher order terms get already absorbed by lower order parameters. For example, consider the  first three terms of the bias expansion in terms of cumulant parameters:
\begin{align}
    f &= 1 + \beta_1 \delta + \frac{1}{2} (\beta_2 + \beta_1^2) (\delta^2 - \sigma^2) \nonumber \\
     & \quad \quad + \frac{1}{6} (\beta_3 + 3 \beta_1 \beta_2 + \beta_1^3) (\delta^3 - \delta \sigma^2) + ...
\end{align}
Here $\beta_1$ and $\beta_2$ contribute also to the third term and to infinitely many higher order terms. If we write out the polynomial to high orders, but set $\beta_3 = 0$, then the leading order error we would make is
\begin{align}
    \sigma_{\epsilon,\beta_3 \delta^3}^2 &= \left( \frac{\beta_3}{6} \right)^2 \expect{(\delta^3 - 3\delta\sigma^2)^2}
\end{align}
which is identical to the error term from the canonical bias terms, but with a different fore-factor. Error estimates of these types are indicated as dashed lines in Figure \ref{fig:bias_relevance}. Strikingly these are significantly smaller than the canonical terms and it seems now plausibly to reach good convergence at $k_{\mathrm{d}} = 0.15 h \mathrm{Mpc}^{-1}$. E.g. to reach $\sigma_{\epsilon} \leq 0.1$ for the strongly biased set we would only need to consider the terms ($\beta_1$, $\beta_2$, $\beta_L$).

It is note-worthy that the contribution of $b_{(\nabla \delta)^2}$ is not much smaller than that of $b_L$ so that it might make sense to include this parameter more commonly in the bias expansion at scales $k_{\mathrm{d}} \gtrsim 0.1 h \mathrm{Mpc}^{-1}$. However, we remind the reader that there are some other possibly important parameters here which we have not considered, so that we cannot present a comprehensive picture of the relevance of bias parameters.

We conclude that it may have significant advantages to phrase the bias expansion in terms of the cumulant parameters $\beta_n$ instead of the canonical bias parameters -- especially so when considering highly biased tracers or small smoothing scales where the rate of convergence of the canonical expansion is excruciatingly slow. 

\section{Conclusions} \label{sec:conclusions}

In this article we have presented a method to measure bias parameters through moments of the galaxy environment distribution $p(\delta | g)$. We have shown that such estimators can be derived both for scalar bias parameters like $b_1$ and $b_L$ and for tensorial ones like $b_{K^2}$ and $b_{(\nabla \delta)^2}$ and that they can be made scale independent by considering spatial corrections at various orders. We have verified the reliability of these estimators by recovering well established relations between known parameters and we have additionally used our estimators to measure so far unknown terms like $b_{3 \equiv 3}$ and $b_{(\nabla \delta)^2}$  -- where the latter appears to be particularly large so that it might be worth to include it more commonly in the bias expansion. 

The main benefits of \rev{the} method are its simplicity and its generality. It is only required to evaluate the linear field of a simulation at the Lagrangian locations of a set of tracers -- e.g. given by the most bound id in a halo catalogue -- and to take simple expectation values. This measurement uses a minimal number of discretization steps -- making it numerically robust and easy to understand. \rev{While a similar approach has previously been used by } \citet{paranjape_2013a, paranjape_2013b}, \rev{we have shown that it is possible to incorporate spatial corrections at any order to get accurate reconstructions of large scale bias parameters even at fairly large damping scales $k_{\mathrm{d}}$.}  \rev{In contrast to measurement approaches that depend on forward models, e.g. of power spectra or correlation functions, measured bias parameters are significantly more independent of each other in the method at hand. For example,} the measurement of $b_1$ is completely independent of the assumed order of the expansion in powers of $\delta$. This makes it generally unlikely that parameters numerically compensate for the absence of other terms. Finally, we note that all bias parameters can be evaluated at a low computational cost with a single simulation and that it is possible to use arbitrary filtering methods with the slight alterations explained in Appendix \ref{app:filtering}.

However, to avoid confusion we want to point out a few limitations of our method for estimating biases. First of all, we note that our method assumes that the background distribution is known analytically. Therefore, it does not immediately translate to situations where this is not the case. For example, at third order in the bias expansion, time-derivatives of the tidal tensor appear which depend on the second Lagrangian perturbation theory potential $\phi^{(2)}$ which is not a Gaussian random field. Therefore, the formalism can only be translated to the associated bias parameters if the distribution of $\phi^{(2)}$ and its derivatives can be written (or at least adequately approximated) analytically. Further, the formalism cannot directly be applied to the Eulerian galaxy environment distribution $p^E(\delta|g)$,  since the non-linear Eulerian matter distribution $p^E(\delta)$ does also not follow Gaussian statistics. 

Beyond the novel bias estimators, we have also newly introduced the cumulant bias parameters 
\begin{align}
    \beta_n &= \left. \frac{\partial^n}{\partial \delta_0^n} \log \left( \frac{n_g (\delta_0)}{n_{g,0}} \right) \right|_{\delta_0 = 0}
\end{align}
These cumulant bias parameters are proportional to the cumulants of the galaxy environement distribution\footnote{Except for $\beta_2$, which has an additional contribution.} and they relate to the canonical bias parameters exactly in the same way as cumulants relate to moments. When measuring these parameters, we have found them to be more independent from each other than canonical bias parameters. Further, we have shown that for haloes cumulant biases of order $n \geq 3$ are very close to zero. This has several intriguing consequences: 

(1) The galaxy environment distribution and the bias function of haloes are very well approximated by a Gaussian. This motivates the usage of a Gaussian bias model at order $n=2$ and it motivates e.g. to expand around a Gaussian bias model for orders $n > 2$. We will discuss the Gaussian bias model in detail in a companion paper and show that it has several desirable properties \citepgaus. 

(2) Coevolution relations of haloes at order $n \geq 3$ are equivalent to $\beta_n = 0$. Therefore, one may predict such relations between canonical bias parameters at very high orders by combining $\beta_n=0$ with the mapping between canonical and cumulant bias parameters. However, this also suggests that such coevolution relations may not carry much physical significance, but are rather the artefacts of a suboptimal basis choice. An expansion around a Gaussian model would naturally incorporate all of these relations at any order.

(3) The bias expansion may have significantly improved convergence when phrased in terms of cumulants instead of canonical bias parameters. At the field level the convergence of the canonical expansion seems already questionable at scales $k_{\mathrm{d}} \gtrsim 0.15 h \mathrm{Mpc}^{-1}$, since parameters may easily grow as their order grows. However, when phrased in terms of cumulant biases, we expect neglected higher order terms to be significantly less important, as they are already well captured through lower order parameters. Therefore, we expect cumulant biases to be most beneficial wherever the canonical expansion shows poor convergence -- e.g. at high masses, at late times or at small smoothing scales.

\begin{acknowledgements}
The authors thank Simon White, Oliver Hahn and Fabian Schmidt for helpful discussions and comments to the draft. JS thanks Oliver Philcox for helpful discussions. We acknowledge funding from the Spanish Ministry of Science and Innovation through grant number PID2021-128338NB-I00. 
RV acknowledges the support of the Juan de la Cierva fellowship (FJC2021-048002-I). MPI is supported by STFC consolidated grant no. RA5496.
\end{acknowledgements}

%-----
% for the bibliography, at the end
\bibliographystyle{aa} % style aa.bst
\bibliography{aa51176-24} % your references archive.bib

%-------------------------------------------------------------------
\begin{appendix} 

\section{Estimators with different filters} \label{app:filtering}
In Section \ref{sec:theory}, we have shown derivations under the assumption of a sharp $k$-space filter. For this choice all considered bias functions and bias parameters relate trivially to the `separate universe' parameters that are commonly referred to in the literature. However, in general other filtering kernels could also be chosen. For example if we define the smoothed density as
\begin{align}
    \delta_l(k) &= W(k) \delta(k)
\end{align}
where $\delta_l$ is the smoothed density and $\delta$ the unsmoothed one. If the filter $W$ is for example chosen to be a Gaussian kernel
\begin{align}
    W_{\mathrm{g}}(k) = \exp \left(-\frac{k^2}{2 k_{\mathrm{d}}^2} \right)
\end{align}
instead of a sharp $k$ filter, then some considerations have to be adapted. In principle, all of the formula from Section \ref{sec:theory} could still be applied, but they would refer to a different set of bias parameters that describe the response to perturbations of a different shape. Rather than introducing a new set of bias parameters, it is more convenient to directly write the commonly used parameters in terms of the new basis. Therefore, we will show here how to directly measure the `separate universe' bias parameters from the moments of the galaxy environment distribution that was obtained with a density field that was filtered with any kernel.

First of all, we consider how a bias function on very large scales $f_l(\delta_l)$ should relate to the bias function on some much smaller scale $f_s(\delta_s)$. If the PBS is still valid on the smaller scale, then the probability of forming a galaxy, given knowledge of $\delta_s$ and $\delta_l$ depends only on the value of $\delta_s$:
\begin{align}
    p(\mathrm{g} | \delta_s, \delta_l) &= p(\mathrm{g} | \delta_s) \,\,.
\end{align}
Under this assumption, we can re-express the large scale bias function through the small scale function marginalized over the conditional probability of the small scale densities:
\begin{align}
    f_l(\delta_l) = \frac{p(\mathrm{g} |\delta_l)}{p(\mathrm{g})} &= \int \frac{p(\mathrm{g} | \delta_s)}{p(g)} p(\delta_s | \delta_l)  d \delta_s \nonumber \\
                   &= \int f_s(\delta_s) p(\delta_s | \delta_l)  d \delta_s \,\,.
\end{align}
Since, the joint distribution of $\delta_s$ and $\delta_l$ is a multivariate Gaussian, the conditional distribution corresponds to a multivariate Gaussian that is conditioned on the value of one variable, which itself is a Gaussian with modified mean and covariance:
\begin{align}
        p(\delta_s | \delta_l) &= N(\delta_s, \mu_*, \sigma_*) \\
        \mu_* &= \alpha \delta_l \\
        \sigma_*^2 &= \sigma_{ss}^2 - \alpha^2 \sigma_{ll}^2
\end{align}
where $\alpha = \sigma_{ls}^2 / \sigma_{ll}^2$ corresponds to the correlation between smoothed large scale and small scale density field,  $\sigma_{ll}^2 = \expect{\delta_l \delta_l}$, $\sigma_{ls}^2 = \expect{\delta_l \delta_s}$ and $\sigma_{ss} = \expect{\delta_s \delta_s}$. 

If we define the renormalized bias function $F_\alpha$ through the bias function $f_l$ as in Section \ref{sec:theory}, then we find
\begin{align}
    F_\alpha(\delta_0) &:= \int f_l(\delta_l) p(\delta_l - \delta_0) \mathrm{d}\delta_l \nonumber \\
                  &= \int f_s(\delta_s) p_{ss}(\delta_{ss} - \alpha \delta_0)  d \delta_s
\end{align}
where $p_{ss}$ is simply the background distribution of the small scale densities. This can be shown either by evaluating the integrals (corresponding to convolutions) explicitly or by considering the large scale limit of $f_l$ where $\sigma_* \rightarrow \sigma_{ss}$. 

We can now relate the renormalized bias functions obtained with different filtering methods, since we may assume that $f_s$ is independent of the filter employed on large scales. Therefore,
\begin{align}
    F_{\alpha_1} \left(\frac{\delta_0}{\alpha_1} \right) = F_{\alpha_2} \left(\frac{\delta_0}{\alpha_2} \right)
\end{align}
where we have indicated the usage of different filters through different $\alpha$ in the index. Therefore, different filters simply lead to a rescaling of the renormalized bias function. The value of $\alpha$ depends on the filter and the smoothing scale. If we assume that $\delta_s$ was obtained with some small scale filter $W_s$ and $\delta_l$ with $W_l$, we have
\begin{align}
    \sigma_{ls}^2 &= \expect{\int \delta(\myvec{k}_1) W_s(k_1) \mathrm{d}^3\myvec{k}_1 \int \delta(\myvec{k}_2) W_l(k_2) \mathrm{d}^3\myvec{k}_2 } \nonumber \\
                  &= \int \int \expect{\delta(\myvec{k}_1) \delta(\myvec{k}_2)} W_s(k_1) W_l(k_2) \mathrm{d}^3\myvec{k}_1 \mathrm{d}^3\myvec{k}_2 \nonumber \\
                  &= \int \int P(k_1) \delta_D(\myvec{k}_1 - \myvec{k}_2) W_s(k_1) W_l(k_2) \mathrm{d}^3\myvec{k}_1 \mathrm{d}^3\myvec{k}_2 \nonumber \\
                  &= \int P(k) W_s(k) W_l(k) \mathrm{d}^3\myvec{k} \\
                  &\approx \int P(k) W_l(k) \mathrm{d}^3\myvec{k}
\end{align}
where the last approximation assumes that $W_s$ is very close to 1 over the support of $W_l$, because $W_l$ corresponds to a much larger scale smoothing than $W_s$. Further, we have
\begin{align}
    \sigma_{ll}^2 &= \int P(k) W_l^2(k) \mathrm{d}^3\myvec{k} \\
    \alpha &= \frac{\int P(k) W_l(k) \mathrm{d}^3\myvec{k}}{\int P(k) W_l^2(k) \mathrm{d}^3\myvec{k}} \,\,.
\end{align}
It is easy to see that $\alpha = 1$ for any sharp Fourier space filter, independent of the filtering scale. However, for other filters it is important to note that $\alpha$ may be generally larger than $1$, for example $\alpha \sim 2$ for a Gaussian filter at scales of $k_{\mathrm{d}} \sim 0.1 h \mathrm{Mpc}^{-1}$. Further, $\alpha$ depends generally on the filtering-scale. The renormalized bias function does therefore only correspond to a well-defined limit at arbitrary large scales if $\alpha$ is constant (as for the sharp $k$ filter).

Since the case $\alpha=1$ seems clearly to be the simplest and leads to a sensibly scale-independent renormalization, we consider it as the reference case and simply label its renormalized bias function $F$, and we have focused in the main-text on the discussion of this case. However, we can still use the relations derived in the main-text to measure the properties of $F$ directly through measuring aspects of some filter-dependent function $F_{\alpha}$. 

We will directly show this for the general multivariate case here by adapting the considerations from Section \ref{sec:multivariate_cumulants}. For the multivariate case the renormalized bias functions relate as
\begin{align}
    F_{\mat{A}}(\myvec{x}_0) &= F(\mat{A} \myvec{x}_0)
\end{align}
where
\begin{align}
    \mat{A} &= \mat{C}_{ls} \mat{C}_{ll}^{-1} \\
    \mat{C}_{ls} &= \expect{ \myvec{x}_l \otimes \myvec{x}_s } \\
    \mat{C}_{ll} &= \expect{ \myvec{x}_l \otimes \myvec{x}_l }
\end{align}
where $\mat{C}_{ls}$ may more conveniently be measured as $\expect{ \myvec{x}_m \otimes \myvec{x}_m }$ where $\myvec{x}_m$ is evaluated in the linear fields filtered with $\sqrt{W_l}$.

The cumulant-generating function of the galaxy environment distribution with $\delta_l$ is again given through equation \eqref{eqn:multivariate_cumgen}, just with $F_{\mat{A}}$ instead of $F$:
\begin{align}
    K_{\mat{A}}(\myvec{t}) &= \frac{1}{2} \myvec{t}^T \mat{C}_{ll} \myvec{t} + \log \left(  F_{\mat{A}}(\mat{C}_{ll} \myvec{t}) \right) \nonumber \\
                           &= \frac{1}{2} \myvec{t}^T \mat{C}_{ll} \myvec{t} + \log \left(  F(\mat{C}_{ls} \myvec{t}) \right)
\end{align}
For measuring the cumulant bias parameters, we may then define a more convenient variable
\begin{align}
    \myvec{u}_* &= \mat{C}_{ls} \myvec{x}_0
\end{align}
which has the cumulant generating function
\begin{align}
    K_{\myvec{u}_*}(\myvec{t}) &= \frac{1}{2} \myvec{t}^T  \mat{C}_{ls}^{-1}  \mat{C}_{ll} \mat{C}_{ls}^{-1} \myvec{t} + \log \left(  F(\myvec{t}) \right)
\end{align}
so that the cumulant bias parameters can be inferred by differentiating, leading to:
\begin{align}
    \myvec{\beta}_{ijk...} &= \begin{cases}
        \kappa_{u_*,ijk...}  & \mathrm{\quad if \quad} i + j + k + ...  \neq 2  \\
        \kappa_{u_*,ijk...} - C^{-1}_{*,ab} & \mathrm{\quad if \quad} i + j + k + ...  = 2
    \end{cases}
\end{align}
where $\mat{C}_*^{-1} = \mat{C}_{ls}^{-1}  \mat{C}_{ll} \mat{C}_{ls}^{-1}$. We can therefore measure the separate universe bias parameters with an arbitrarily filtered distribution in a very similar manner to the one using a sharp $k$-space filter. It is worth noting that the inferred relations show also an important property: If $\beta_n$ vanish for $n > 2$, that automatically implies that the corresponding cumulants vanish for any filtered galaxy environment distribution. Therefore, if the bias function is Gaussian for one filter, then it will also appear Gaussian for any other filter.

\section{Isotropic tensors}
\FloatBarrier

\subsection{Tensors with three or more symmetry groups} \label{app:tens3sym}

\begin{figure}
    %\centering
    \includegraphics[width=0.8 \columnwidth]{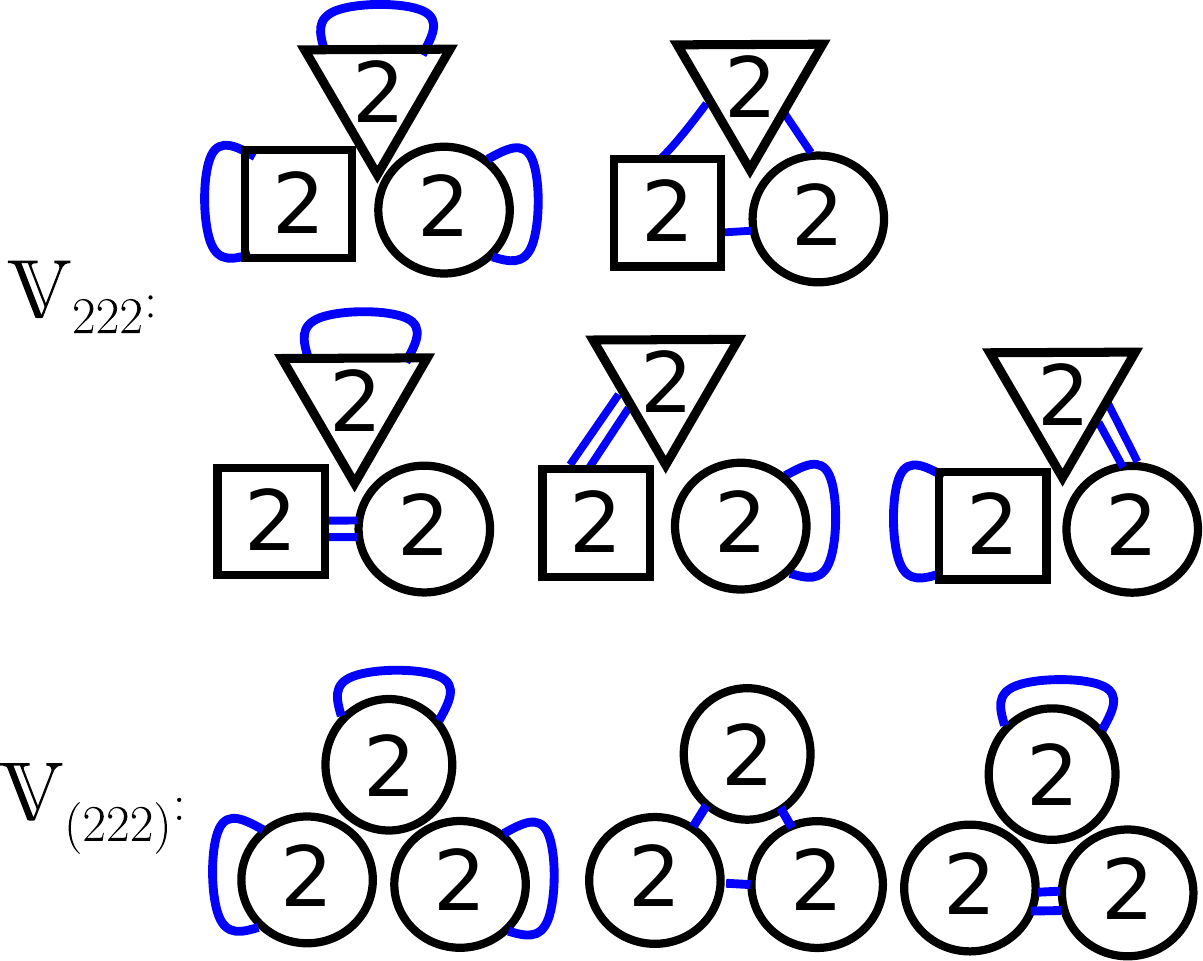}
    \caption{Illustration of the basis tensors of $\mathds{V}_{222}$ versus those of $\mathds{V}_{(222)}$. Top: without the symmetry requirement between groups (indicated by usage of different symbols for each group) there are five basis tensors. Bottom: With the additional symmetry requirement between groups (indicated by usage of the same symbol for each group), since the latter three tensors of  $\mathds{V}_{222}$ get all united to a single symmetrized term (bottom).}
    \label{fig:v222_diagram}
\end{figure}

In the main text we have only explained, how to obtain basis tensors for tensor spaces with at most two symmetry groups. However, for higher order biases like e.g. the equivalent of $b_3$ the isotropic tensor decomposition requires consideration of additional symmetries. We define $\mathds{V}_{222}$ as the space of tensors that are symmetric under permutations of the first two, second two and third two indices. Further, we define $\mathds{V}_{(222)}$ as the space of tensors that are additionally invariant to exchanges of these symmetry groups. These spaces have different number of basis tensors
\begin{align}
    \mathds{V}_{222} &= \myspan{\{\mat{I}_{222}, \mat{I}_{2-2-2-}, \mat{I}_{2=22}, \mat{I}_{22=2}, \mat{I}_{222=}\}} \\
    \mathds{V}_{(222)} &= \myspan{\{\mat{I}_{222}, \mat{I}_{2-2-2-}, \mat{I}_{(2=22)}\}}
\end{align}
where
\begin{align}
    \mat{I}_{(2=22)} &= \frac{1}{3} (\mat{I}_{2=22} + \mat{I}_{22=2} + \mat{I}_{222=})
\end{align}
is additionally symmetrized between exchanges between the symmetry groups and the other tensors are given by
\begin{align}
     I_{222,ijklmn} &= S_{222}(\delta_{ij}\delta_{kl}\delta_{mn}) \\
     I_{2-2-2-,ijklmn} &= S_{222}(\delta_{ik}\delta_{lm}\delta_{nj}) \\
     I_{2=22,ijklmn} &= S_{222}(\delta_{ik}\delta_{jl}\delta_{mn}) \\
     I_{22=2,ijklmn} &= S_{222}(\delta_{ij}\delta_{km}\delta_{ln}) \\
     I_{222=,ijklmn} &= S_{222}(\delta_{im}\delta_{kl}\delta_{jn})
\end{align}
The basis tensors of $\mathds{V}_{(222)}$ are visualized through a diagram in Figure \ref{fig:v222_diagram}. The third derivative tensor of that would appear in the bias expansion can be decomposed in this space
\begin{align}
    \left. \frac{\partial^3 F}{\partial \mat{T} \partial \mat{T} \partial \mat{T}} \right|_{\mat{T} = 0}   \in \mathds{V}_{(222)}
\end{align}

\subsection{Covariance of potential derivatives} \label{app:covariances}
To find an explicit form for the distribution of the derivatives of the tidal tensor and other higher order derivatives of the potential, we need to find their covariance matrix and invert it.

For this we first consider the covariance between any derivatives of the potential. This is given by
\begin{align}
    C_{ij...,ab...} &= \langle \phi_{ij...} \phi_{ab...} \rangle \nonumber \\
                    &= \frac{1}{(2 \pi)^6} \left\langle \int \int \frac{(ik_i)(ik_j) ... (ik_a) (ik_b) ...}{k^4} \right. \nonumber \\
                    &\quad \quad \quad \quad \quad \left. \vphantom{\int} e^{-i (\myvec{k_1} - \myvec{k_2}) \myvec{x}} \delta_k^*(\myvec{k}_1) \delta_k(\myvec{k}_2) \mathrm{d}\myvec{k}_1 \mathrm{d}\myvec{k}_2  \right\rangle
\end{align}
where $\delta_k$ is the Fourier transform of density field $\delta$ and we have used that $\delta$ is a real field so that $\delta_k(\myvec{k}) = \delta_k^*(-\myvec{k})$. Using the definition of the power spectrum 
\begin{align}
\langle \delta_k^*(\myvec{k}_1) \delta_k(\myvec{k}_2)  \rangle &= (2 \pi^3) P(\myvec{k}) \delta_\mathrm{D}^3(\myvec{k}_1 - \myvec{k}_2)
\end{align}
where $\delta_\mathrm{D}$ is the Dirac delta function, we can simplify
\begin{align}
    C_{ij...,ab...} &= \frac{1}{(2 \pi)^3} \int \frac{(ik_i)(ik_j) ... (ik_a) (ik_b) ...}{k^4} P(k) \mathrm{d}^3\myvec{k} \label{eqn:covariance_potderiv}
\end{align}
We can see that this expression does not depend on the order of the indices nor whether they stem from the first or the second potential term. Therefore, we just count the total number $n_1$ of occurrences of $k_x$, $n_2$ of $k_y$ and $n_3$ of $k_z$. Without loss of generality we can order the coordinates so that $n_1 \geq n_2 \geq n_3$. Further, we can switch to spherical coordinates and replace the cosine of the azimutal variable:
\begin{align}
    \myvec{k} &= \begin{pmatrix} k \mu \\ k (1 - \mu)^2 \cos \phi \\ k (1 - \mu)^2 \sin \phi )^T \end{pmatrix}
\end{align}
so that we have
\begin{align}
    \Sigma(ij...ab) &= \frac{\sigma_M^2}{4 \pi} \int_{-1}^{1} \mu^{n_1} \mathrm{d} \mu  \int_{0}^{2 \pi} \cos^{n_2}(\phi) \sin^{n_3}(\phi) \mathrm{d} \phi \\
         \sigma_M^2 &= \frac{1}{2 \pi^2} \int P(k) k^{2M} \mathrm{d}k \\
                  M &= \frac{n_1 + n_2 + n_3 - 4}{2} \,\,.
\end{align}
These integrals can easily be evaluated by hand or with a computer algebra system. We list a few examples of the covariances in Table \ref{tab:covariances} Note that only terms where each $n_1$, $n_2$ and $n_3$ are even are non-zero. Note that all uneven terms are zero, so that e.g. the 2 second derivatives of the potential have zero covariance with the third derivatives, but a non-zero covariance with the fourth derivatives.
\begin{table}
    \caption{Covariances of various derivatives of the potential. }
    \label{tab:covariances}
    \centering
    \begin{tabular}{c|c|c|c|c}
Example Terms & $n_1$ & $n_2$ & $n_3$ & Value \\
\hline
$\left\langle \phi_{11}^2 \right\rangle$ & 4 & 0 & 0 & $\frac{\sigma_{0}^{2}}{5}$ \\
$\left\langle \phi_{11} \phi_{22} \right\rangle$, $\left\langle \phi_{12}^2 \right\rangle$ & 2 & 2 & 0 & $\frac{\sigma_{0}^{2}}{15}$ \\
\hline
$\left\langle \phi_{1111} \phi_{11} \right\rangle$, $\left\langle \phi_{111}^2 \right\rangle$ & 6 & 0 & 0 & $\frac{\sigma_{1}^{2}}{7}$ \\
$\left\langle \phi_{1111} \phi_{22} \right\rangle$, $\left\langle \phi_{1112} \phi_{12} \right\rangle$, $\left\langle \phi_{112}^2 \right\rangle$ & 4 & 2 & 0 & $\frac{\sigma_{1}^{2}}{35}$ \\
$\left\langle \phi_{1123}\phi_{23} \right\rangle$, $\left\langle \phi_{1122}\phi_{33} \right\rangle$, $\left\langle \phi_{123}^2 \right\rangle$ & 2 & 2 & 2 & $\frac{\sigma_{1}^{2}}{105}$ \\
\hline
$\left\langle \phi_{11111} \phi_{111} \right\rangle$, $\left\langle \phi_{1111}^2 \right\rangle$ & 8 & 0 & 0 & $\frac{\sigma_{2}^{2}}{9}$ \\
$\left\langle \phi_{11111} \phi_{122} \right\rangle$, $\left\langle \phi_{1112}^2 \right\rangle$ & 6 & 2 & 0 & $\frac{\sigma_{2}^{2}}{63}$ \\
$\left\langle \phi_{11112} \phi_{222} \right\rangle$, $\left\langle \phi_{1111} \phi_{2222} \right\rangle$, $\left\langle \phi_{1122}^2 \right\rangle$
& 4 & 4 & 0 & $\frac{\sigma_{2}^{2}}{105}$ \\
$\left\langle \phi_{11112} \phi_{233} \right\rangle$, $\left\langle \phi_{1112}\phi_{1233} \right\rangle$, $\left\langle \phi_{1123}^2 \right\rangle$ & 4 & 2 & 2 & $\frac{\sigma_{2}^{2}}{315}$ \\
    \end{tabular}
    \tablefoot{
    The first column is an (incomplete) list of examples to give an idea of what kind of terms fall under the same combination of $n_1$, $n_2$ and $n_3$. All uneven terms are omitted, because they are zero and terms containing 0th and first derivatives are omitted, because the are not Galilei-invariant.
    }
\end{table}
\subsection{The distribution of the tidal tensor} \label{app:tidal_distribution}
From equation \eqref{eqn:covariance_potderiv} it can be seen that the covariance tensor of any derivatives of the potential has to be an isotropic tensor with full symmetry in all indices. Therefore, the covariance matrix of the tidal tensor is $\mat{C}_{\mat{T}} \in \mathds{V}_4$ so that it has to be proportional to $\mat{I}_4$ with a coefficient that we can read off from table \ref{tab:covariances}:
\begin{align}
    \mat{C}_{\mat{T}} &= \expect{\mat{T} \otimes \mat{T}} \nonumber \\
            &= \frac{\sigma^2}{5} \mat{J}_4 \,\,.
\end{align}
This tensor is not invertible, since e.g. the $T_{12}$ component is perfectly degenerate with the $T_{21}$ component. However, in such cases the distribution of $\mat{T}$ can still be inferred by constructing a generalized inverse $\mat{C}_{\mat{T}}^{+}$ which has the property
\begin{align}
    \mat{C} \mat{C}_{\mat{T}}^{+} \mat{C} &= \mat{C}_{\mat{T}} \label{eqn:pseudo_inverse}
\end{align}
It then holds 
\begin{align}
    p(\mat{T}) &= N \exp \left( -\frac{1}{2} \mat{T}^T \mat{C}_{\mat{T}}^+ \mat{T} \right)
\end{align}
To find the generalized inverse we may assume that $\mat{C}_{\mat{T}}^+ \in \mathds{V}_{22}$ so that we can make the Ansatz
\begin{align}
    \mat{C}_{\mat{T}}^{+} &= A_{22} \mat{J}_{22} + A_{2=2} \mat{J}_{2=2}
\end{align}
We can use the decomposition $\mat{J}_{4} = \frac{5}{9} \mat{J}_{22} + \frac{2}{3} \mat{J}_{2=2}$ (determined computationally) and evaluate equation \eqref{eqn:pseudo_inverse} term by term:
\begin{align}
    \mat{C}_{\mat{T}} \mat{C}_{\mat{T}}^{+} &= A_{22} \frac{\sigma^{2}}{9} (\mat{J}_{22} \mul{2} \mat{J}_{22}) + A_{2=2} \frac{2 \sigma^{2}}{15} (\mat{J}_{2=2} \mul{2} \mat{J}_{2=2}) \\ 
        &= 3 A_{22} \frac{\sigma^{2}}{9} \mat{J}_{22} + A_{2=2} \frac{2 \sigma^{2}}{15} \mat{J}_{2=2}\\
    \mat{C}_{\mat{T}} \mat{C}_{\mat{T}}^{+} \mat{C}_{\mat{T}} &= 9 A_{22} \left( \frac{\sigma^{2}}{9} \right)^2 \mat{J}_{22} + A_{2=2} \left(\frac{2 \sigma^{2}}{15} \right)^2 \mat{J}_{2=2}\\
     &\equiv \mat{C}_{\mat{T}} = \frac{\sigma^{2}}{9} \mat{J}_{22} + \frac{2 \sigma^{2}}{15} \mat{J}_{2=2}
\end{align}
where we have evaluated the appearing tensor products computationally. By identifying coefficients, we find $A_{22} = \frac{1}{\sigma^2}$ and $A_{2=2} = \frac{15}{2 \sigma^2}$
and therefore have the generalized inverse of the covariance matrix:
\begin{align}
    \mat{C}_{\mat{T}}^+ &= \frac{1}{\sigma^{2}} \mat{J}_{22} + \frac{15}{2 \sigma^{2}} \mat{J}_{2=2}
\end{align}
and the distribution of the tidal tensor is given by
\begin{align}
    p(\mat{T}) &= N \exp \left(-  \frac{2 \mat{T}^T  \mat{J}_{22} \mat{T}  + 15 \mat{T}^T  \mat{J}_{2=2} \mat{T} }{4 \sigma^2} \right) \\
               &= N \exp \left(-  \frac{2 \delta^2 + 15 K^2}{4 \sigma^2} \right) \,\,.
\end{align}

\subsection{The distribution of third derivatives} \label{app:distr_thirdderiv}
The third derivative tensor of the potential $\mat{S}$ has the covariance matrix
\begin{align}
    \mat{C}_{\mat{S}} &= \expect{\mat{S} \otimes \mat{S}} \\
            &= \frac{\sigma_1^2}{7} \mat{J}_6\\
            &= \frac{3 \sigma_{1}^{2}}{25} \mat{J}_{3-3} + \frac{2 \sigma_{1}^{2}}{35} \mat{J}_{3\equiv3} \,\,.
\end{align}
Analogously to the calculation in Section \ref{app:tidal_distribution}, we find the generalized inverse
\begin{align}
   \mat{C}_{\mat{S}}^+ &= \frac{3}{\sigma_1^{2}} \mat{J}_{3-3} + \frac{35}{2 \sigma_1^{2}} \mat{J}_{3\equiv3}
\end{align}
where we have again used numerical representations to evaluate products between isotropic tensors when evaluating equation \eqref{eqn:pseudo_inverse}.
Therefore, the distribution of $S$ is given by
\begin{align}
    p(\mat{S}) &= N \exp\left(-\frac{35  \mat{S} \mat{J}_{3 \equiv 3} \mat{S}  - 15 \mat{S} \mat{J}_{3-3} \mat{S}  }{4 \sigma_{1}^{2}}\right) \nonumber \\
               &=: N \exp\left(-\frac{35 S_{3 \equiv 3} - 15 S_{3-3} }{4 \sigma_{1}^{2}}\right)\,\,.
\end{align}
where in the last line we have defined some new scalar variables through the given contractions, where $S_{3-3} = (\nabla \delta)^2$.

\subsection{Joint distribution of second and fourth derivatives} \label{app:distr_fourthderiv}

We can write the covariance matrix of fourth and second derivatives as a block matrix
\begin{align}
    \mat{C}_{\mat{T}\mat{R}} &= \begin{pmatrix}
                        \expect{\mat{T} \otimes \mat{T}} & \expect{\mat{T} \otimes \mat{R}} \\
                        \expect{\mat{R} \otimes \mat{T}} & \expect{\mat{R} \otimes \mat{R}}
                    \end{pmatrix} \\ 
                    &= \begin{pmatrix}
                        \frac{\sigma^2}{5} \mat{J}_4 & \frac{\sigma_1^2}{7} \mat{J}_{6} \\
                        \frac{\sigma_1^2}{7} \mat{J}_{6} & \frac{\sigma_2^2}{9} \mat{J}_8
                    \end{pmatrix} \\
&= \begin{pmatrix}\frac{\sigma_{0}^{2}}{9} \mat{J}_{22} + \frac{2 \sigma_{0}^{2}}{15} \mat{J}_{2=2} & \frac{\sigma_{1}^{2}}{15} \mat{J}_{24} + \frac{4 \sigma_{1}^{2}}{35} \mat{J}_{2=4}\\\frac{\sigma_{1}^{2}}{15} \mat{J}_{42} + \frac{4 \sigma_{1}^{2} }{35} \mat{J}_{4=2} & \frac{\sigma_{2}^{2} }{25} \mat{J}_{44} + \frac{24 \sigma_{2}^{2} }{245} \mat{J}_{4=4} + \frac{8 \sigma_{2}^{2} }{315} \mat{J}_{4==4}\end{pmatrix}\,\,.
\end{align}
For its inverse we make the Ansatz
\begin{align}
    \mat{C}_{\mat{T}\mat{R}}^+ &= 
                    \begin{pmatrix}
                        A \mat{J}_{22} + B \mat{J}_{2=2} &  C\mat{J}_{24} + D\mat{J}_{2=4} \\
                        C\mat{J}_{42} + D\mat{J}_{4=2} & E 
                        \mat{J}_{44} + F \mat{J}_{4=4} + G \mat{J}_{4==4}
                    \end{pmatrix}
\end{align}
and we find the solution to its generalized inverse by solving
\begin{align}
    \mat{C}_{\mat{T}\mat{R}} = \mat{C}_{\mat{T}\mat{R}} \mat{C}_{\mat{T}\mat{R}}^+ \mat{C}_{\mat{T}\mat{R}}
\end{align}
with a computer algebra system and find
\begin{align}
\mat{C}_{\mat{T}\mat{R}}^+ &= \left[\begin{matrix}\frac{15 \sigma_{2}^{2} \mat{J}_{2=2}}{2 \sigma_{0}^{2} \sigma_{2}^{2} - 2 \sigma_{1}^{4}} + \frac{\sigma_{2}^{2} \mat{J}_{22}}{\sigma_{0}^{2} \sigma_{2}^{2} - \sigma_{1}^{4}} & - \frac{15 \sigma_{1}^{2} \mat{J}_{2=4}}{2 \sigma_{0}^{2} \sigma_{2}^{2} - 2 \sigma_{1}^{4}} - \frac{\sigma_{1}^{2} \mat{J}_{24}}{\sigma_{0}^{2} \sigma_{2}^{2} - \sigma_{1}^{4}}\\- \frac{15 \sigma_{1}^{2} \mat{J}_{4=2}}{2 \sigma_{0}^{2} \sigma_{2}^{2} - 2 \sigma_{1}^{4}} - \frac{\sigma_{1}^{2} \mat{J}_{42}}{\sigma_{0}^{2} \sigma_{2}^{2} - \sigma_{1}^{4}} & \frac{15 \sigma_{0}^{2} \mat{J}_{4=4}}{2 \sigma_{0}^{2} \sigma_{2}^{2} - 2 \sigma_{1}^{4}} + \frac{\sigma_{0}^{2} \mat{J}_{44}}{\sigma_{0}^{2} \sigma_{2}^{2} - \sigma_{1}^{4}} + \frac{315 \mat{J}_{4==4}}{8 \sigma_{2}^{2}}\end{matrix}\right]
\end{align}
The joint distribution of second and fourth derivatives of the potential can be written as
\begin{align}
    p(\mat{T}, \mat{R}) &= N \exp \left( - \frac{1}{2} \begin{pmatrix}
        \mat{T} & \mat{R} \end{pmatrix} \mat{C}_{\mat{T}\mat{R}}^+ \begin{pmatrix}
        \mat{T} \\ \mat{R} \end{pmatrix} \right) \,\,.
\end{align}

\subsection{Tidal estimator with spatial correction} \label{app:tidal_estim_with_corr}
We find the tidal estimator with spatial corrections of order two as 
\begin{align}
    b_{\mat{J}_{2=2}} &= \expectgal{\frac{1}{p(\mat{T}, \mat{R})} \frac{\partial^2 p(\mat{T}, \mat{R})}{\partial \mat{T} \partial \mat{T}} \mul{4} \frac{\mat{J}_{2=2}}{\norm{ \mat{J}_{2=2} }^2}} \\
                      &= \frac{15}{4 \sigma_{*}^{8}} \expectgal{3 K^2 \sigma_{2}^{4} + 6 \phi_{2=4} \sigma_{1}^{2} \sigma_{2}^{2} + 3 \phi_{4=4} \sigma_{1}^{4} - 2 \sigma_{2}^{2} \sigma_{*}^{4}} \label{eqn:bk2_o2_estimator_app}
\end{align}
where
\begin{align}
    \phi_{2=4} &:= \mat{T} J_{2=4} \mat{R} = \sum_{ij} \partial_i \partial_j \phi \partial_i \partial_j \delta - \frac{1}{3} \delta \nabla^2 \delta \\
    \phi_{4=4} &:= \mat{T} J_{4=4} \mat{R} = \sum_{ij} (\partial_i \partial_j \delta)^2 - \frac{1}{3} (\nabla^2 \delta)^2
\end{align}
While we do not list them here, we have verified that if we analogously evaluate the expressions for $b_{\mat{J}_{2}}$, $b_{\mat{J}_{4}}$, $b_{\mat{J}_{22}}$ and $b_{\mat{J}_{24}}$ that they are identical to the estimators for $b_1, b_L, b_2$ and $b_{\delta, L}$ respectively from equations \eqref{eqn:b1_o2} - \eqref{eqn:bdl_o2}.

\end{appendix}

\end{document}